\documentclass[aip, pop, amsmath,amssymb, reprint,floatfix]{revtex4-1}
\usepackage{color}
\usepackage[table]{xcolor} % FOR COLOR TABLE
\usepackage{graphicx}% Include figure files
\usepackage{dcolumn}% Align table columns on decimal point
\usepackage{bm}% bold math
\usepackage{psfrag}
\usepackage{epstopdf}
\usepackage{subcaption}
\usepackage{textcomp}
\usepackage{hyperref}
\usepackage{comment}
\usepackage{multirow}
\usepackage{amstext} % NEED FOR IOP

\usepackage{sidecap}
\newcommand{\Ip}{$I$}
\newcommand{\Bt}{$B$}
\newcommand{\betat}{$\beta_{T}$}
\newcommand{\betan}{$\beta_{N}$}
\newcommand{\qnf}{$q_{95}$}
\newcommand{\elong}{$\kappa$}
\newcommand{\triavg}{$\left| \delta_{avg} \right|$}
\newcommand{\PoverPLH}{$P_{net}/P_{LH08}$}
\newcommand{\PLH}{$P_{LH08}$}
\newcommand{\Ptot}{$P_{tot}$}
\newcommand{\Pech}{$P_{ECH}$}
\newcommand{\Pnet}{$P_{net}$}
\newcommand{\Poh}{$P_{OH}$}
\newcommand{\Paux}{$P_{aux}$}
\newcommand{\Prad}{$P_{rad}$}
\newcommand{\Pfus}{$P_{fus}$}
\newcommand{\Tinj}{$T_{inj}$}
\newcommand{\IaB}{$IaB$}
\newcommand{\In}{$I/aB$}
\newcommand{\lawson}{$\left<p \right>$$\tau$}
\newcommand{\snyder}{$\left<p \right>$$\tau/IaB$}
\newcommand{\HH}{$H_{H98y2}$}
\newcommand{\HL}{$H_{L89}$}
\newcommand{\GH}{$G_{H98y2}$}
\newcommand{\GL}{$G_{L89}$}
\newcommand{\Galt}{$G_{alt}$}
\newcommand{\pres}{$\left<p \right>$}
\newcommand{\nustar}{$\nu_e^*$}
\newcommand{\taue}{$\tau$}
\newcommand{\pped}{$p_{ped}$}
\newcommand{\peped}{$p_{e,ped}$}
\newcommand{\ppedratio}{$p_{ped}$/$\left<p\right>$}
\newcommand{\noverng}{$\left<n_e \right>$/$n_{G}$}

\newcommand{\nsepoverng}{$n_{e,sep}$/$n_{G}$}
\newcommand{\navg}{$\left<n_e \right>$}
\newcommand{\nsep}{$n_{e,sep}$}
\newcommand{\nped}{$n_{e,ped}$}
\newcommand{\Zeff}{$Z_{eff}$}
\newcommand{\frad}{$f_{rad}$}
\newcommand{\Vcore}{$V_{core}$}

\newcommand{\echratio}{$P_{ECH}/P_{tot}$}

\newcommand{\rmp}{RMP}
\newcommand{\QH}{QH}
\newcommand{\EDA}{EDA-H}
\newcommand{\negd}{Neg-D}%$\delta$}
\newcommand{\Imode}{I-mode}
\newcommand{\Lmode}{L-mode}
\newcommand{\Hmode}{H-mode}

\begin{document}

% ----IOP CHANGE #4 -----
% ---- IOP FORMATTING ----
\begin{comment}

\title[]{Plasma Performance and Operational Space without ELMs in DIII-D} % Invert?

\author{C. Paz-Soldan$^{1,2}$ and the DIII-D Team}

\address{$^1$General Atomics, San Diego, CA 92186-5608, USA}
\address{$^2$Department of Applied Physics \& Applied Mathematics, Columbia University, New York, New York 10027, USA}

%\reprint{}
\end{comment}

% ---- APS -------

%\begin{comment}
\preprint{}

\title[]{Plasma Performance and Operational Space without ELMs in DIII-D} % Invert?

\author{C. Paz-Soldan}
\affiliation{General Atomics, San Diego, California 92186-5608, USA}
\affiliation{Department of Applied Physics \& Applied Mathematics, Columbia University, New York, New York 10027, USA}
\email{carlos.pazsoldan@columbia.edu}

\author{the DIII-D team}

%\end{comment}

% ----- REST OF DOCUMENT ----

\date{\today}

\begin{abstract}
A database of DIII-D plasmas without edge-localized modes (ELMs) compares the operating space and plasma performance of stationary no-ELM regimes found in conventional tokamaks: ELM suppression with resonant magnetic perturbations (\rmp{}s), quiescent H-mode (QH, including wide-pedestal variant), improved confinement mode (\Imode{}), enhanced D-alpha H-mode (\EDA{}), conventional L-mode, and negative triangularity L-mode (\negd{}). Operational space is documented in terms of engineering and physics parameters, revealing divergent constraints for each regime. Some operational space discriminants (such as pedestal collisionality) are well known, while others, such as low torque \& safety factor, or high power \& density, are less commonly emphasized. Normalized performance (confinement quality and normalized pressure) also discriminate the no-ELM regimes and favor the regimes tolerant to power in DIII-D: \rmp{}, \QH{}, and \negd{}. Absolute performance (volume-averaged pressure \pres{}, confinement time \taue{}, and triple product \lawson{}) also discriminates no-ELM regimes and is found to rise linearly with \IaB{} (a metric for the magnetic configuration strength, the product of current \Ip{}, minor radius $a$, and field \Bt{} that has units of force), and also benefits from tolerance to power.  The highest normalized performance using the metric \snyder{} is found in \QH{} and \rmp{} regimes. Focusing on ITER-shaped no-ELM plasmas, Q=10 at 15 MA scaled global performance is met with two out of four considered metrics, but only thus far at high torque. Though comparable \QH{} and \rmp{} performance is found, the pedestal pressure (\pped{}) is very different. \pped{} in \rmp{} plasmas is relatively low, and the best performance is found with a high core \pres{} fraction alongside high core rotation, consistent with an ExB shear confinement enhancement. \pped{} of \QH{} plasmas is significantly higher than \rmp{}, and \QH{} performance does not correlate with core rotation. However, the highest \QH{} \pped{} are found with high carbon fraction. While normalized performance of \negd{} plasmas is comparable to \QH{} and \rmp{} plasmas, the absolute confinement of \negd{} is lower, owing to low elongation and low \pped{} achieved thus far. Considering integration with electron cyclotron heating (ECH), the operational space for \rmp{} and \QH{} plasmas narrow, while that for \EDA{} plasmas open, and the high \pped{}/\pres{} regimes (\EDA{} and \QH{}) preserve the highest performance. Only the \EDA{}, \negd{}, and \Lmode{} scenarios have approached divertor-compatible high separatrix density conditions, with \negd{} preserving the highest performance owing to its compatibility with both high power and density. Comparison to highlighted ELMing plasmas finds a clearer pedestal correlation to plasma performance, but also reveals that the peak performance of plasmas without ELMs is significantly lower in DIII-D, owing to limits in operational space accessed so far without ELMs.
\end{abstract}

%\pacs{52.30.Cv, 52.35.Py, 52.55.Tn, 52.55.Fa} % GOOD PACS
%\keywords{plasma, ELM, lawson, performance}

\maketitle

%%%%%%%%%%%%% %%%%%%%%%%%%%%%%%%%%%%%%%%%%%%%%% %%%%%%%%%%%%%%%%% %%%%%%%%
%%%%% %%%%%%%%%%%%%  %%%%%%%%%%% %%%%%%% MAIN TEXT%%%%% %%%%%%%%%%%%%% %%%%%%%%% 
%%%%%%%% %%%%%%%%%%% %%%%%%%%% %%%%%%%%%%%% %%%%%%%%%%%%%%%%%%%% %%%%%%%%%%

% ---- ---- ---- ---- -------- ----  ---- ---- ---- ---- ---- ---- ---- ---- ----
% ---- ---- ---- ---- -------- ----  INTRODUCTION ---- ---- ---- ---- ---- ----
% ---- ---- ---- ---- -------- ----  ---- ---- ---- ---- ---- ---- ---- ---- ----

\section{Introduction and Motivation}
\label{sec:intro}

A grand challenge for the tokamak approach to fusion energy production is the achievement of a high-performance fusion core integrated with a boundary solution compatible with available first-wall materials. A key issue at this interface is the edge-localized mode (ELM), an instability arising from edge gradients which if uncontrolled delivers an impulsive heat and particle load to the first-wall \cite{Zohm1996,Leonard2014}. Above a threshold size, the ELM not only impacts the integrity and lifetime of plasma-facing components \cite{Loarte2003a,Federici2003,Dux2011,Pitts2013} but also imposes limits on the tokamak operating space (and achievable plasma performance) by driving excessive material influx to the plasma\cite{Kallenbach2005,Beurskens2014,Kim2018}. Developing scenarios and techniques to mitigate or eliminate the ELM is thus a topic of well-recognized importance in the fusion community. Addressing the ELM challenge has also motivated significant actuator investment in upcoming fusion-grade tokamak projects such as ITER \cite{Loarte2014,Maingi2014,Garofalo2020}.

% NEW PARAGRAPH HERE FURTHER MOTIVATING?
In mid-scale tokamaks with carbon walls (such as DIII-D) the operating space is not generally limited by the ELM. As such scenario development work often proceeds without enforcing a constraint on ELM virulence. Peak plasma performance in DIII-D is found in ELMing plasmas \cite{Lazarus1997,Luce2003,Snyder2019}. This is perhaps natural, as the ELM is a pressure-limiting instability. In parallel, several regimes without ELMs have been discovered and advanced - all of which can be accessed in DIII-D. These regimes are described in Sec. \ref{sec:regimes} and summarized in Tab. \ref{tab:regimes}. Though detailed studies of individual no-ELM regime access, sustainment, and performance against various parametric dependencies abound (see Tab. \ref{tab:regimes}), little work exists in the literature to allow systematic comparison of the performance and access of the various no-ELM regimes against both each other and their ELMing counterparts.

\begin{table*}
  \begin{center}
    \begin{tabular}{|c|c|c|c|c|c|c|c|c|} % <-- Alignments: 1st column left, 2nd middle and 3rd right, with vertical lines in between
      \hline
      \textbf{Regime} & \textbf{AUG} & \textbf{C-Mod}  & \textbf{DIII-D} &  \textbf{EAST} & {    }\textbf{JET}{    } & \textbf{JT-60U} &\textbf{KSTAR} & \textbf{DB Color} \\
      \hline
      \textbf{\EDA{}} & \onlinecite{Gil2020} & \onlinecite{Greenwald2000} & \onlinecite{Mossessian2003} & \onlinecite{Sun2019} & \onlinecite{Stober2005}$^{\dagger{}*}$ & N/A & N/A & \cellcolor{yellow}   \\
      \textbf{\Imode{}} & \onlinecite{Ryter1998}$^\dagger{}$ \onlinecite{Happel2017},\onlinecite{Ryter2017}& \onlinecite{Whyte2010} & \onlinecite{Marinoni2015} & \onlinecite{Liu2020b} & \onlinecite{Andrew2004}$^{\dagger{}*}$ & N/A & N/A & \cellcolor{green}    \\
      \textbf{\negd{}} & \onlinecite{Happel2020} & N/A & \onlinecite{Austin2019} & N/A & N/A & N/A & N/A & \cellcolor{red}   \\
      \textbf{\QH{}} & \onlinecite{Suttrop2003}$^\dagger{}$ & N/A & \onlinecite{Greenfield2001} & \onlinecite{Qian2020IAEA} & \onlinecite{Suttrop2005}$^\dagger{}$ & \onlinecite{Sakamoto2004} & \onlinecite{Lee2020} & \cellcolor{black}   \\
      \textbf{\rmp{}} & \onlinecite{Nazikian2016IAEA},\onlinecite{Suttrop2018} & N/A & \onlinecite{Evans2004} & \onlinecite{Sun2016} & N/A & N/A & \onlinecite{Jeon2012} & \cellcolor{blue}   \\
      \hline
    \end{tabular}
 % TCV?
 \vspace{-5 pt}
    \caption{No-ELM regime access in MA-class conventional tokamaks. $^\dagger{}$: access prior to metal wall installation. $^{*}$: non-stationary or ambiguous access. DB Color indicates the icon color used in subsequent figures, with L-modes also shown in magenta.}
    \label{tab:regimes}
  \end{center}
\end{table*}

% GOALS
This work is thus motivated to fill this gap and more fundamentally to understand key stationary plasma performance drivers {\it{with a no-ELM constraint}}. The DIII-D tokamak has nearly two decades of experience in accessing and understanding the various stationary no-ELM regimes listed in Sec. \ref{sec:regimes}. Notwithstanding this, no comparative study has been reported that directly compares the observed operating space and plasma performance achieved thus far in each regime (beyond reporting the normalized confinement quality H-factor\cite{Fenstermacher2013}). At a minimum, this work seeks to provide documentation of what has been achieved in DIII-D and what gaps remain, using easily evaluated metrics that can be compared between tokamaks. A related goal of this study is to facilitate comparative assessment of no-ELM plasma performance (in DIII-D and beyond), moving beyond simply reporting the H-factor. Additionally, a goal is to provide relative assessment of gaps to a viable reactor scenario for all no-ELM regimes, both in terms of performance and integration. An investigation of the possible performance penalty associated with removing the ELM is also desired, here achieved by comparing the performance of ELMing and no-ELM plasmas. A final goal is to provide a foundation from which future extensions of this work to other tokamaks can be performed. As will be discussed, extrapolation to reactor conditions is best achieved by considering the combined experience of several tokamaks, as present-day devices can each only reach a subset of the expected operating conditions of a reactor.

\subsection{Stationary Tokamak Operating Regimes without ELMs}
\label{sec:regimes}
% Describe regimes

The no-ELM regimes considered in this work are briefly described in this section. It is not within the scope of this work to present a detailed discussion of the pedestal stability, micro-turbulence modes, and resultant understanding of each no-ELM regime.  This task is well-covered in Refs. \onlinecite{Oyama2006,Evans2015,Viezzer2018}. In alphabetical order, the regimes are the:
\begin{itemize}

\item The {\it{Enhanced D-alpha H-mode}} (\EDA{}) \cite{Greenwald1997,Greenwald1999a, Greenwald2000}, a regime possessing a fully formed density and temperature pedestal, typified by a ``quasi-coherent mode'' electro-magnetic fluctuation in the $\mathcal{O}$(100 kHz) range that drives additional transport\cite{LaBombard2014,Golfinopoulos2018}.

\item The {\it{Improved energy confinement mode}} (\Imode{}) \cite{Ryter1998,Whyte2010,Hubbard2016}, a regime with a temperature but not a density pedestal, and improved energy confinement but not particle confinement as compared to \Lmode{} levels \cite{Hubbard2017}. A 100 kHz range ``weakly-coherent mode'' coupled to a geodesic acoustic mode (GAM) is ubiquitously observed that drives additional transport\cite{Cziegler2013,Manz2015}.

\item Operation with a {\it{Negative Triangularity}} plasma shape (\negd{})\cite{Moret1997,Camenen2007,Austin2019,Marinoni2019}. \negd{} plasmas have an \Lmode{} edge (no or only partial pedestal) but the shaping yields improved core turbulent transport \cite{Merlo2015} and suppresses the \Hmode{} transition to powers well above that expected by established scalings \cite{Marinoni2020IAEA,Martin2008}. 

\item The {\it{Quiescent H-mode}} (\QH{})\cite{Greenfield2001,Burrell2001}, a regime with fully formed density and temperature pedestal. \QH{} plasmas are typified by either a coherent electromagnetic ``edge-harmonic oscillation'' (EHO)\cite{Chen2016,Wilks2018,Pankin2020} or an incoherent ``broadband MHD'' oscillation\cite{Guo2015} both in the $\mathcal{O}$(10s kHz) range that drive additional transport. A recently emphasized broadband MHD \QH{} variant, the {\it{wide-pedestal \QH{}-mode}}\cite{Burrell2016,Chen2020} is also included. As compared to standard \QH{} plasmas, wide-pedestal \QH{}-modes are found to have more favorable access and performance qualities at low torque and with dominant electron heating \cite{Ernst2016,Ernst2018IAEA}.

\item ELM suppression via {\it{Resonant Magnetic Perturbations}} (\rmp{})\cite{Evans2004,Evans2006,Jeon2012,Kirk2013c,Kirk2015,Sun2016,Suttrop2018}. Application of $\approx$0.1\% non-axisymmetric fields from nearby coils maintain a fully formed pedestal yet prevent the ELM, but only in narrow ``resonant windows'' of the edge safety factor. Additional transport is provided by: parallel transport across macroscopic pedestal-top magnetic islands arising from a large degree of island formation \cite{Waelbroeck2009,Nazikian2015,Hu2019,Fitzpatrick2020a}; increased turbulent fluctuation levels directly arising from field penetration \cite{McKee2013,Lee2016,Lee2019}; increased turbulence arising indirectly from penetration-induced changes to the radial electric field structure \cite{Sung2017b,Taimourzadeh2019}; or a combination of these effects. The additional transport has also been proposed to be due to mechanisms such as neoclassical effects \cite{Callen2012b,Huijsmans2015} or localized peeling-ballooning instabilities driven by 3D equilibrium modifications \cite{Bird2013,Willensdorfer2017,Ryan2019,Kim2020a} which do not require significant field penetration.

\end{itemize}

For completeness the default {\it{Low-confinement}} \Lmode{} is also included in the study, though it is not listed in Tab. \ref{tab:regimes}. Note all no-ELM regimes are subject to enhanced edge transport, either via enhanced turbulent fluctuations (\Lmode{}, \negd{}, \QH{}, \rmp{}), (quasi-)coherent fluctuations (\EDA{}, \Imode{}, \QH{}), or imposed laboratory-frame non-axisymmetry (\rmp{}). Furthermore only \negd{} and \rmp{} impose well-defined design considerations on the tokamak: the former dictates the overall tokamak shape, and the latter requires inclusion of nearby non-axisymmetric coils (as is planned for ITER \cite{Evans2013,Daly2013}).

% ---- ---- ---- ---- -------- ----  ---- ---- ---- ---- ---- ---- ---- ---- ---- ----
% ---- ---- ---- ---- -------- ----  Definition Old Appendix   ---- ---- ---- ---- ---- ----
% ---- ---- ---- ---- -------- ----  ---- ---- ---- ---- ---- ---- ---- ---- ---- ----

%\subsection{Database Criteria and Caveats}
%\label{sec:dbdef}

\subsection{Criteria for Inclusion in Database}
\label{sec:criteria}
% --- Define Database ---
This work presents a study of specific intervals within DIII-D plasmas. In order to define the intervals used to populate this database, several selection criteria were applied, enumerated here:
\begin{enumerate}
\item Intervals must not have ELMs of any type. This excludes even some regimes where the ELM may be benign. However, as extrapolation of ELM heat loads is a challenging topic, this can be considered the most stringent criteria for a viable ELM solution.  Future work may expand this study to candidate benign ELM regimes, such as the ``grassy-ELM'' regime \cite{Oyama2006,Viezzer2018,Xu2019}, and to ELMs mitigated via RMPs or high-frequency pellet pacing \cite{Loarte2014}.
\item Intervals must be stationary, as defined by the rate of rise of the average density (\navg{}) being below $1.5\times10^{19}/s$. This threshold is somewhat arbitrary but is functionally effective for DIII-D, with little dependence in key metrics seen below this value. This criteria also served to exclude ELM-free phases.
\item Intervals must at least 300 ms long (about two energy confinement times, \taue{}). Longer durations (up to several seconds) are utilized if stationary conditions are maintained.
\item All main control parameters (shape, \Ip{}, \Bt{}) must be fixed during the interval. All intervals either fix total power, or regulate fixed pressure via power modulation. Gas or density may evolve, according to the stationarity condition described above. Discharges with ramping down \Bt{} to increase the off-axis current drive are excluded.
\item The plasma cross-section or near-SOL must not interfere with present day divertor baffles, allowing present-day discharge reproduction in principle. This limited the study to DIII-D shot numbers above 100700 (c. 2000) \cite{Reis1996,Bozek1999}.  Some discharges prior to shot 127300 (c. 2005) are incompatible with the present-day lower DIII-D divertor and thus are excluded. This criteria only excluded very early high-elongation \QH{} plasmas \cite{Burrell2001} and early low triangularity \rmp{} discharges \cite{Evans2006a}.
\item The main ion species must be deuterium.
\end{enumerate}
All datapoints in all figures come with a shot number in small font, enabling single discharges to be followed throughout this paper.

\subsection{Caveats}
\label{sec:caveats}
% --- Caveats --- (Too long?)
Several caveats of this study deserve up front mention:
\begin{enumerate}
\item DIII-D experimental run-time awarded to explore some regimes (namely \rmp{} and \QH{}) vastly exceeds that awarded to others (namely \EDA{}, \Imode{}, and \negd{}). Further, efforts in some regimes have been more focused on plasma performance than others, and high-performance is more commonly pursued in ELMing regimes. In addition, the specifics of the DIII-D tokamak (such as covariance of torque and power at high power) can selectively limit performance and access in ways that are not generic. Simply, this is a snapshot of ongoing work on a single tokamak, and with a sampling bias. 

%Future discharges may also surpass the operational limits or plasma performance presented herein. Indeed, this should be a welcome development.
\item This caveat concerns the completeness of the database, which was assembled through a combination of manual and automated techniques. Automated techniques were useful in identifying discharges near limits of (easily definable) engineering parameters. However, owing fundamentally to the difficulty of robustly separating a stationary no-ELM state from either small-ELM plasmas or transient ELM-free states, the bulk of the database compilation was done manually. This included consulting with relevant experts, compiling existing single-regime databases, surveying the published literature, and re-analyzing sessions targeting a specific operational direction. As such, while a best effort was made, it is impossible to guarantee that a relevant discharge was not missed.
\item ELMs were often present in discharges preceding (or following) the selected time interval. As such, the presented time intervals may not be accessible without ELMs earlier in the discharge evolution. As such they might not be realizable in a tokamak that cannot tolerate any ELMs. The fully formed pedestal regimes (\QH{}, \rmp{}, \EDA{}) are especially prone to spurious ELMs, while the no or partial pedestal regimes (\negd{}, \Imode{}, \Lmode{}) are robustly without ELMs.
\item Correlation is not causation, a statement which is especially salient when considering a database study. No trends observed in the database should be considered causal based on these findings alone. Further, apparent correlations in the database can contradict controlled scans which isolate variation to a single parameter.
\item A general word of caution must be expressed regarding extrapolation of results from a mid-scale tokamak. For reasons described throughout the text, these results do not directly map to burning plasma conditions. Notwithstanding this, the results shown provide ample inspiration for future targeted experiments (emphasizing model validation) and also highlight key gaps for further multi-machine study.
\end{enumerate}

 %Important conclusions and their impact on extrapolation to a reactor will be discussed in Sec. \ref{sec:disc}.

% -- Structure ----
\subsection{Structure of Paper}
The structure of this paper is as follows: the basic operating space of stationary no-ELM plasmas is discussed in Sec. \ref{sec:ops}. The achieved plasma performance (as measured by a variety of metrics) is then presented in Sec. \ref{sec:perf}. Correlations with plasma performance in each regime are given in \ref{sec:corr}. A comparison of no-ELM to ELMing plasma performance then follows in Sec. \ref{sec:ELMs}, repeating key figures including high performance ELMing plasma datapoints. No-ELM plasma integration considerations with electron heating and dissipative divertor are then shown in Sec. \ref{sec:integ}. Discussions and conclusions follow in Sec. \ref{sec:disc}. Appendix \ref{sec:metrics} compares performance metrics against eachother.

% ---- ---- ---- ---- -------- ----  ---- ---- ---- ---- ---- ---- ---- ---- ----
% ---- ---- ---- ---- -------- ----    OP SPACE   ---- ---- ---- ---- ---- ----
% ---- ---- ---- ---- -------- ----  ---- ---- ---- ---- ---- ---- ---- ---- ----

\begin{figure*}
\begin{subfigure}{1\textwidth}
\begin{subfigure}{0.325\textwidth}
\centering
\includegraphics[width=1\textwidth]{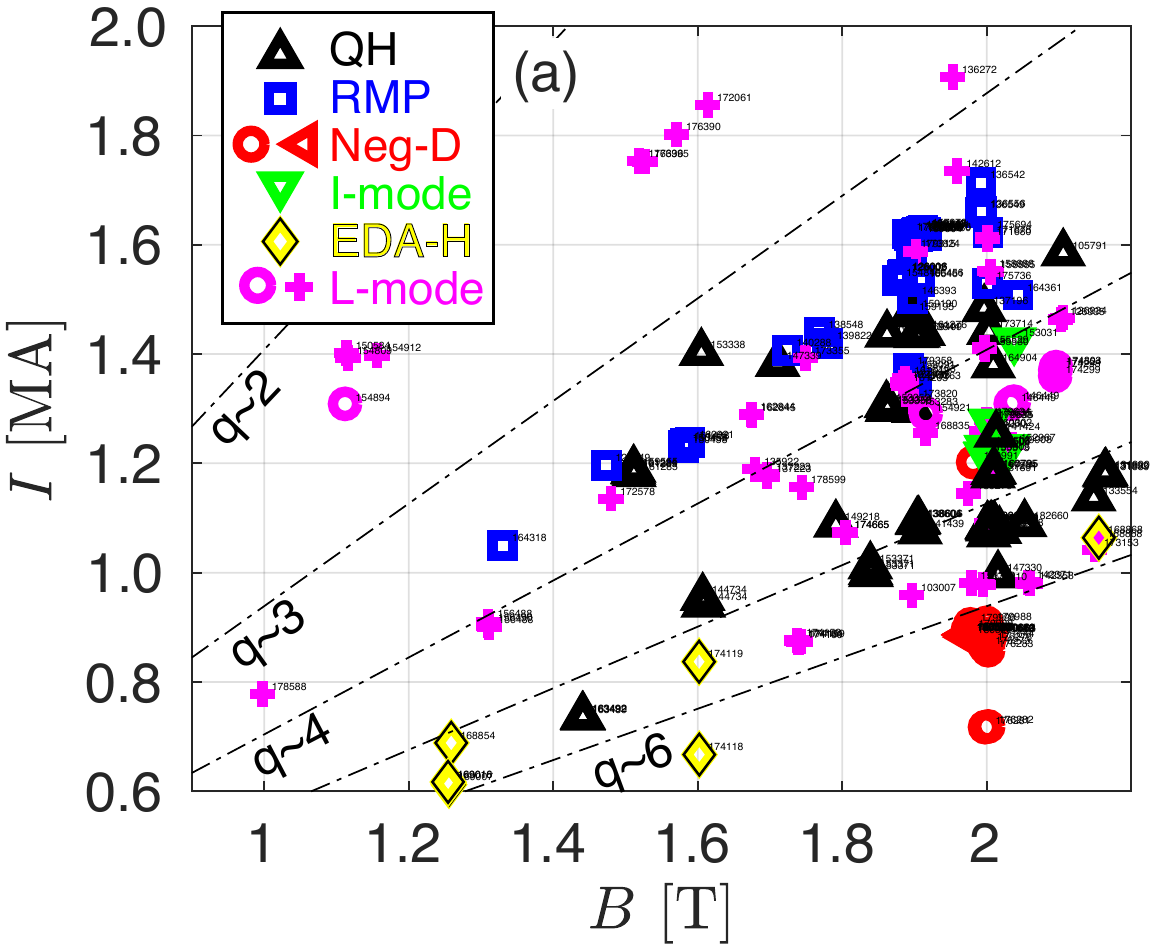}
\end{subfigure}
\begin{subfigure}{0.325\textwidth}
\centering
\includegraphics[width=1\textwidth]{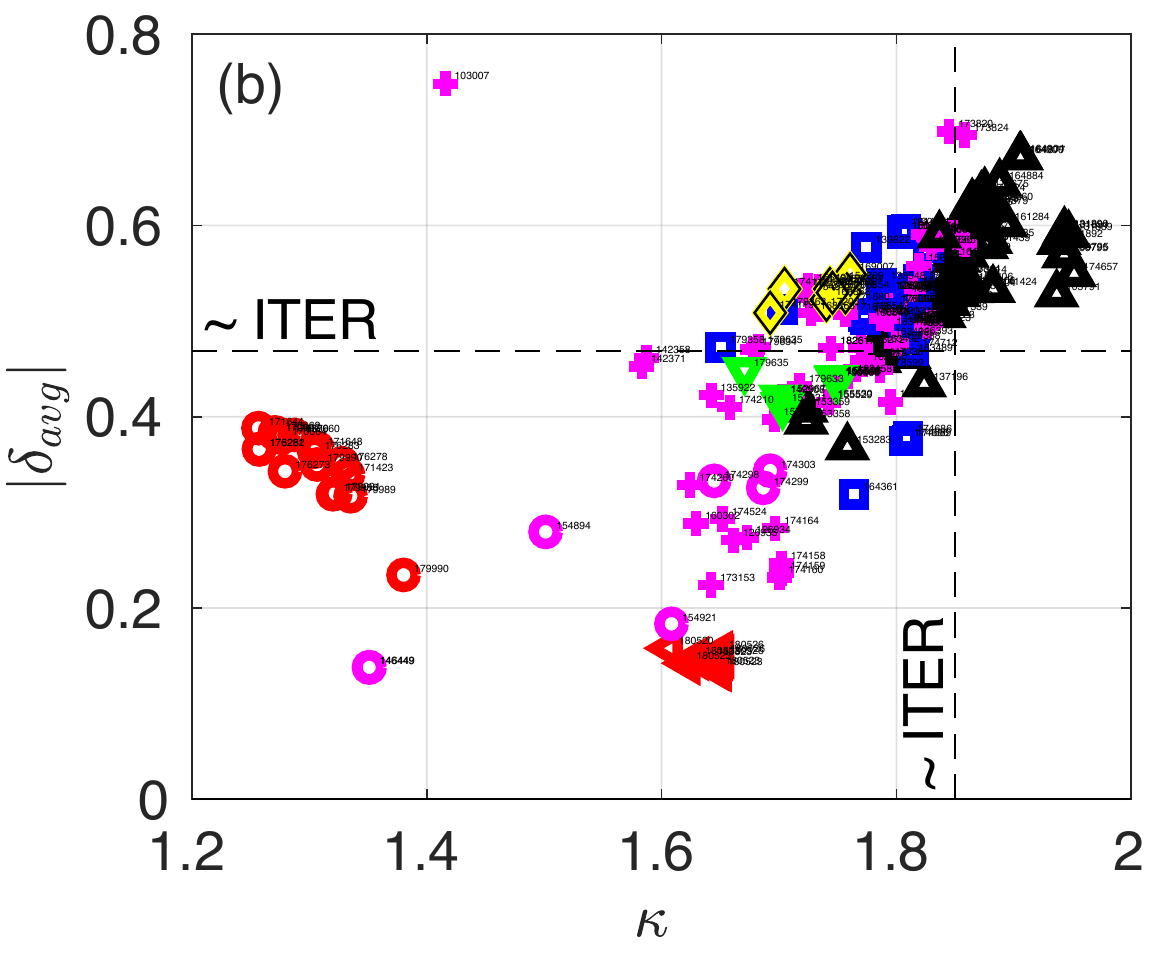}
\end{subfigure}
%\vspace{-5 pt}
\begin{subfigure}{0.325\textwidth}
\centering
\includegraphics[width=1\textwidth]{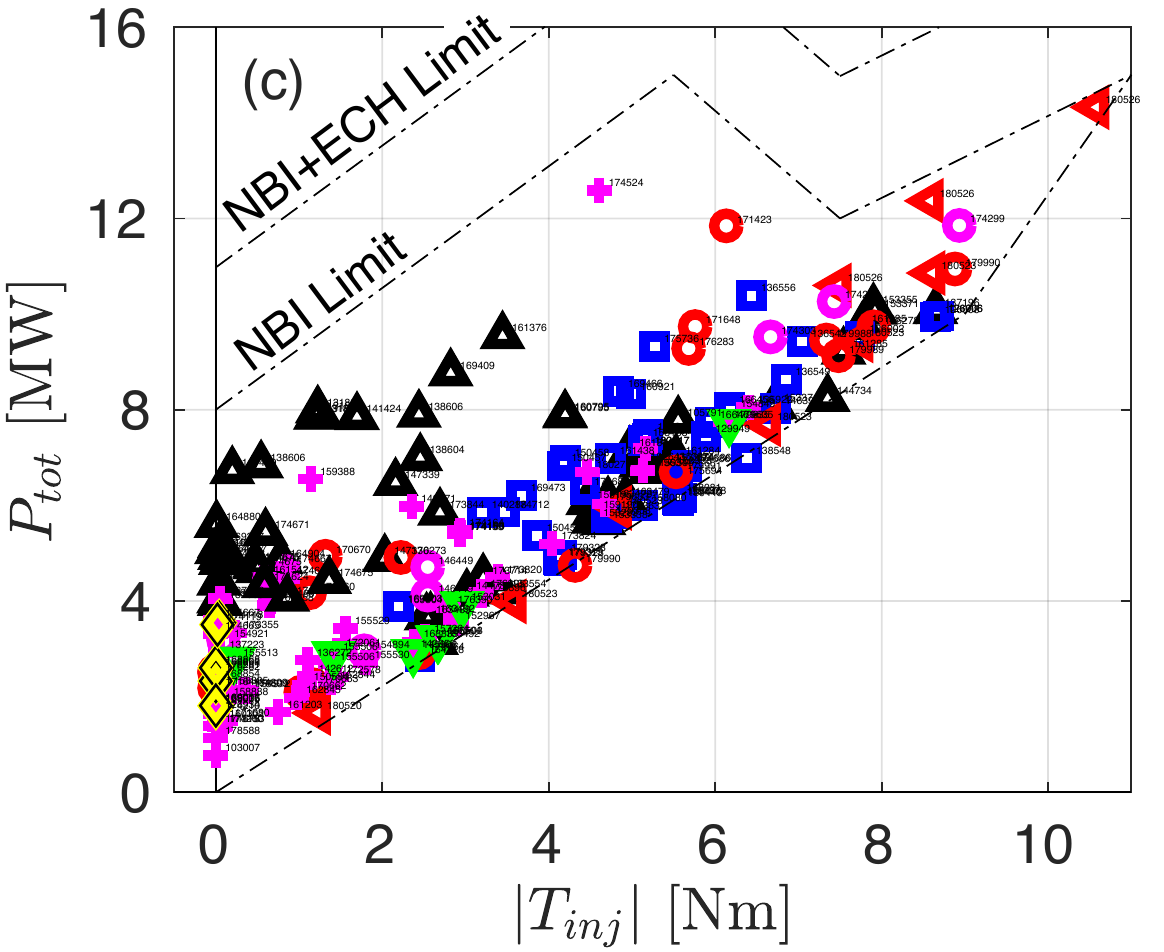}
\end{subfigure}
\begin{subfigure}{0.325\textwidth}
\centering
\includegraphics[width=1\textwidth]{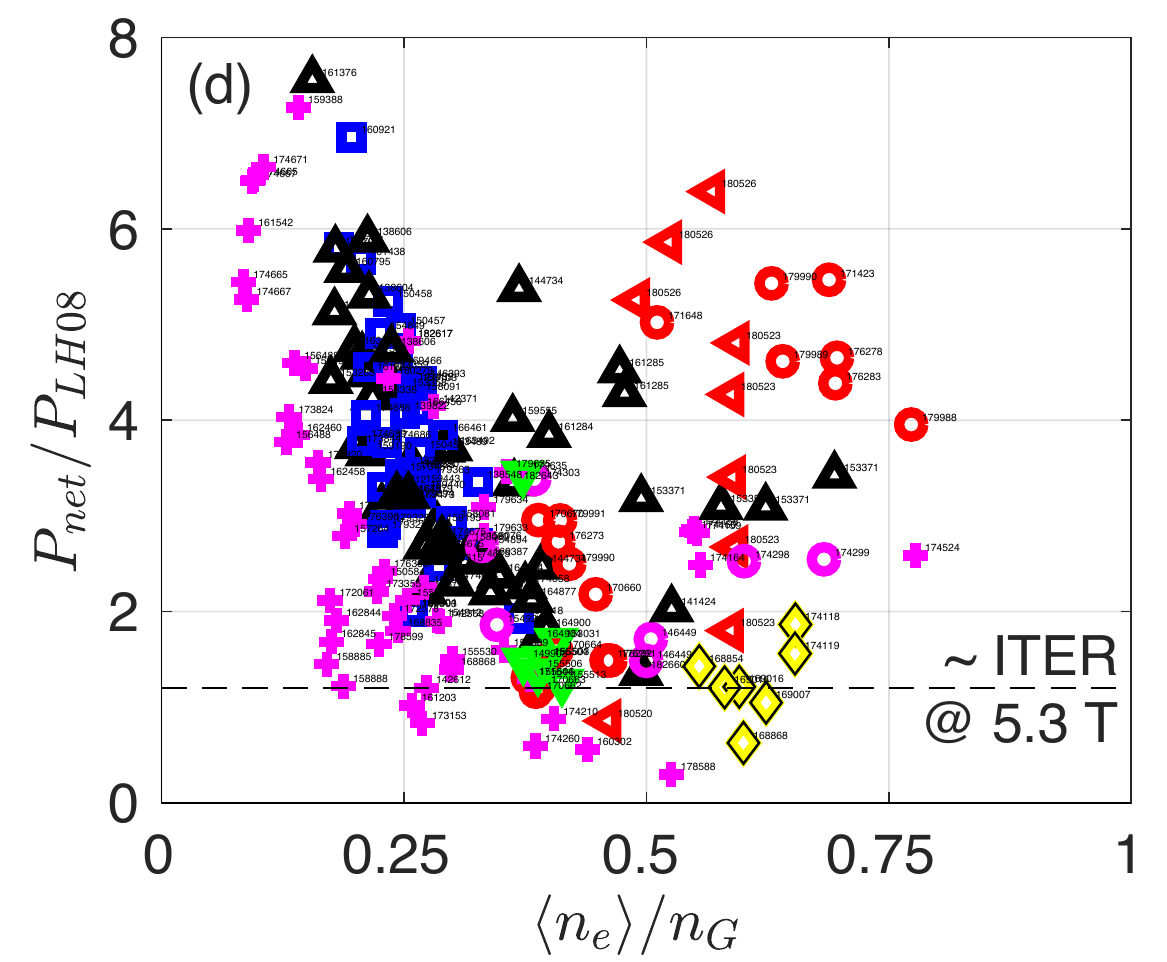}
\end{subfigure}
\begin{subfigure}{0.325\textwidth}
\centering
\includegraphics[width=1\textwidth]{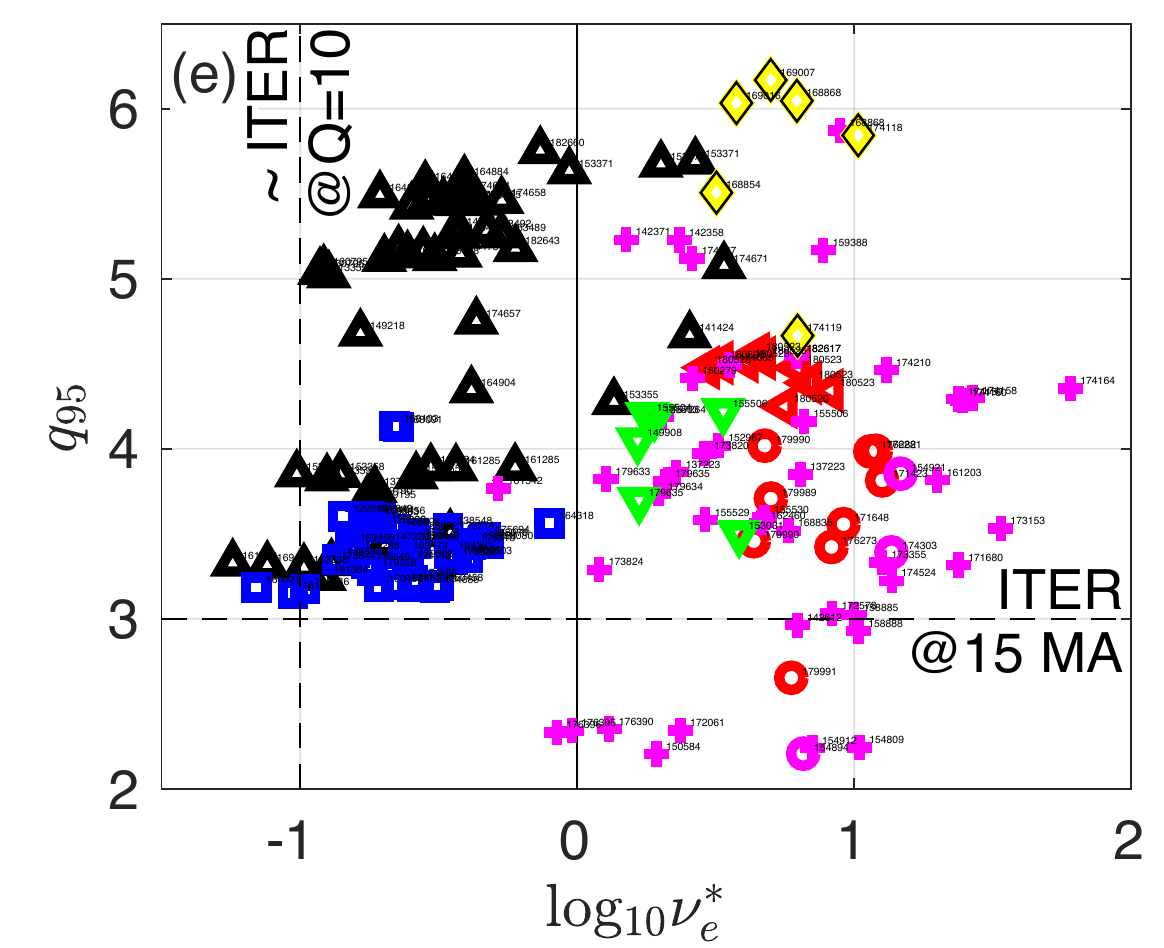}
\end{subfigure}
\begin{subfigure}{0.325\textwidth}
\centering
\includegraphics[width=1\textwidth]{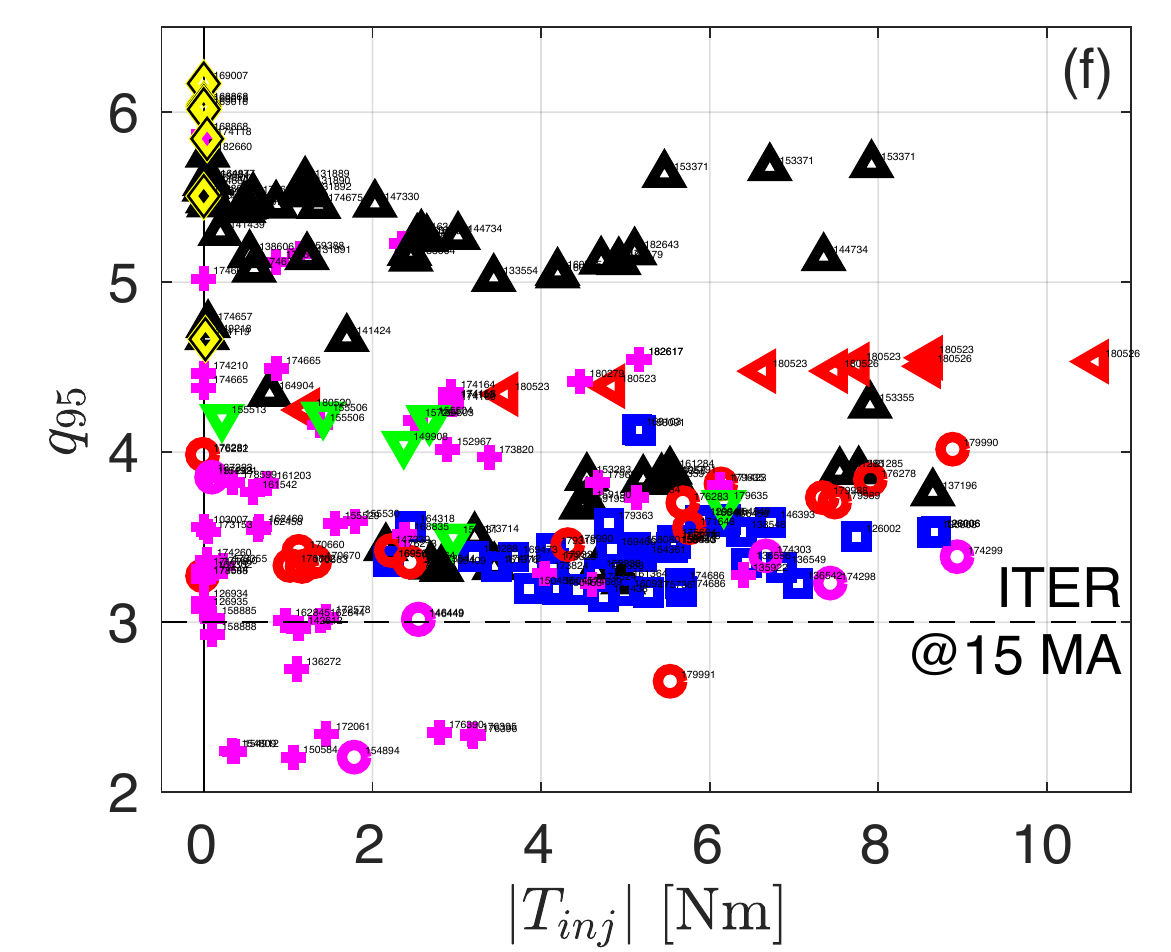}
\end{subfigure}
\end{subfigure}
 \vspace{-5 pt}
\caption{DIII-D no-ELM operational space in terms of basic machine and plasma parameters:  (a) Toroidal current (\Ip{}) and toroidal magnetic field (\Bt{}), dash-dot lines indicate approximately constant \qnf{}. (b) elongation (\elong{}) and absolute average triangularity (\triavg{}), (c) total power (\Ptot{}) to injected torque (\Tinj{}), dash-dot lines indicate approximate range accessible to the NBI system, (d) ratio of line-average density to the Greenwald density (\noverng{}) and ratio of net power (subtracting core radiative losses) to scaled L-H threshold power (\PoverPLH{}), (e) edge collisionality (\nustar{}) and edge safety factor (\qnf{}), (f) \Tinj{} and \qnf{}. ITER baseline targets are indicated if applicable. In these and all figures, different colors/symbols are used for each regime. Limiter plasmas are given a circular symbol.}
\label{fig:basics}
\end{figure*}

\section{Operational Space}
\label{sec:ops}

Stationary plasmas without ELMs are found only in a subset of the accessible DIII-D operating space. Figure \ref{fig:basics} presents the no-ELM operational space against a variety of parameters. In some axes, the operating space is defined by the return of the ELM, while in others, it is defined by hardware, and for some regimes it is simply unexplored. Consideration of the operating space is essential as it directly relates to the achievable absolute plasma performance (though not necessarily the normalized performance).

% IP and BT
\paragraph*{Plasma current (\Ip{}) and toroidal field (\Bt{}):} Operating space in terms of \Ip{} and \Bt{} is shown in Fig. \ref{fig:basics}(a). This most basic representation already reveals important features. \rmp{} plasmas are found only in a narrow range of \qnf{} from 3-4, owing to well known resonant window requirements \cite{Evans2008,Suttrop2018,Hu2020b}, though other devices have found \rmp{} resonant windows in different ranges of \qnf{} \cite{In2019,Sun2016}. Note \rmp{} plasmas at \qnf{}=7 reported in Ref. \cite{Fenstermacher2008} are found to have small residual ELMs and thus do not meet the criteria discussed in Sec. \ref{sec:criteria}.  \QH{} plasmas are found across a wide range of \Ip{} and \Bt{}. \Lmode{} plasmas are in principle found for all \Ip{} and \Bt{}, with only a subset presented. Notable are \Lmode{} discharges operating with \qnf{} just above 2 \cite{Piovesan2014,Hanson2014}. The less explored regimes of \EDA{}, \negd{}, and \Imode{} have as yet only been observed in a subset of \Ip{} \& \Bt{} operational space, though DIII-D findings with \EDA{} and \Imode{} are consistent with those reported in other devices \cite{Greenwald2000,Hubbard2017,Gil2020}.

%  Shape
\paragraph*{Plasma shaping:} The DIII-D shaping flexibility as manifest in the no-ELM operating space is shown in Fig. \ref{fig:basics}(b), represented as absolute average triangularity (\triavg{}) and elongation (\elong{}). \QH{} plasmas are found to operate over a wide range of shapes, including the most strongly shaped plasmas. \rmp{} plasmas are not tolerant of the strongest shapes \cite{Shafer2020IAEA}. \negd{} plasmas are systematically weak in shape, possessing either low \elong{} or \triavg{}, owing to their poor fit to the DIII-D vacuum vessel cross-section. Weak shapes (low \triavg{} and \elong{}) naturally transition to limiter configurations (indicated by circles). \Lmode{} plasmas in principle can also exist everywhere in this space. Later sections will focus on ITER-like plasma shapes, indicated by dash-dot lines.

% Power and Torque
\paragraph*{Input power (\Ptot{}) and injected torque (\Tinj{}):} DIII-D can access to a wide range of powers (\Ptot{}) and torques (\Tinj{}), as shown in Fig. \ref{fig:basics}(c). \Ptot{} is defined as \Paux{} + \Poh{}, where \Paux{} is auxiliary heating power (NBI and ECH) and \Poh{} is ohmic heating power (with \Poh{}$<$2 MW). Several notable features can be found. First, no stationary plasma without ELMs has been observed while injecting the maximum balanced (\Tinj{}=0) power, with \QH{} plasmas coming the closest. \Lmode{} plasmas are less frequently found at high \Ptot{}, with notable examples being high radiation fraction (\frad{}) plasmas and limited plasmas (circle symbols). \rmp{} plasmas are not tolerant of significant counter-NBI\cite{Moyer2017,PazSoldan2019a}, but do exist over a wide range of \Ptot{}. \negd{} plasmas have uniquely accessed the maximum \Ptot{}, and as will be discussed have thus far not found a core MHD limit.

% LH and Greenwald
\paragraph*{Power normalized to scaled LH threshold power \cite{Martin2008} (\PoverPLH{}) and Greenwald density \cite{Greenwald1988} (\noverng{}):} Tolerance of a no-ELM regime to net exhausted power (\Pnet{}) is partially set by \Hmode{} threshold physics, as is the virulence of any ELMs encountered \cite{Eich2017,Knolker2018}. \Pnet{} is defined as \Ptot{}-\Prad{}, where \Prad{} here is radiated power inside the separatrix. Fig. \ref{fig:basics}(d) shows that high \Pnet{} \Lmode{} plasmas are exclusively found at low \noverng{}, with ELMing \Hmode{} obtained at higher density. \EDA{} and \Imode{} regimes are found at intermediate \noverng{} but at low \PoverPLH{}. \rmp{} and \QH{} plasmas are able to access high \PoverPLH{}, and are uniquely limited by core MHD as opposed to the return of the ELM. \rmp{} plasmas exhibit a strict upper limit in pedestal-top density and thus \noverng{} \cite{Evans2006a,Petrie2011,PazSoldan2019}. \QH{} plasmas can access higher \noverng{} than \rmp{} plasmas, owing to their improved compatibility with strong shapes and current-limited pedestals at high density \cite{Solomon2014,Garofalo2015a,Solomon2016}, though in Sec. \ref{sec:integ} the ability of high \noverng{} to raise the separatrix density in \QH{} plasmas will be shown to be less favorable. \negd{} plasmas are found to be uniquely able to access high \PoverPLH{} and \noverng{}.

% Collisionality and Safety Factor
\paragraph*{Edge collisionality (\nustar{}) and safety factor (\qnf{}):} Plasma regimes without ELMs are known to occupy distinctive points in the pedestal (or edge) collisionality (\nustar{}) and edge safety factor (\qnf{})\cite{Oyama2006,Viezzer2018}. Indeed the no-ELM regimes found in DIII-D are well-separated by these parameters, as shown in Fig. \ref{fig:basics}(e). \rmp{} and \QH{} plasmas are found below \nustar{} of about 1, while all other regimes are at higher \nustar{}. As discussed in Fig. \ref{fig:basics}(a), \rmp{} plasmas are only found at low \qnf{} on DIII-D, and \EDA{} are only found at high \qnf{}. \Imode{} is only observed thus far at intermediate \qnf{} $\approx$ 4. Other regimes are observed over a range of \qnf{}. In DIII-D, low \nustar{} access is found together with high \PoverPLH{}, whereas a reactor can be expected to operate with low \PoverPLH{} and low \nustar{}, as will be further discussed in Sec. \ref{sec:disc}.

% Torque and Safety Factor
\paragraph*{Torque (\Tinj{}) and safety factor (\qnf{}):} While less emphasized in the literature, DIII-D no-ELM plasmas clearly exhibit limitations in simultaneous operation of low \Tinj{}, low \qnf{}, and low \nustar{}. This is likely the edge stability regime expected in ITER, with the important caveat that extrapolating \Tinj{} depends on what rotation-dependent mechanism is of interest \cite{Garofalo2011}. As shown in Fig. \ref{fig:basics}(f), \rmp{} plasmas match \qnf{} and \nustar{} \cite{Wade2015}, but find the return of the ELM at low \Tinj{} \cite{Moyer2017,PazSoldan2019a}. \QH{} plasmas so far have matched two of low \nustar{}, \qnf{}, and \Tinj{}, but not all three together owing to the onset of core MHD or the ELM \cite{Garofalo2015a}. This finding has motivated extensive DIII-D experimentation to understand low \nustar{} + \qnf{} + \Tinj{} limits and their extrapolation \cite{Garofalo2020}, and this regime will soon be accessible in the world research program as new capabilities come online \cite{Stober2020,Shirai2017}. Note the low \qnf{} \& low \Tinj{} limit is not found for the high \nustar{} regimes, though even these have not been demonstrated at high power or pressure (\betan{}$>$1.2). \Lmode{} plasmas suffer a return to ELMing \Hmode{} at higher \betan{}, while \negd{} has not yet been attempted in this low \qnf{} + \Tinj{} part of parameter space. Access to low \qnf{} (specifically high \Ip{}) and high \Ptot{} will be shown in the next section to strongly impact the achievable performance without ELMs in DIII-D.

% ---- ---- ---- ---- -------- ----  ---- ---- ---- ---- ---- ---- ---- ---- ----
% ---- ---- ---- ---- -------- ----   PERFORMANCE   ---- ---- ---- ---- ----
% ---- ---- ---- ---- -------- ----  ---- ---- ---- ---- ---- ---- ---- ---- ----

\section{Plasma Performance without ELMs}
\label{sec:perf}

Plasma performance of DIII-D plasmas without ELMs is now presented using metrics previously proposed in the literature. This section begins with a survey of normalized performance (normalized pressure and confinement quality), then moves to absolute performance without constraints, and finally explores plasma performance within ITER shaping constraints. Comparison of ELMing and no-ELM plasma performance is left to Sec. \ref{sec:ELMs}.

\begin{figure*}
\begin{subfigure}{1\textwidth}
\begin{subfigure}{0.325\textwidth}
\centering
\includegraphics[width=1\textwidth]{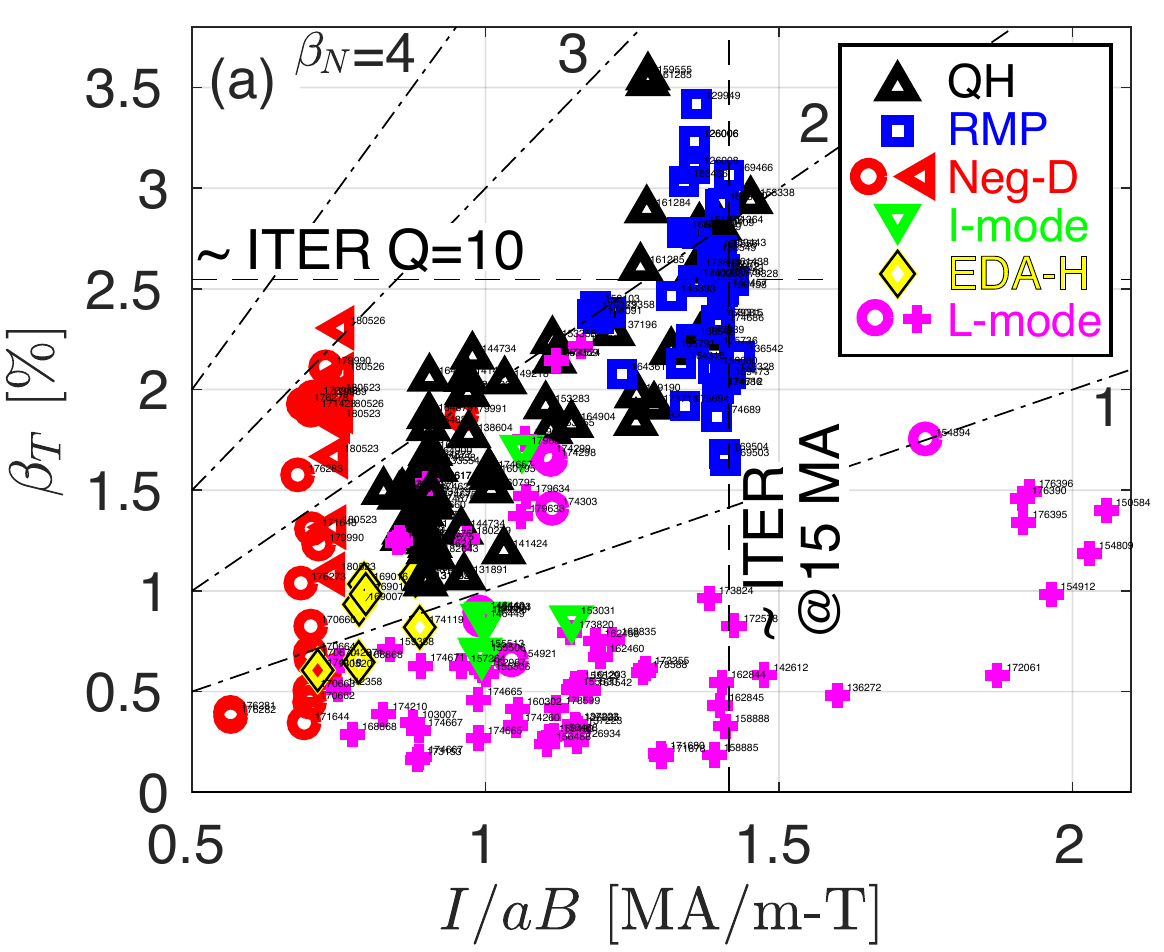}
\end{subfigure}
\begin{subfigure}{0.325\textwidth}
\centering
\includegraphics[width=1\textwidth]{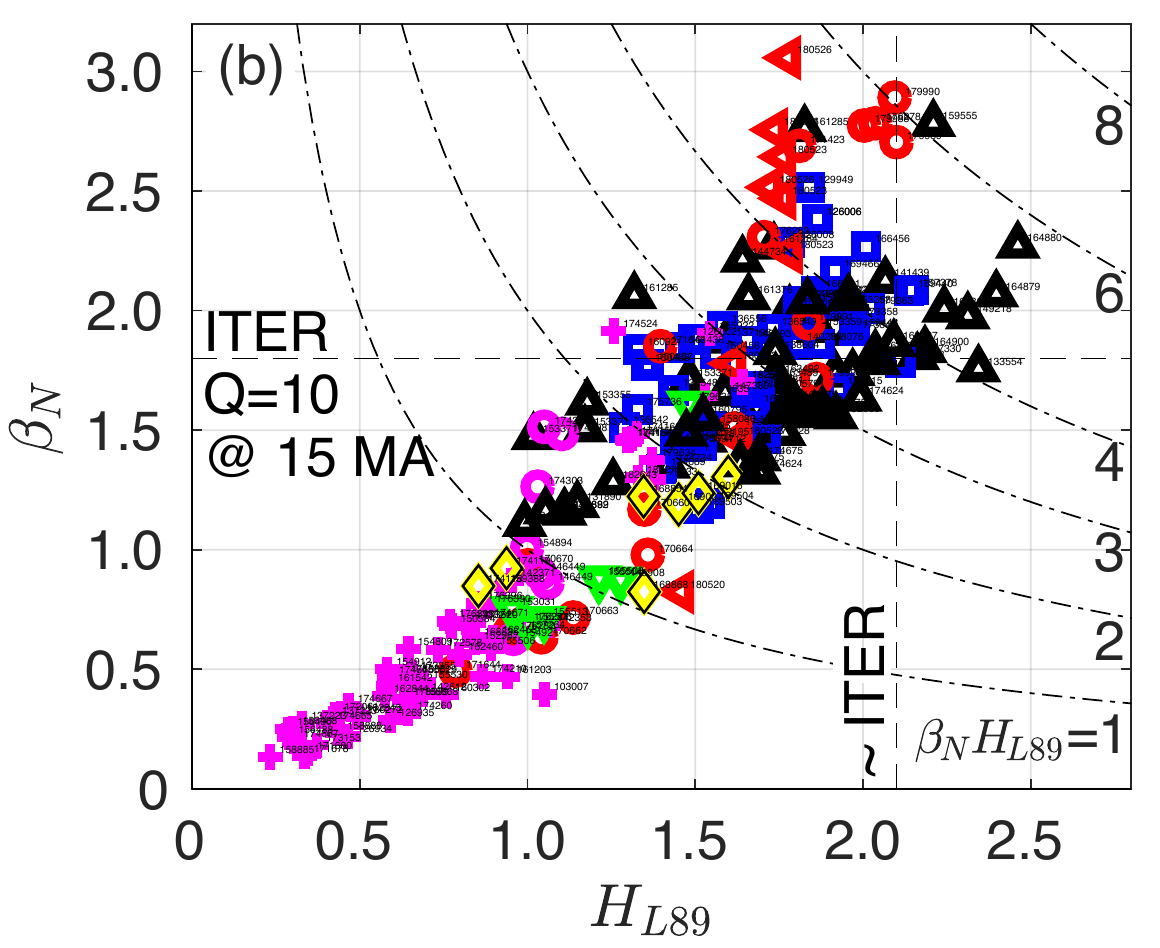}
\end{subfigure}
\begin{subfigure}{0.325\textwidth}
\centering
\includegraphics[width=1\textwidth]{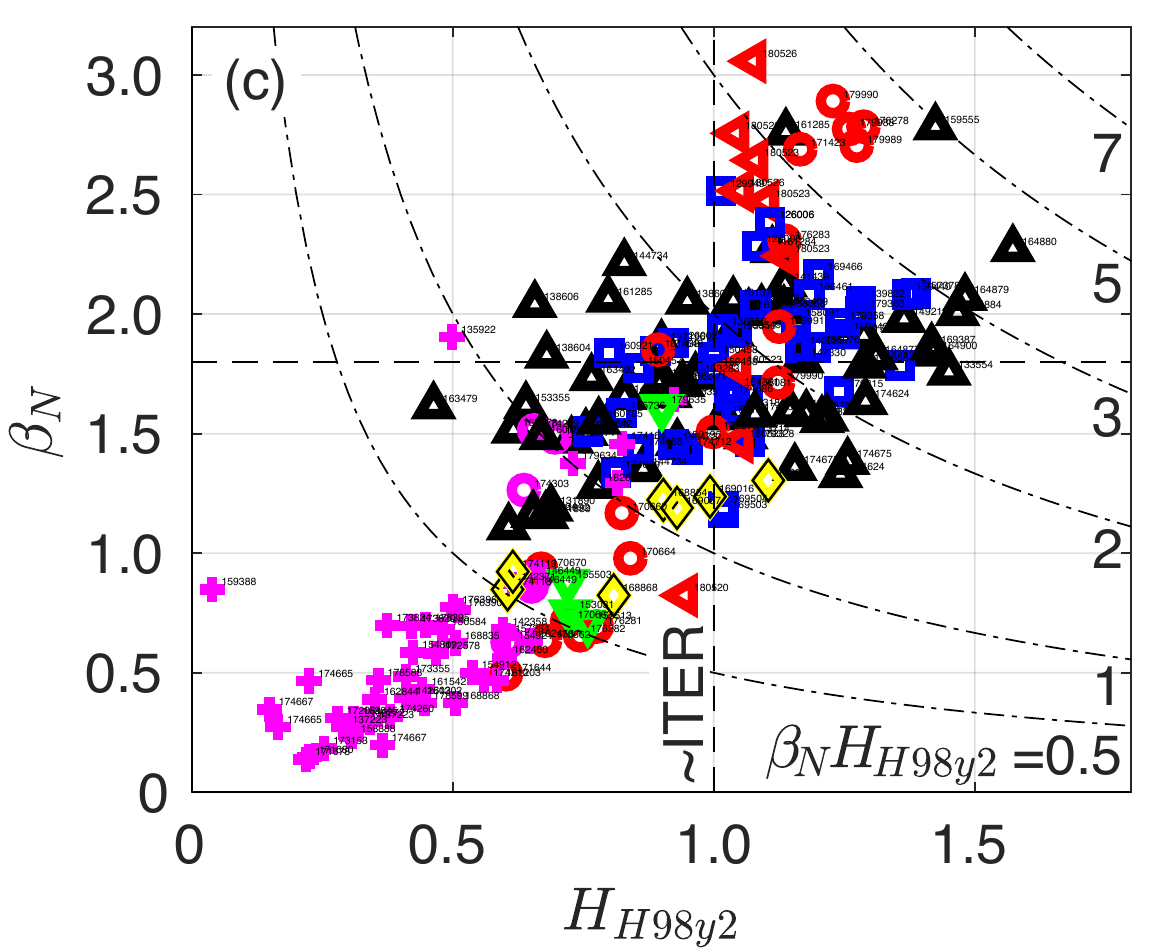}
\end{subfigure}
\end{subfigure}
 \vspace{-5 pt}
\caption{(a) Troyon core stability space in terms of toroidal beta (\betat{}) and normalized current (\In{}), dash-dot lines indicate constant normalized pressure (\betan{}). Confinement quality factors (b) \HL{} and (c) \HH{} as a function \betan{}, with dash-dot lines at a constant product of $H$ and \betan{}. Dashed lines indicate ITER 15 MA Q=10 targets.}
\label{fig:Hfactors}
\end{figure*}

\subsection{Normalized Performance}

% Troyon
Discussion of normalized performance begins with global stability. The Troyon stability diagram \cite{Troyon1984} shown in Fig. \ref{fig:Hfactors}(a) indicates the normalized current (\In{}), pressure normalized to toroidal field (\betat{}), and concurrently the normalized beta [\betan{}=\betat{}/(\In{})]. The different no-ELM regimes well-separate in this diagram. \In{} access has already been discussed in relation to Fig. \ref{fig:basics}(a), with \Lmode{} uniquely accessing high \In{}. The \betan{} accessible is clearly higher for \QH{}, \rmp{}, and \negd{} plasmas owing to the increased tolerance of power discussed in relation to Fig. \ref{fig:basics}(c). Peak \betan{} of \rmp{} and \QH{} plasmas are limited by core instability (generally tearing modes), while \negd{} is confinement-limited, and the other regimes by the return of the ELM. Note the wide-pedestal \QH{} variant has not yet found a \betan{} limit (present limit \betan{} 2.3) and finds confiment is not degraded by raising \Ptot{} \cite{Chen2020}. Interestingly, the highest stationary \betan{} without ELMs is found for \negd{} plasmas, despite the expectation of degraded core stability\cite{Medvedev2015}. The reason for this is not yet understood, but is speculated to be related to \negd{} plasmas operating at higher \noverng{} than \rmp{} and \QH{} plasmas, as shown in Fig. \ref{fig:basics}(d), which may reduce the pressure peaking instability drive \cite{Ferron2005}. It is also clear from Fig. \ref{fig:Hfactors}(a) that \QH{} and \rmp{} access the highest \betat{}, as high \betan{} and high \IaB{} are simultaneously achieved. Accessing high \betan{} at high \In{} has not yet been attempted for the \negd{} regime in DIII-D.

% H-factor describe
Plasma performance is commonly described in terms of the confinement quality `H-factors', defined as the ratio of the measured energy confinement time (\taue{}=$W/$\Ptot{}, where $W$ is the plasma kinetic stored energy) to expectations of scalings derived from multiple tokamaks. The confinement quality H-factors have been derived for type-I ELMing \Hmode{} plasmas (\HH{}\cite{IPBch2}), and L-mode plasmas (\HL{}\cite{Yushmanov1990}). Note that as no-ELM plasmas are often found at low \noverng{} [See Fig. \ref{fig:basics}(d)], the fast ion content is sometimes not negligible, motivating presentation of both \HL{} (which includes fast ion pressure) and \HH{} (which explicitly removes it). For all cases, fast ion pressure is evaluated automatically with between-shot codess using the formulas presented in Ref. \cite{Heidbrink1994}.

% Discuss H-vs-Beta
The performance against these H-factors for the database described in Sec. \ref{sec:ops} is shown in Fig. \ref{fig:Hfactors}(b,c), plotted against \betan{}.  As can be seen, these axes are also effective to separate the no-ELM regimes. Furthermore a good correlation exists between H-factor and \betan{} across all regimes, despite widely varying \Ptot{}. This correlation is consistent with \Ptot{} being overly penalized in the confinement scalings. Considering each regime, \Lmode{} plasmas are found at low normalized performance, \rmp{} and \QH{} plasmas are found at higher H-factor and \betan{}, while \Imode{} and \EDA{} plasmas are found at an intermediate level of normalized performance. \negd{} plasmas span a wide range of performance, from \Lmode{} like (low H and \betan{}) to normalized performance on par or better than \rmp{} and \QH{} regimes (high H and \betan{}). The highest H-factors are uniquely found for \QH{} plasmas.

\begin{figure*}
\begin{subfigure}{1\textwidth}
\begin{subfigure}{0.325\textwidth}
\centering
\includegraphics[width=1\textwidth]{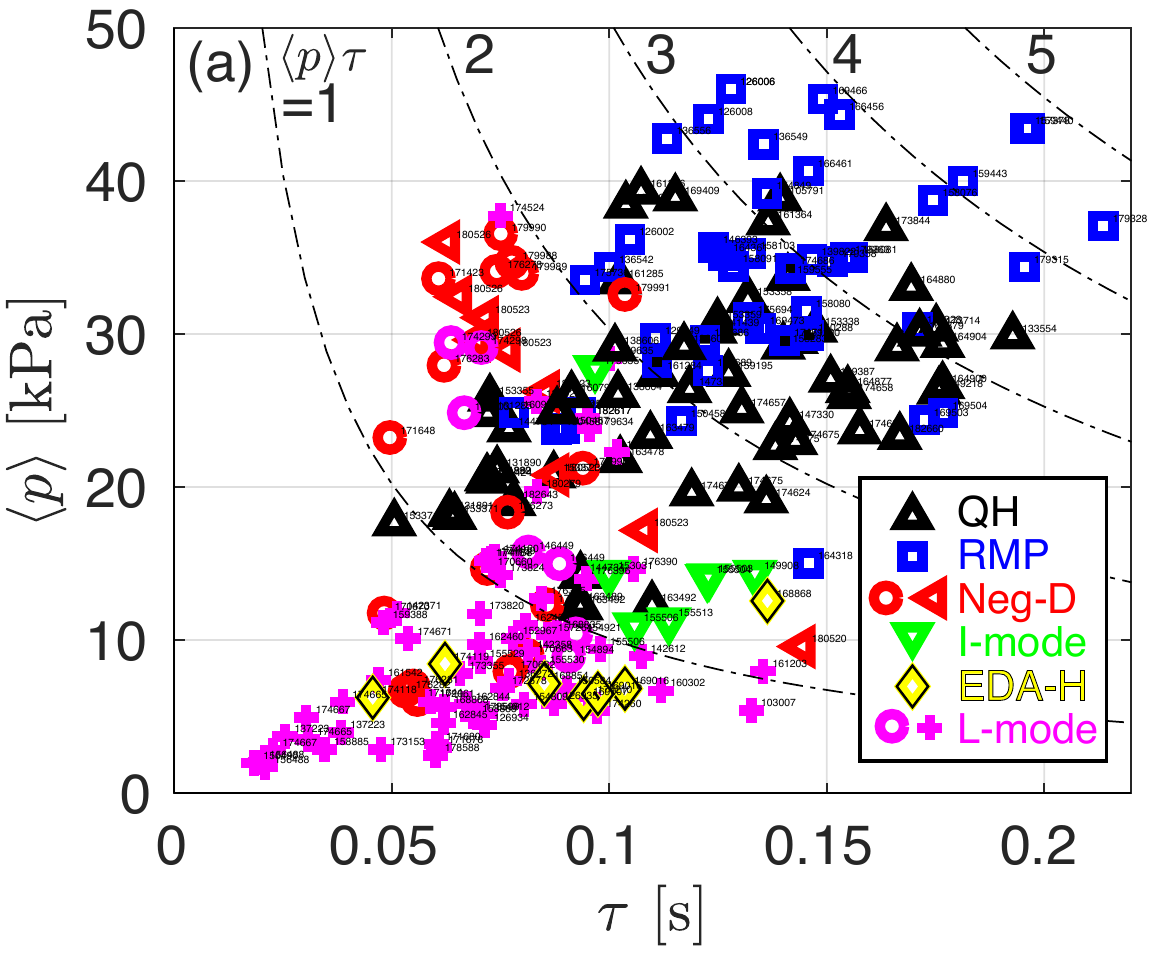}
\end{subfigure}
\begin{subfigure}{0.325\textwidth}
\centering
\includegraphics[width=1\textwidth]{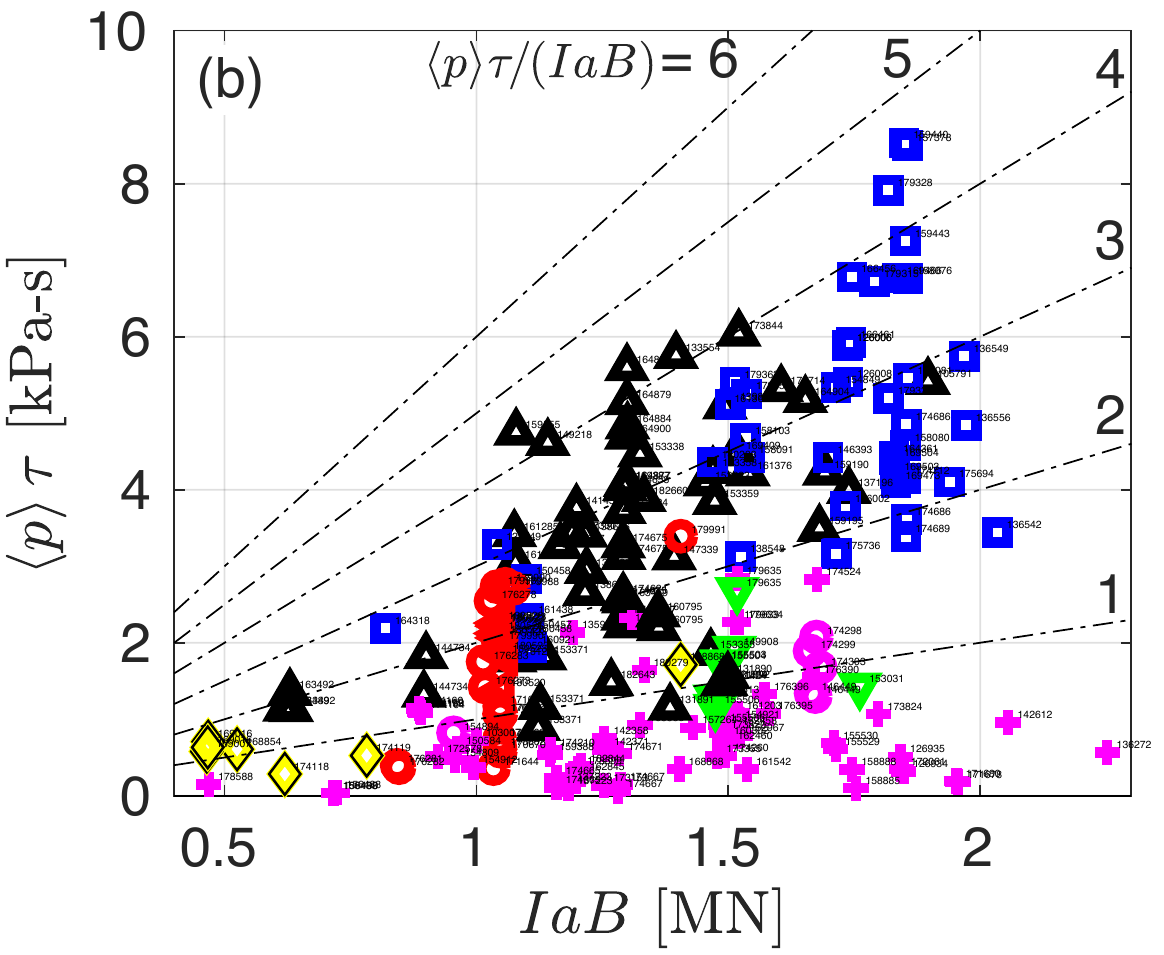}
\end{subfigure}
\begin{subfigure}{0.325\textwidth}
\centering
\includegraphics[width=1\textwidth]{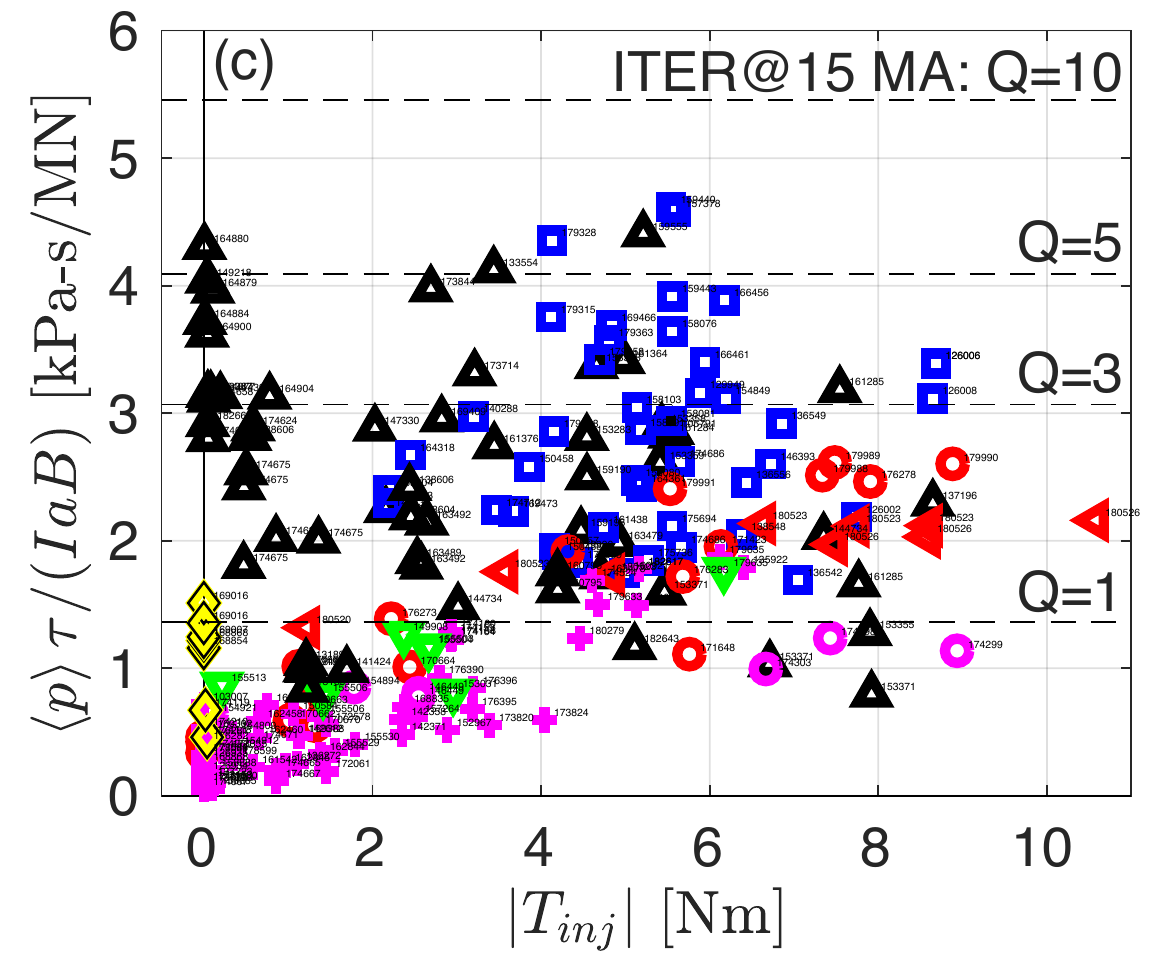}
\end{subfigure}
\end{subfigure}
 \vspace{-5 pt}
\caption{Absolute plasma performance in the DIII-D tokamak without ELMs, represented by: (a) average pressure (\pres{}) and confinement time (\taue{}), with dash-dot lines at constant triple product (\lawson{}).  (b) Triple product \lawson{} referenced to \IaB{}, showing an increasing trend. (c) \lawson{} normalized to \IaB{} referenced to \Tinj{}, with dashed lines referencing fusion gain Q values at 15 MA in ITER.}
\label{fig:abs}
\end{figure*}

\subsection{Absolute Performance}

% Lawson
Absolute performance has long been measured in terms of Lawson's triple product\cite{Lawson1957} [\lawson{}, where \pres{} is the volume averaged pressure $W$/(1.5*volume)]. Triple product decomposed into constituent \pres{} and \taue{} is shown in Fig. \ref{fig:abs}(a). No-ELM regime separation is again found. \QH{} and \rmp{} plasmas both reach high \lawson{} with a roughly equal mix of \pres{} and \taue{}, while \EDA{} and \Imode{} plasmas emphasize \taue{} over \pres{}, and \negd{} plasmas emphasize \pres{} over \taue{}. Note that most NBI heated DIII-D plasmas have $T_i/T_e > 1$, but \pres{} does not separate ion and electron contributions (despite only ions fusing). Some \Lmode{} datapoints occupy the same absolute performance as \negd{}, \QH{}, and \rmp{}. The discrepant view of \negd{} in normalized performance (Fig. \ref{fig:Hfactors}) and absolute performance (Fig. \ref{fig:abs}) is reflective of the weaker shape and lower \Ip{} of the \negd{} plasmas created thus far, and perhaps due to the H-factors over-penalizing input power. As can be seen, the highest \lawson{} without ELMs on DIII-D is found in \rmp{} plasmas, a surprising result given the significant pedestal degradation encountered in \rmp{} plasmas. These high \lawson{} \rmp{} plasmas feature near-threshold RMP current values or below-threshold values employing hysteresis, sometimes using active RMP current feedback control \cite{Laggner2020}. The role of the pedestal will be further explored in Sec. \ref{sec:corr}.

% Snyder
As described in Ref. \cite{Snyder2019}, an effective normalization for \lawson{} is \IaB{}, the product of the plasma current (\Ip{}), minor radius ($a$), and toroidal field (\Bt{}). \IaB{} has units of force (MN), and is a simple metric for the strength of the tokamak magnetic configuration. As shown in Fig. \ref{fig:abs}(b), peak \lawson{} tends to rise linearly with \IaB{}, supporting the value of \IaB{} as a normalization parameter for \lawson{} as will be soon introduced. The DIII-D tokamak (with its present divertor baffling) is capable of \IaB{} of around 2.5 MN by operating full-field strongly shaped plasmas (high \elong{} and $\Delta_{avg}$) at high \Ip{}. Without ELMs however, \IaB{} is found to not exceed around 2 MN (excepting \Lmode{}). This indicates a straightforward path to raise \lawson{} is to develop stationary no-ELM plasmas at the maximum achievable \IaB{}, though this implies overcoming the operational limits associated with high \In{} (low \qnf{}) and strong shapes presented in Sec. \ref{sec:ops}.

% Motivation of Snyder
Supported by the trends of \lawson{} with \IaB{} shown in Fig. \ref{fig:abs}(b), for the subsequent sections the main metric to consider plasma performance will be \snyder{}. This metric is meant to show the quality of fusion performance across a range of $T_i$, with an advantage of only requiring simple engineering inputs without imposing any scaling relationships. Said differently, this metric is agnostic as to the importance of factors such as plasma shape and density, and rather lets the triple product speak for itself. Note standard type-I ELMing \Hmode{} performance across DIII-D, Alcator C-mod, and JET was found in Ref. \cite{Snyder2019} to yield comparable \snyder{}, despite very different  \Ip{}, $a$, and \Bt{}. Further discussion of the \snyder{} metric, alongside comparison to other metrics can be found in Appendix \ref{sec:metrics}.

% Snyder results
\snyder{} is now plotted against \Tinj{} in Fig. \ref{fig:abs}(c). A clear distinction is seen at low \Tinj{}, where only \QH{} plasmas are able to achieve high \snyder{}, consistent with only \QH{} tolerating high \Ptot{} at low \Tinj{} [shown in Fig. \ref{fig:basics}(c)]. All other regimes (except \negd{}, which has not been explored) are limited by the ELM at low \Tinj{} and high \Ptot{}. Performance of \negd{} plasmas in terms of \snyder{} is below \QH{} and \rmp{} plasmas for all \Tinj{}, possibly due to the weak shaping or to the absence of a pedestal. Finally, use of \snyder{} allows for a comparison of performance scaled to ITER Q=10 at 15 MA, namely \snyder{} $\approx$ 5.5\cite{Snyder2019}. While Sec. \ref{sec:ELMs} will show this can be met and exceeded for ELMing plasmas, it is not yet met in DIII-D for stationary plasmas without ELMs, even when ignoring all ITER operational constraints.

\begin{figure*}
\begin{subfigure}{1\textwidth}
\begin{subfigure}{0.325\textwidth}
\centering
\includegraphics[width=1\textwidth]{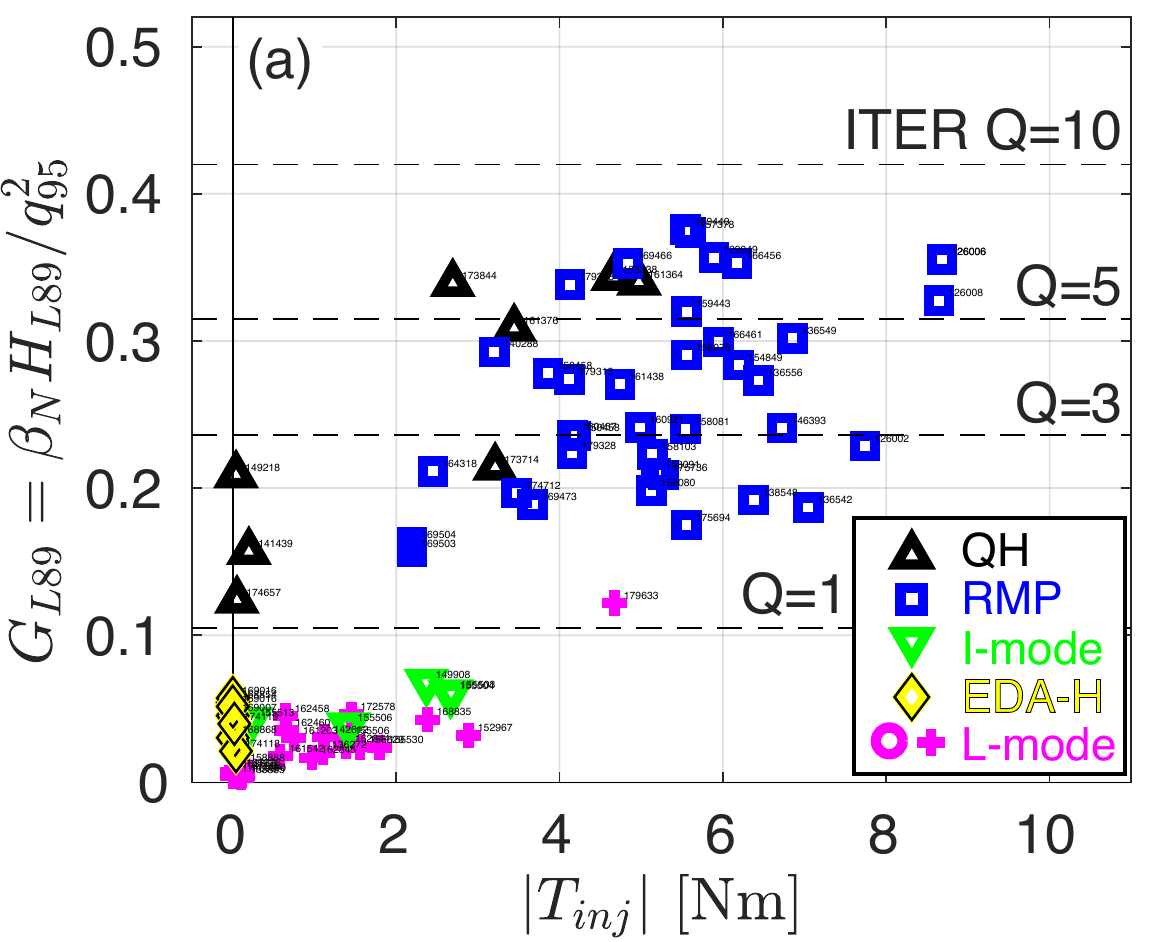}
\end{subfigure}
\begin{subfigure}{0.325\textwidth}
\centering
\includegraphics[width=1\textwidth]{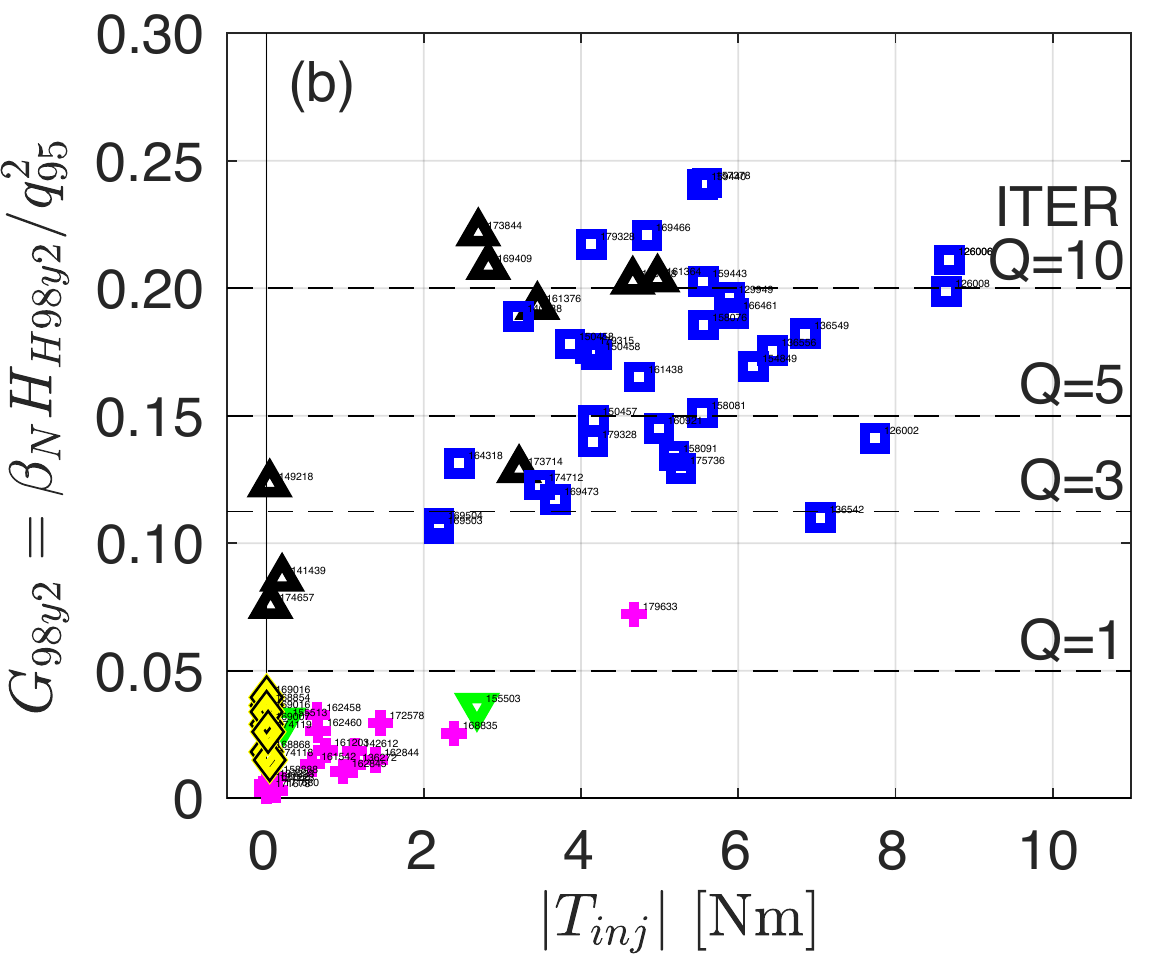}
\end{subfigure}
\begin{subfigure}{0.325\textwidth}
\centering
\includegraphics[width=1\textwidth]{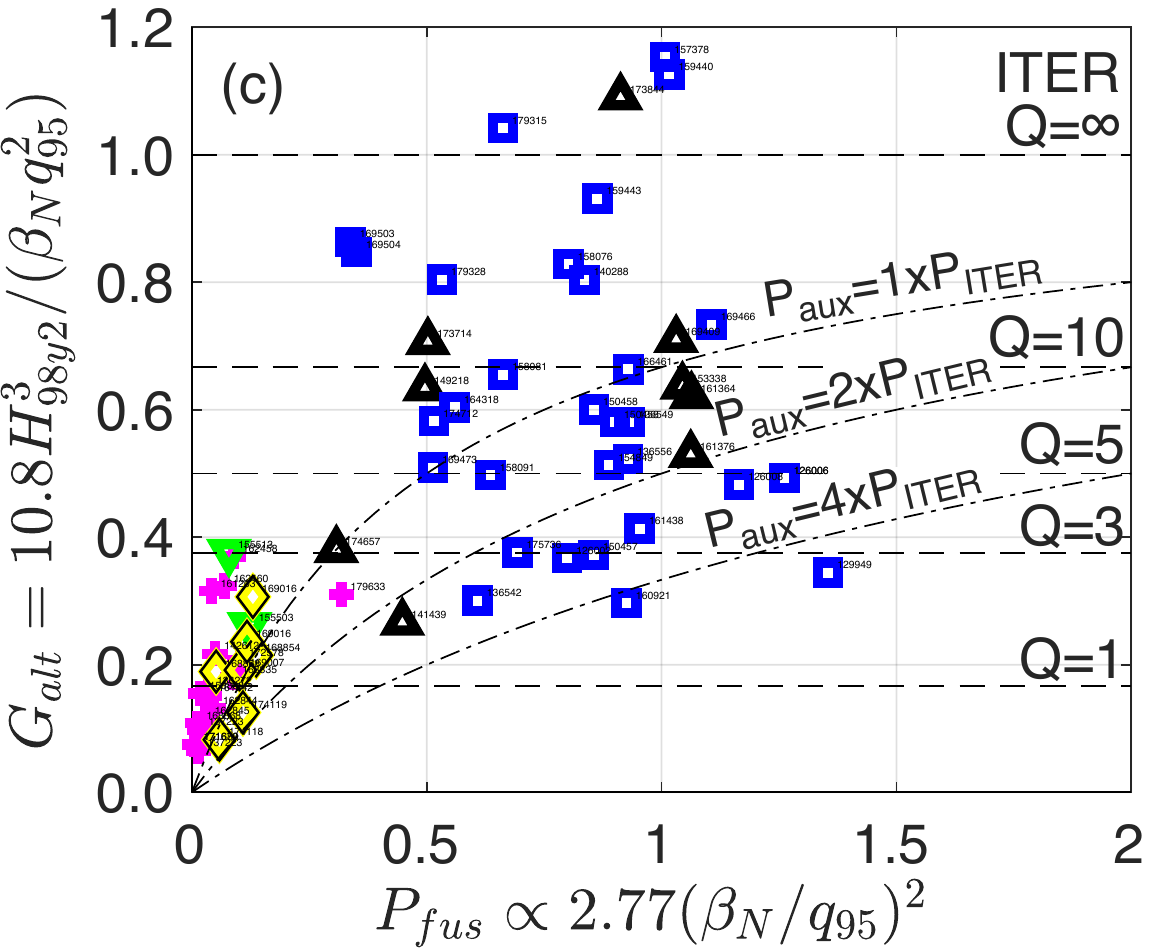}
\end{subfigure}
\end{subfigure}
 \vspace{-5 pt}
\caption{Gain metrics meant to estimate ITER performance for discharges with ITER-like plasma shapes. Gain factors based on Ref. \cite{Luce2005} for (a) L-mode confinement (\HL{}) and (b) \Hmode{} confinement (\HH{}) plotted against \Tinj{}. (c) Gain factor based on Ref. \cite{Peeters2007}, allowing comparison to the \Paux{} available in ITER, showing many datapoints inaccessible with ITER's \Paux{}.}
\label{fig:ITER}
\end{figure*}

\subsection{Plasma Performance with ITER-like Shapes}

% Introduce Problem
Plasma performance focusing on ITER-like plasma shapes is now presented, with the ITER shape only loosely specified as \triavg{}=0.4-0.6, \elong{}=1.7-1.95 \cite{Aymar2001}. The goal is to exclude strongly shaped double-null plasmas as well as other scenarios clearly inaccessible to ITER, such as \negd{}. As such only scenarios that in could in principle be realized in ITER are retained. Results are first presented as a function of \Tinj{}, highlighting the special challenge of achieving both ELM control and high performance with relatively small external momentum input in ITER \cite{Garofalo2020}. Extrapolating \Tinj{} depepends on what rotation-dependent mechanism is of interest, but attempting to match the dimensional rotation in ITER yields an ITER-equivalent torque of $\approx$ 0.5 Nm \cite{Garofalo2011}. Section \ref{sec:integ} will discuss other integration constraints that apply to all burning plasmas, including ITER. For all metrics considered the \rmp{} and \QH{} plasma performance will be clearly superior to other regimes, though this is largely due to the unique tolerance of \rmp{} and \QH{} plasmas to low \qnf{} (15 MA-equivalent in ITER) and high \Ptot{} in DIII-D.

% Results of Luce's G
Plasma fusion gain factors \GL{}=\HL{}\betan{}/\qnf{}$^2$ and \GH{}=\HH{}\betan{}/\qnf{}$^2$ presented in Refs. \cite{Luce2004,Luce2005} are presented in Fig. \ref{fig:ITER}(a)-(b) against \Tinj{}. Both G-factors are the product of what was shown in Fig. \ref{fig:Hfactors}(b,c), but now divided by the square of the normalized current (\qnf{}). The G-factors prescribe a target value for ITER Q=10 operation at any current. Both \rmp{} and \QH{} plasmas are found to reach the ITER Q=10 target in terms of \GH{} but fall just short for \GL{}. Peak performance of both \QH{} and \rmp{} plasmas thus far achieved is lower at low \Tinj{}. As seen in Fig. \ref{fig:basics}(c,f) no \rmp{} plasma exists at \Tinj{}$<$2 Nm, and \rmp{} plasma performance will be shown in Sec. \ref{sec:corr} to correlate with rotation. \QH{} results in Fig. \ref{fig:ITER} are not reflective of a confinement degradation with rotation \cite{Ernst2016} but rather highlight the two different \qnf{} \QH{} plasma populations shown in Fig. \ref{fig:basics}(f). The high \Tinj{} population is at low \qnf{} and high G-factor, but encounters limits as \Tinj{} is reduced, while the \Tinj{}=0 population is at high \qnf{} and lower G-factor. Optimization between these two \qnf{} populations has not yet been given serious effort, and is highlighted as a valuable experimental direction for \QH{} plasmas. The G-factors of Fig. \ref{fig:ITER}(a,b) and \snyder{} [shown in Fig. \ref{fig:abs}(c)] are found to be largely co-linear (not shown), with the main discriminant being \qnf{}. High \qnf{} plasmas are prejudiced in \GL{}/\GH{} as compared to \snyder{}, consistent with low \qnf{} (high \Ip{}) being over-rewarded in the G-factors. This accounts for differences at \Tinj{}=0, where as shown in Fig \ref{fig:basics}(f) only \Lmode{} plasmas have thus far accessed low \qnf{} and \Tinj{}=0 without ELMs. Also worth noting is that the best performing \Tinj{}=0 plasmas of Fig \ref{fig:abs}(c) are in strongly-shaped plasmas, thus they are excluded from Fig. \ref{fig:ITER}. Comparison of \GL{} to \snyder{} is discussed in Appendix \ref{sec:metrics}.

% Describe Peeters' G
As presented in Ref. [\onlinecite{Peeters2007}], a limitation of \GL{}/\GH{} is that no consideration is given to whether the level of auxiliary heating (\Paux{}) required to reach a given performance level is available in ITER. Simply put, operation at higher \betan{} can overcome insufficient H-factor to reach comparable fusion performance only if there is sufficient \Paux{} to reach the stated \betan{}. This consideration gave rise to an alternate G-factor (\Galt{}=\HH{}$^3$/\betan{}\qnf{}$^2$), different from \GH{} by the ratio (\HH{}/\betan{})$^2$, which emphasizes confinement more than pressure and generally favors lower \Paux{} discharges with still moderate H-factor. Note at constant heating power \betan{} $\propto H$, and all discussed G-factors are degenerate.

% Results of Peeters' G
The \Galt{} formulation allows simultaneous consideration of the expected fusion gain (Q), expected fusion power (\Pfus{}), and the heating power (\Paux{}) required to achieve this performance, as shown in Fig. \ref{fig:ITER}(c). The curved dash-dot lines indicate curves of constant ratio of required \Paux{} to available power in ITER. There is insufficient power in ITER to realize many of the presented cases, evidenced by the data below the top curved line. Nearly all of the \Lmode{}, \EDA{}, and \Imode{} plasmas can be realized within ITER's available \Paux{}, but they do not reach Q=10. A large number of datapoints lie well above the available \Paux{} line, but comparison with Fig. \ref{fig:ITER}(a,b) indicate these are all at high  \Tinj{} ($>$ 2 Nm). Looking at \Tinj{} $<$ 2 Nm, only \QH{} plasmas come close to Q=10 performance and only below 500 MW of \Pfus{}, though \qnf{} optimization may improve on these results in the future. If the high \Tinj{} limit is relaxed, many discharges extrapolate to extremely high fusion gain in ITER, including surprisingly ignition. Comparison to ELMing plasmas is given in Sec. \ref{sec:ELMs}.

% ---- ---- ---- ---- -------- ----  ---- ---- ---- ---- ---- ---- ---- ---- ----
% ---- ---- ---- ---- -------- ----    CORRELATIONS   ---- ---- ---- ---- 
% ---- ---- ---- ---- -------- ----  ---- ---- ---- ---- ---- ---- ---- ---- ----

\section{Core and Pedestal Correlations with No-ELM Plasma Performance}
\label{sec:corr}

\begin{figure*} % Could be good as a vertical figure
\begin{subfigure}{1\textwidth}
\begin{subfigure}{0.325\textwidth}
\centering
\includegraphics[width=1\textwidth]{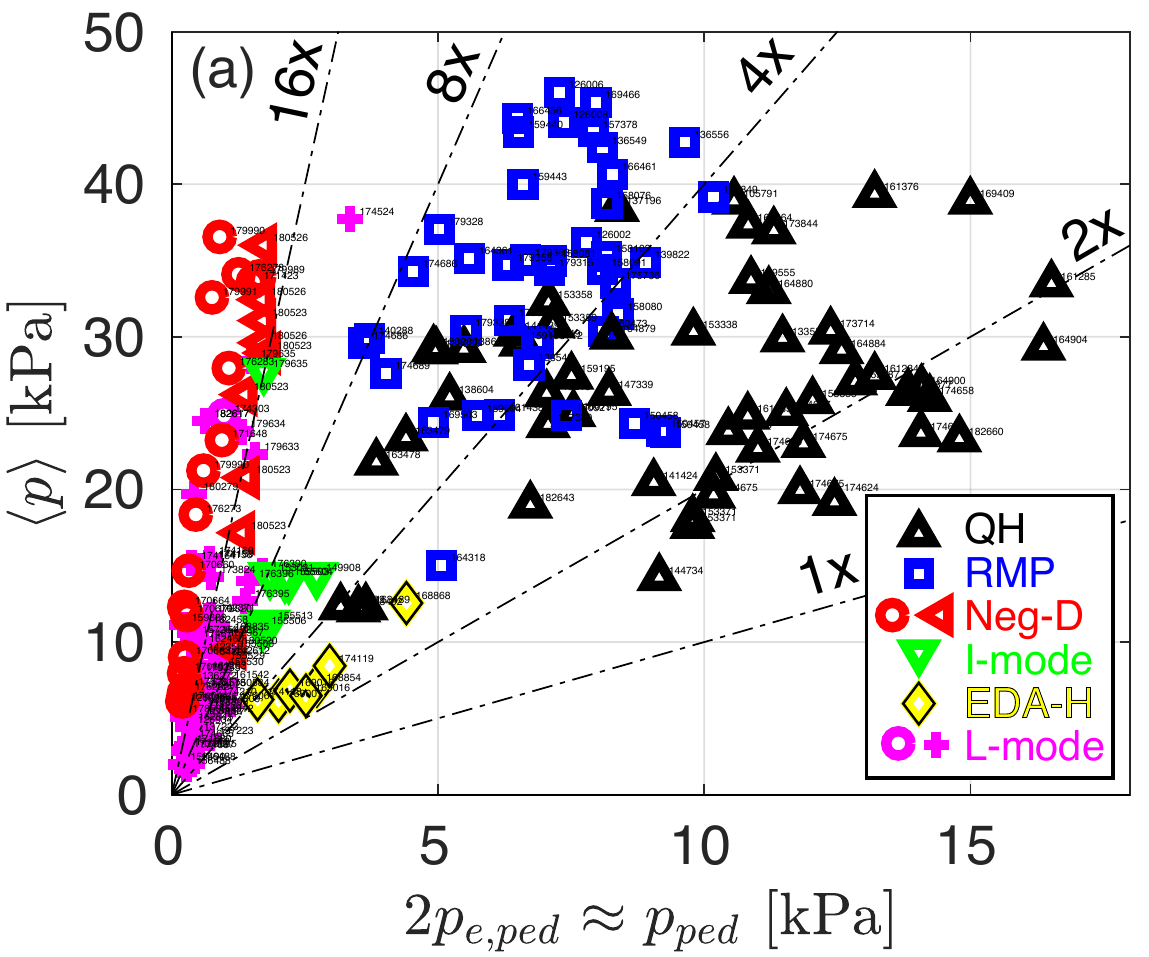}
\end{subfigure}
\begin{subfigure}{0.325\textwidth}
\centering
\includegraphics[width=1\textwidth]{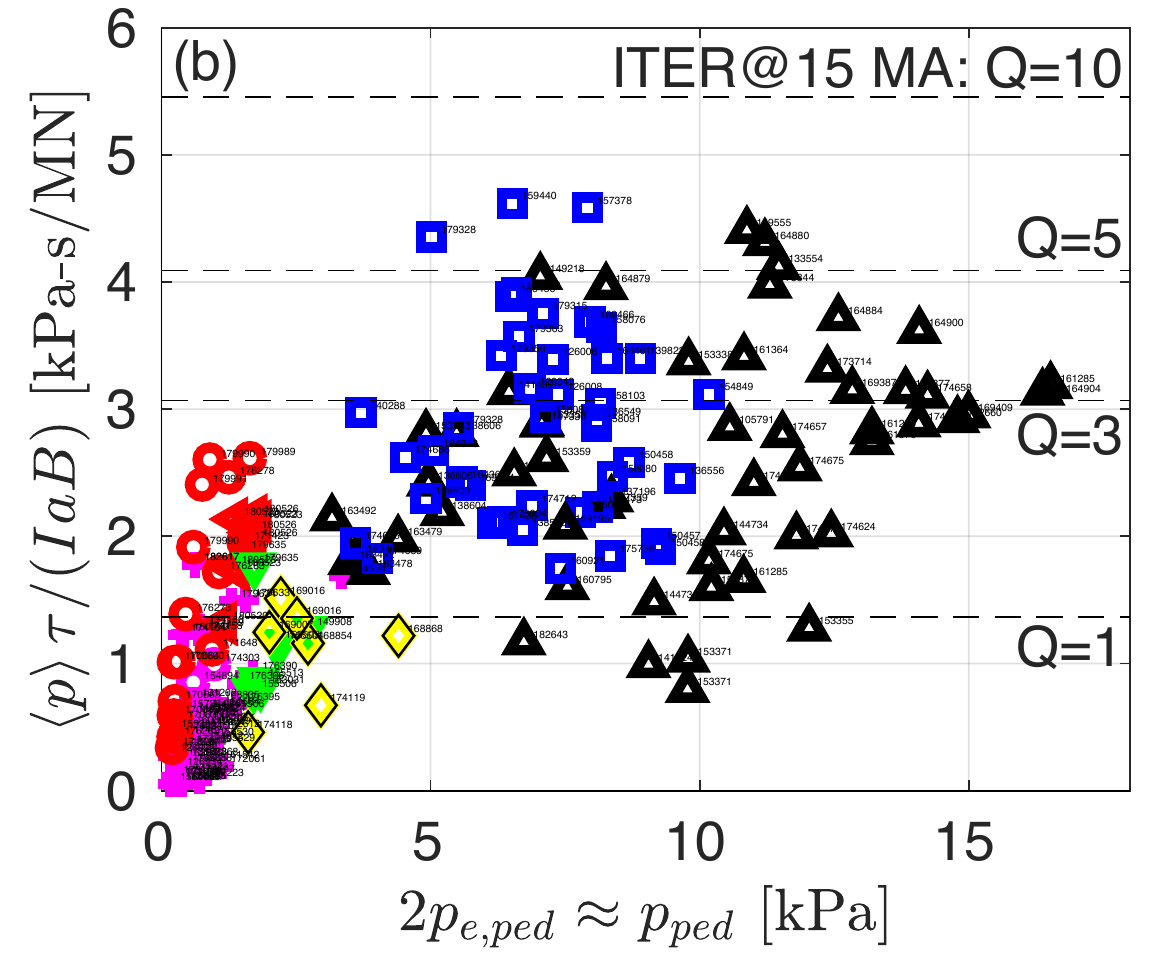}
\end{subfigure}
\begin{subfigure}{0.325\textwidth}
\centering
\includegraphics[width=1\textwidth]{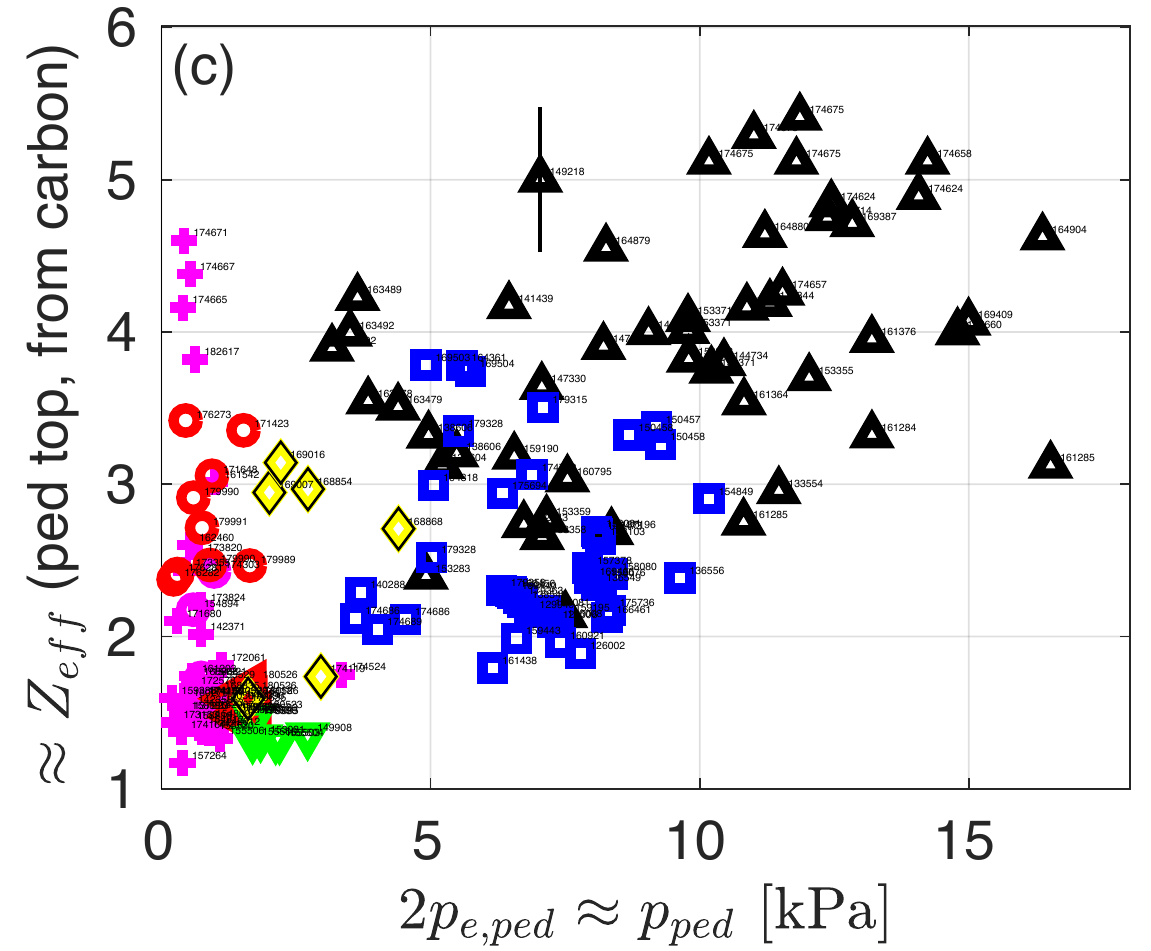}
\end{subfigure}
\end{subfigure}
 \vspace{-5 pt}
\caption{(a) Pedestal pressure (2\peped{}$ \approx $\pped{}) contribution to \pres{}, (b) dependence of plasma performance (quantified using \snyder{}) on \pped{}, (c) correlation of \Zeff{} with \pped{}}
\label{fig:coreped}
\end{figure*}

% Section Intro
The pedestal (or edge) is naturally the key region of interest for understanding access to no-ELM regimes. However, the role of the pedestal in the plasma performance is found to be more nuanced in DIII-D than the simple paradigm of maximizing pedestal pressure (\pped{}) to maximize fusion performance. Additional correlations of the plasma performance with core rotation, impurity content, and total power are identified for some regimes.

% Pedestal Fraction
The contribution of the pedestal pressure \pped{} (here quoted as 2\peped{} and measured by Thomson scattering) to the volume-average pressure (\pres{}) is shown in Fig. \ref{fig:coreped}(a), indicating stark differences between no-ELM regimes. Naturally, the \negd{} and \Lmode{} regimes have almost no pressure contribution from \pped{}, owing to their lack of an edge barrier. The few notable \Lmode{} cases with relatively high \pped{} operate at very high edge density, which despite the very low edge temperatures (10s eV) yields a finite pressure. At least in this representation, the \Imode{} on DIII-D is barely distinguishable from the \Lmode{}. \rmp{}, \EDA{}, and \QH{} plasmas are found to have progressively higher \ppedratio{}. \QH{} plasmas stand alone in the absolute \pped{} (2\peped{}) observed without ELMs in DIII-D. Comparison of no-ELM \pped{} to ELMing regimes appears in Sec. \ref{sec:ELMs}.

% Pedestal and Performance, Carbon Discussion
Plasma performance as quantified by \snyder{} is found to be largest for \rmp{} and \QH{} plasmas with finite \pped{}, however within each regime there is no clear correlation of \snyder{} with \pped{}. This indicates that there are other paths to maximize performance beyond simply maximizing \pped{}, and also that having near-zero \pped{} is a meaningful performance penalty. Prior to moving to other correlations with performance, it bears noting that the uniquely high no-ELM \pped{} observed in \QH{} plasmas is correlated with a uniquely high edge impurity content (particularly the wide-pedestal \QH{} variant), as shown in Fig. \ref{fig:coreped}(c). Here \Zeff{} inferred from CER measurements of carbon density in the outer core (near the pedestal top), and is a more challenging measurement than others presented in this work. As such a larger experimental uncertainty should be ascribed (say $\pm$1 at high \Zeff{}), and furthermore other impurities beyond carbon can contribute to \Zeff{}. The \Zeff{} expected in a fusion reactor is sensitive to the divertor strategy and the atomic number ($Z$) of the selected radiator(s). In DIII-D the dominant impurity is carbon ($Z$=6), and the deuterium fuel concentration decreases linearly from 100\% at \Zeff{}=1 to 0\% at \Zeff{}=6. As such the level of fuel dilution in DIII-D at high \Zeff{} is certainly intolerable. It should be noted however that \Zeff{} in \QH{} plasmas typically exhibit a hollow profile, with on-axis values typically in the range of 2.5-3.5. Dedicated studies have also reported favorable core impurity transport in \QH{} plasmas relative to ELMing plasmas \cite{Chen2020}, despite the fact that lower \Zeff{} values are typically observed in ELMing plasmas (as will shown in Sec. \ref{sec:ELMs}. 

\begin{figure*}
\begin{subfigure}{1\textwidth}
\begin{subfigure}{0.325\textwidth}
\centering
\includegraphics[width=1\textwidth]{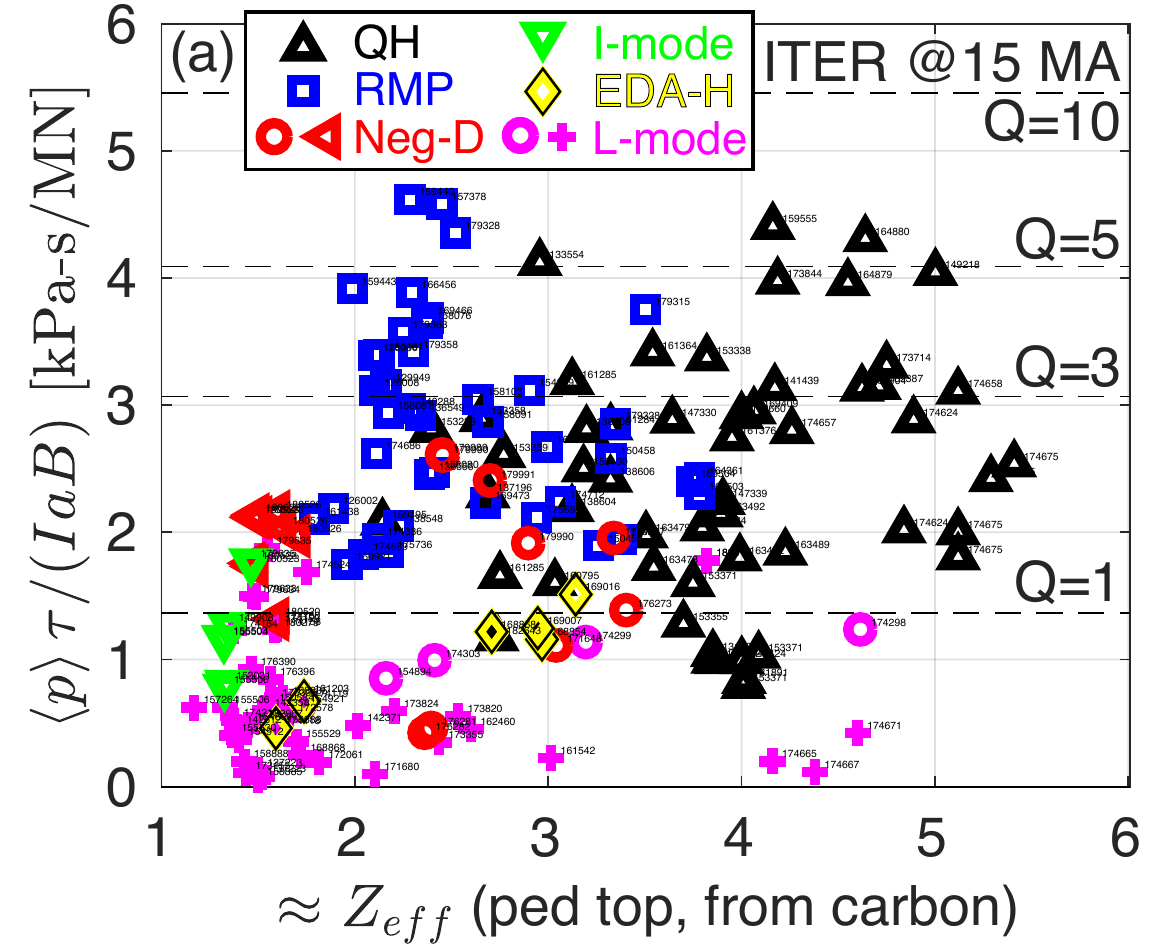}
\end{subfigure}
\begin{subfigure}{0.325\textwidth}
\centering
\includegraphics[width=1\textwidth]{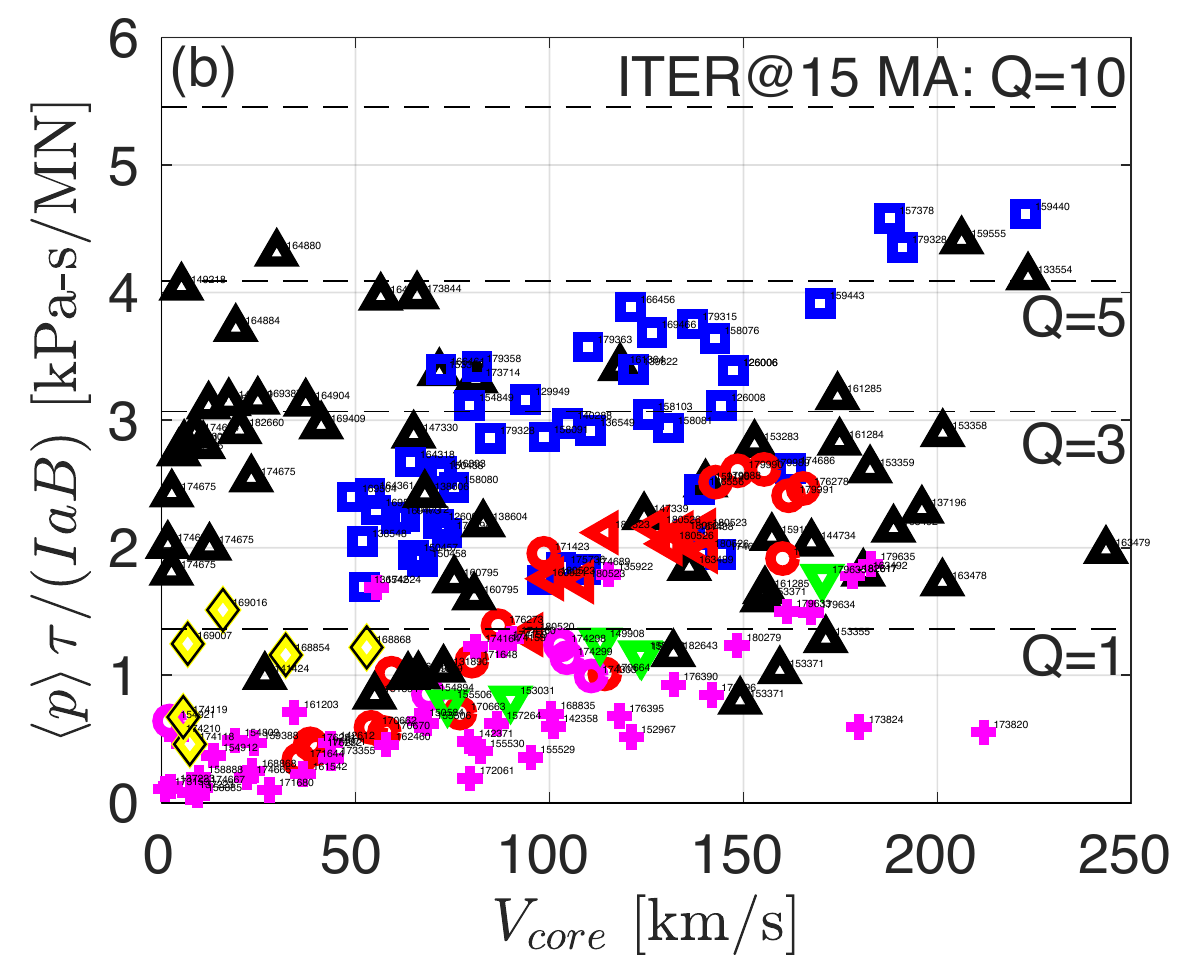}
\end{subfigure}
\begin{subfigure}{0.325\textwidth}
\centering
\includegraphics[width=1\textwidth]{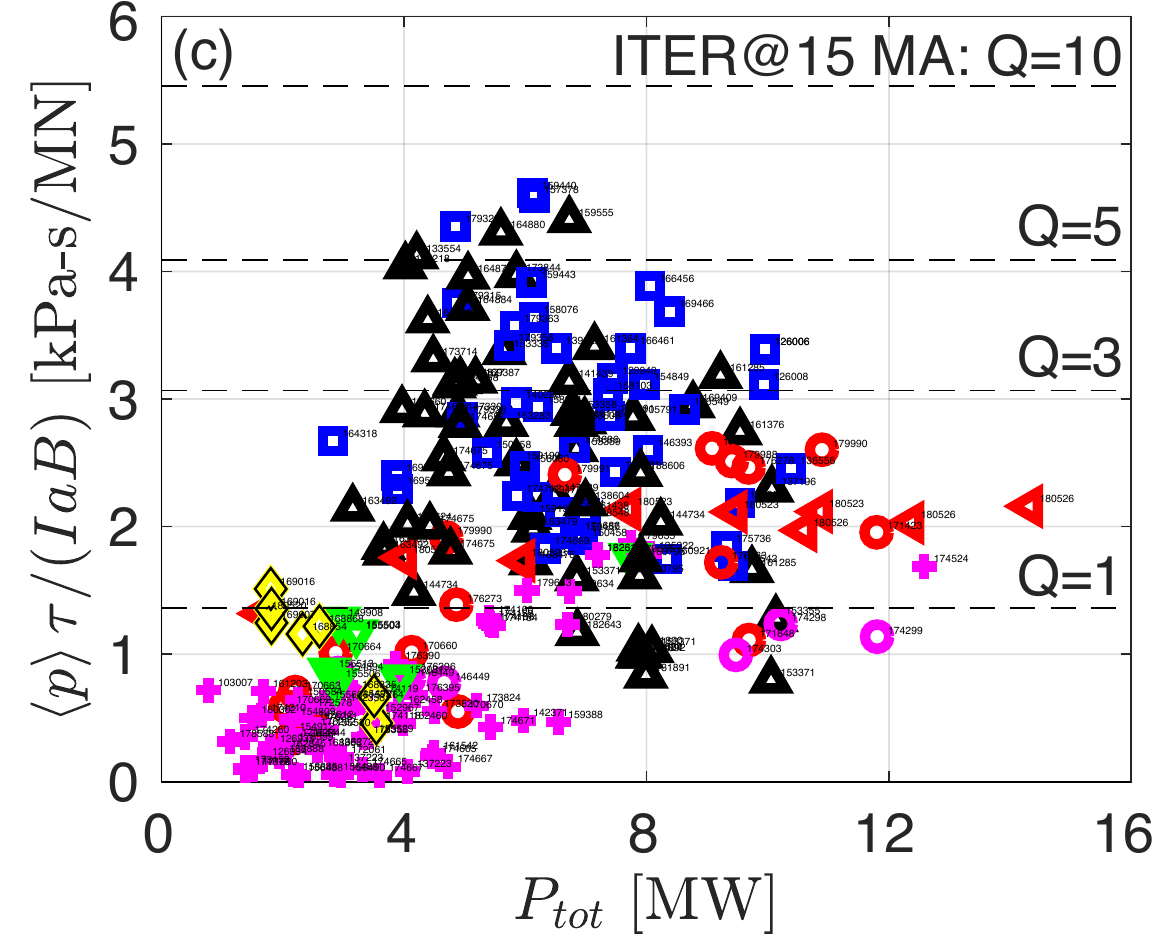}
\end{subfigure}
\end{subfigure}
 \vspace{-5 pt}
\caption{Additional correlations with no-ELM plasma performance (\snyder{}): (a) Carbon impurity content (\Zeff{}), (b) core toroidal rotation (\Vcore{}), (c) total input power (\Ptot{}).}
\label{fig:altcorr}
\end{figure*}

% QH Performance and Carbon
The carbon issue for \QH{} plasmas is further emphasized in Fig. \ref{fig:altcorr}(a) by noting that the highest performance in \QH{} is found with high pedestal-top \Zeff{} values, unlike the other no-ELM regimes on DIII-D (excepting \EDA{}). It is unclear if this correlation has any basis in causality, and past studies focusing on impurity confinement have found reasonably high impurity exhaust for \QH{} plasmas \cite{Grierson2015,Chen2020}. Understanding and reducing the carbon inventory of \QH{} plasmas (while maintaining high performance) is thus an important research direction already underway, and further motivates the establishment of the \QH{} regime in all-metal tokamaks where low \Zeff{} is the default operating condition \cite{Viezzer2019}.

% RMP Discussion
While carbon content is less of an issue for other no-ELM regimes in DIII-D, other correlations that challenge extrapolation are identified. \negd{}, \Lmode{}, and especially \rmp{} plasmas are shown in Fig. \ref{fig:altcorr}(b) to have a strong correlation with the absolute core rotation (\Vcore{}), with only \QH{} plasmas achieving high \snyder{} at low rotation (at least partially due to the larger \pped{}). Considering the relatively narrow \pped{} for \rmp{} plasmas shown in Fig. \ref{fig:coreped}, the performance variations in \rmp{} plasmas are largely due to variations in the core, whose performance is in turn is highly correlated with \Vcore{}. This is consistent with an ExB shear turbulent stabilization mechanism giving rise to high core performance in \rmp{} plasmas, a perhaps unsurprising hypothesis given similar observations in ELMing regimes on DIII-D \cite{Solomon2013,Ding2020}, although alternate mechanisms impacting the confinement of \rmp{} plasmas at fixed \Tinj{} have also been identified \cite{Cui2017}. As shown in Fig. \ref{fig:altcorr}(c), increasing \Ptot{} (and \Tinj{}) to maximal values does not improve the confinement of \rmp{} plasmas, with an optimum in performance at \Ptot{} $\approx$ 6 MW is found (\PoverPLH{} $\approx$ 3), similarly to \QH{} plasmas in the database. While ExB shear confinement improvements are readily found at high \Tinj{}, their extrapolation to future reactors such as ITER is challenging owing to their high moment of inertia as compared to expected \Tinj{} values\cite{Garofalo2011}. However, recent work for ITER finds some ExB shear turbulent suppression can be expected despite this unfavorable scaling\cite{Chrystal2020}. 

%Unfortunately, owing to the intolerance of \rmp{} plasmas in DIII-D to low \Tinj{} as discussed in Fig. \ref{fig:basics}(c), only limited variation of \Tinj{} at fixed \Ptot{} is possible.

% Neg-D Discussion
Interestingly, \negd{} plasma performance is also found to correlate with \Vcore{} as well as with \Ptot{}. The role of ExB shear suppression in the \negd{} performance enhancement has not yet been assessed experimentally, as evidenced by Fig. \ref{fig:basics}(c) the requisite variations in \Tinj{} at constant \Ptot{} have yet to be performed. However, the expected mechanism for enhanced confinement in \negd{} plasmas is thought to be directly related to the shape (triangularity and Shafranov shift) giving rise to a reduction in stiffness \cite{Merlo2015}, supporting a testable hypothesis that the dependence on \Ptot{} is more important than ExB shear.

% Other Regime Discussion
Given the scarcity of \EDA{} and \Imode{} data, any observed correlations may be due to sampling bias. That being said, \EDA{} plasmas appear to have some features of \QH{} plasmas: strong \pped{} contribution, and also feature high \Zeff{} at their highest performance. Similarly, \Imode{} exhibits some features of \rmp{} and \negd{} plasmas: a strong core contribution and possible correlation with \Vcore{} and \Ptot{}. \Lmode{} plasmas also show evidence of a \Ptot{} and \Vcore{} correlation, as would be expected for a regime solely dependent on good core performance.

% ---- ---- ---- ---- -------- ----  ---- ---- ---- ---- ---- ---- ---- ---- ---- ----
% ---- ---- ---- ---- -------- ----  ELMing SECTION   ---- ---- ---- ---- ---- ----
% ---- ---- ---- ---- -------- ----  ---- ---- ---- ---- ---- ---- ---- ---- ---- ----

\section{Comparison to Plasma Performance with ELMs}
\label{sec:ELMs}

A logical interlude at this point is to compare the no-ELM plasma performance and core-pedestal correlations discussed in the preceding sections to DIII-D ELMing plasmas highlighted in the literature.  This section repeats key figures of these two sections but now including ELMing data points. While less effort was undertaken to ensure the highest performance ELMing pulses were found against all abscissa, the ELMing database consists of highlighted `trophy' discharges across the breadth of ELMing regimes studied in DIII-D. Roughly in order of increasing \In{} (decreasing \qnf{}), representative high-performance discharges in high $\beta_{P}$ \cite{Politzer2005,Garofalo2015,Garofalo2018}, high $\ell_{i}$ \cite{Ferron2015}, high $q_{min}$\cite{Holcomb2014,Ferron2011}, steady-state hybrid \cite{Petrie2017,Turco2015}, advanced inductive \cite{Luce2003,Wade2005,Politzer2008a,Solomon2013}, super \Hmode{} \cite{Snyder2019,Knolker2020}, and ITER baseline scenario \cite{Doyle2010,Luce2014IAEA,Luce2018IAEA} are included. This database should not be considered comprehensive in the absolute best of DIII-D ELMing performance, but rather representative of featured high-performance ELMing discharges created since the installation of the divertor baffling structures. Note significantly higher ELMing absolute performance can be found with the higher \elong{}, \triavg{}, and \Ip{} plasmas precluded by the modern DIII-D divertor \cite{Strait1995b,Lazarus1996,Lazarus1997}. Furthermore, no effort was made to include low performance ELMing discharges. The ELMing operating space figure (analogous to no-ELM Fig. \ref{fig:basics}) is not shown, as ELMing plasmas can generally be found everywhere in the DIII-D operating space (excepting low \PoverPLH{}).

\begin{figure*}[!h]
\begin{subfigure}{1\textwidth}
\begin{subfigure}{0.325\textwidth}
\centering
\includegraphics[width=1\textwidth]{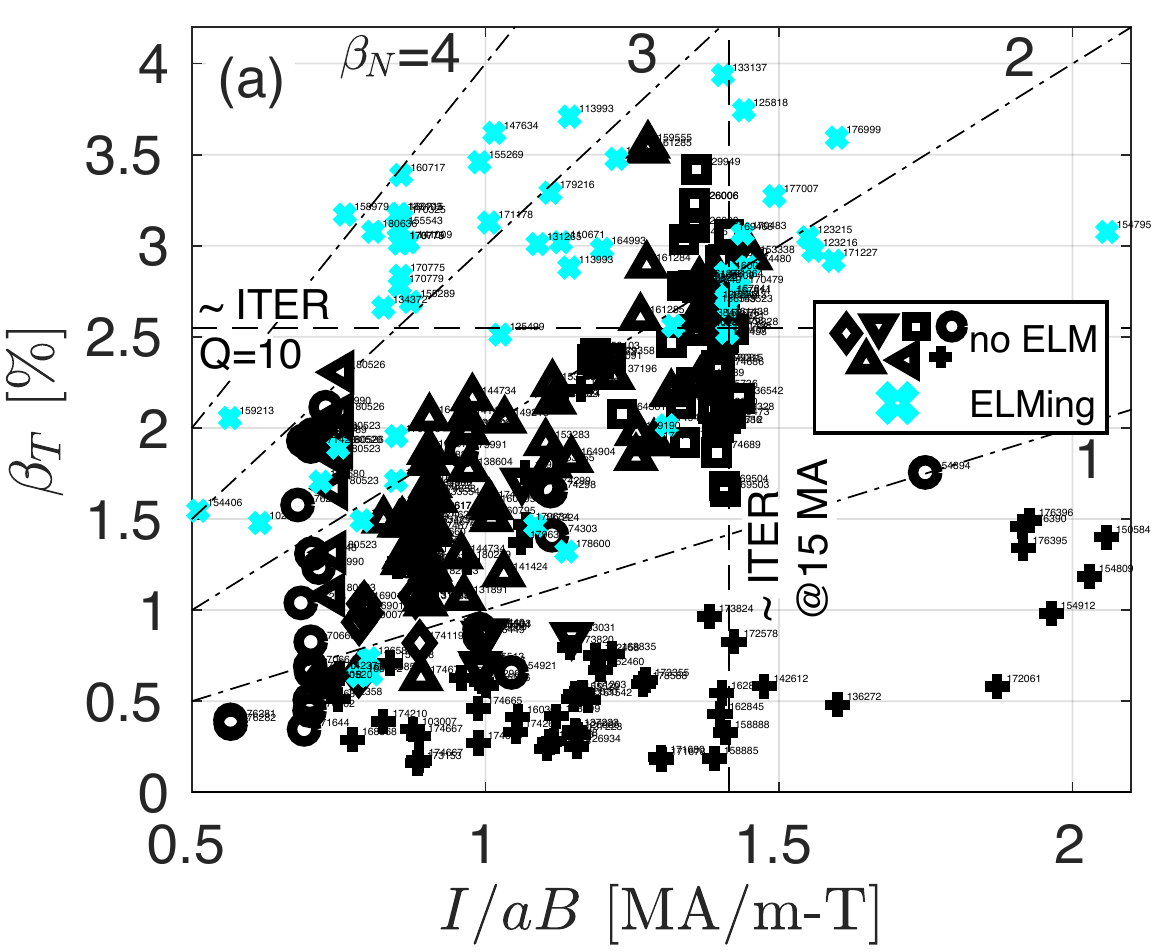}
\end{subfigure}
\begin{subfigure}{0.325\textwidth}
\centering
\includegraphics[width=1\textwidth]{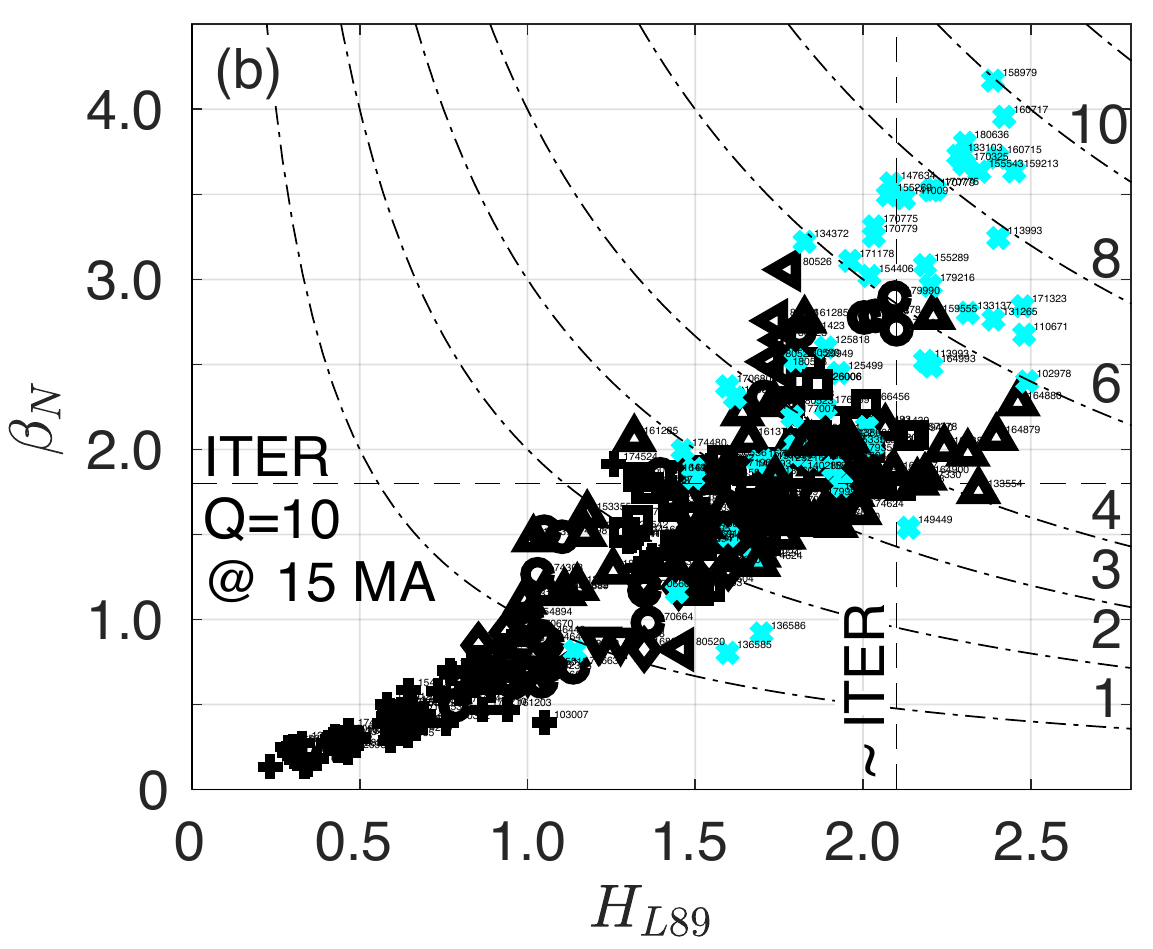}
\end{subfigure}
\begin{subfigure}{0.325\textwidth}
\centering
\includegraphics[width=1\textwidth]{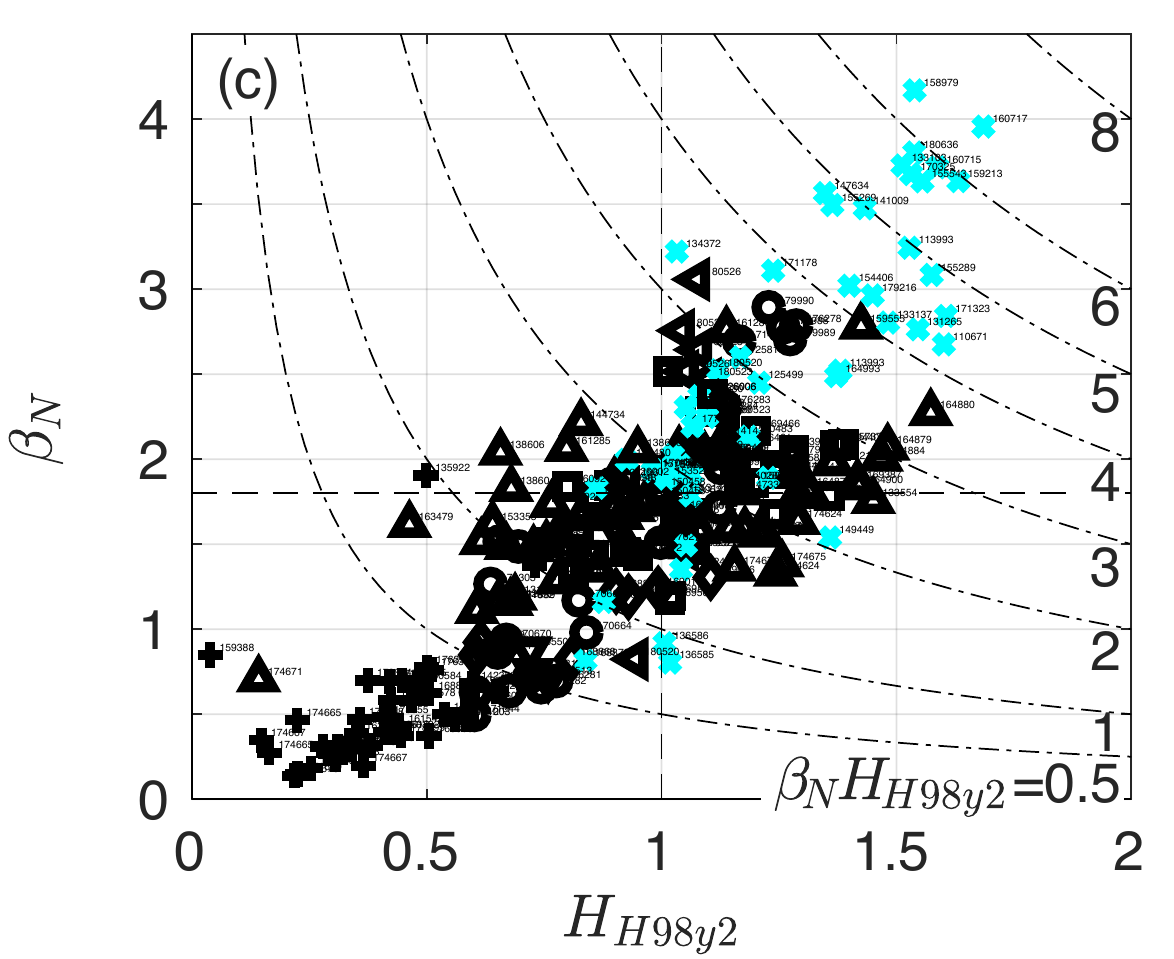}
\end{subfigure}
\end{subfigure}
\vspace{-5 pt}
\caption{Comparison of (a) normalized current (\In{}) and pressure (\betan{}), (b) confinement quality factors for L-mode (\HL{}) and (c) \Hmode{} (\HH{}) against \betan{} for ELMing and no-ELM plasmas.}
\label{fig:ELM-H}
\end{figure*}

Discussion begins with normalized performance, shown in Fig. \ref{fig:ELM-H} (to be compared with no-ELM Fig. \ref{fig:Hfactors}).  The Troyon stability diagram [Fig. \ref{fig:ELM-H}(a)] shows a swath of ELMing datapoints at high \betat{} for various \In{}, generally showing many scenario paths to high performance. Stationary discharges at considerably higher \betan{} can be found (up to above \betan{}=4). Also notable on the Troyon diagram are ELMing data at \In{}=2 (\qnf{} $\approx$ 2.3), which while stationary for over 0.3 s (thus meeting the inclusion criteria of Sec. \ref{sec:criteria}) evolve to tearing instability on a longer timescale. Normalized confinement quality (\HL{} and \HH{}) are found to be comparable to the best no-ELM plasmas, but high H-factor is now sustained to much higher \betan{} and occurs over a wide range of scenarios.

\begin{figure*}[!h]
\begin{subfigure}{1\textwidth}
\begin{subfigure}{0.325\textwidth}
\centering
\includegraphics[width=1\textwidth]{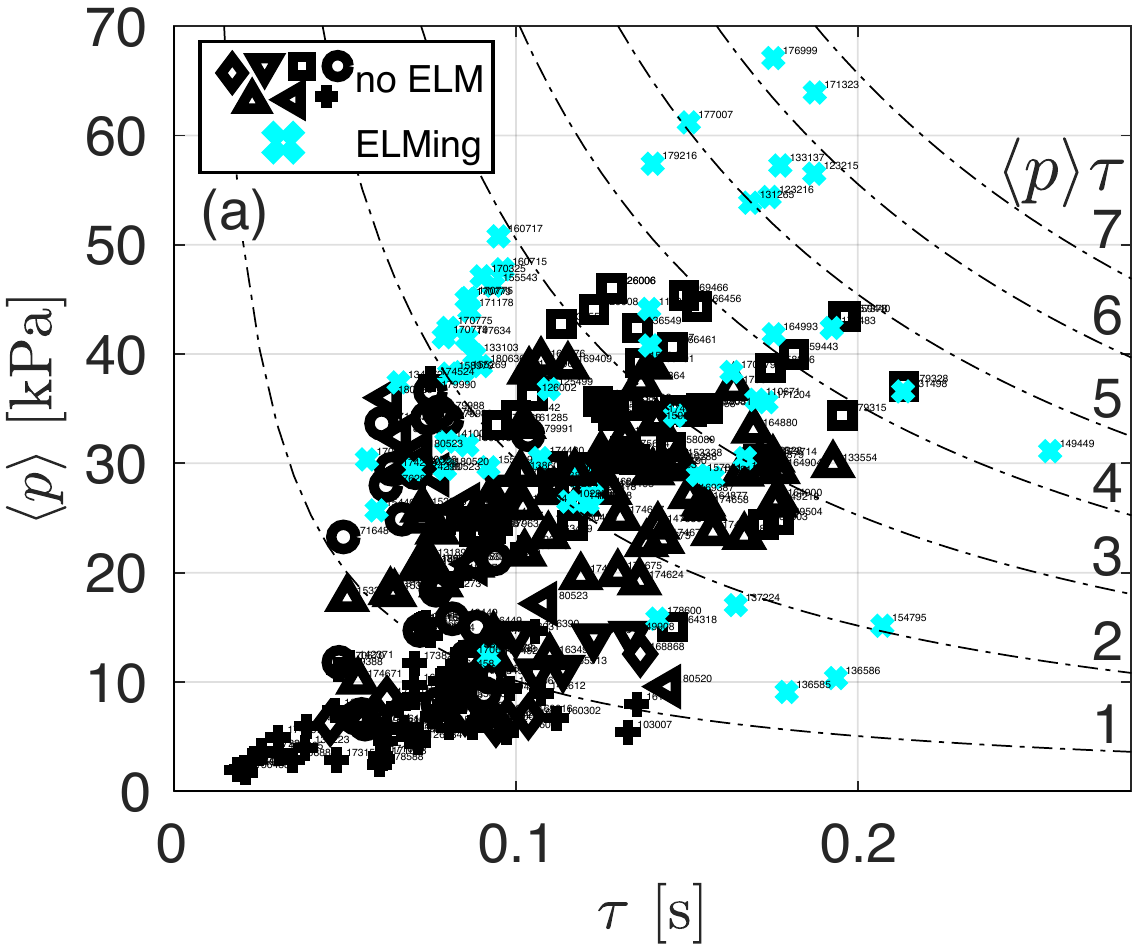}
\end{subfigure}
\begin{subfigure}{0.325\textwidth}
\centering
\includegraphics[width=1\textwidth]{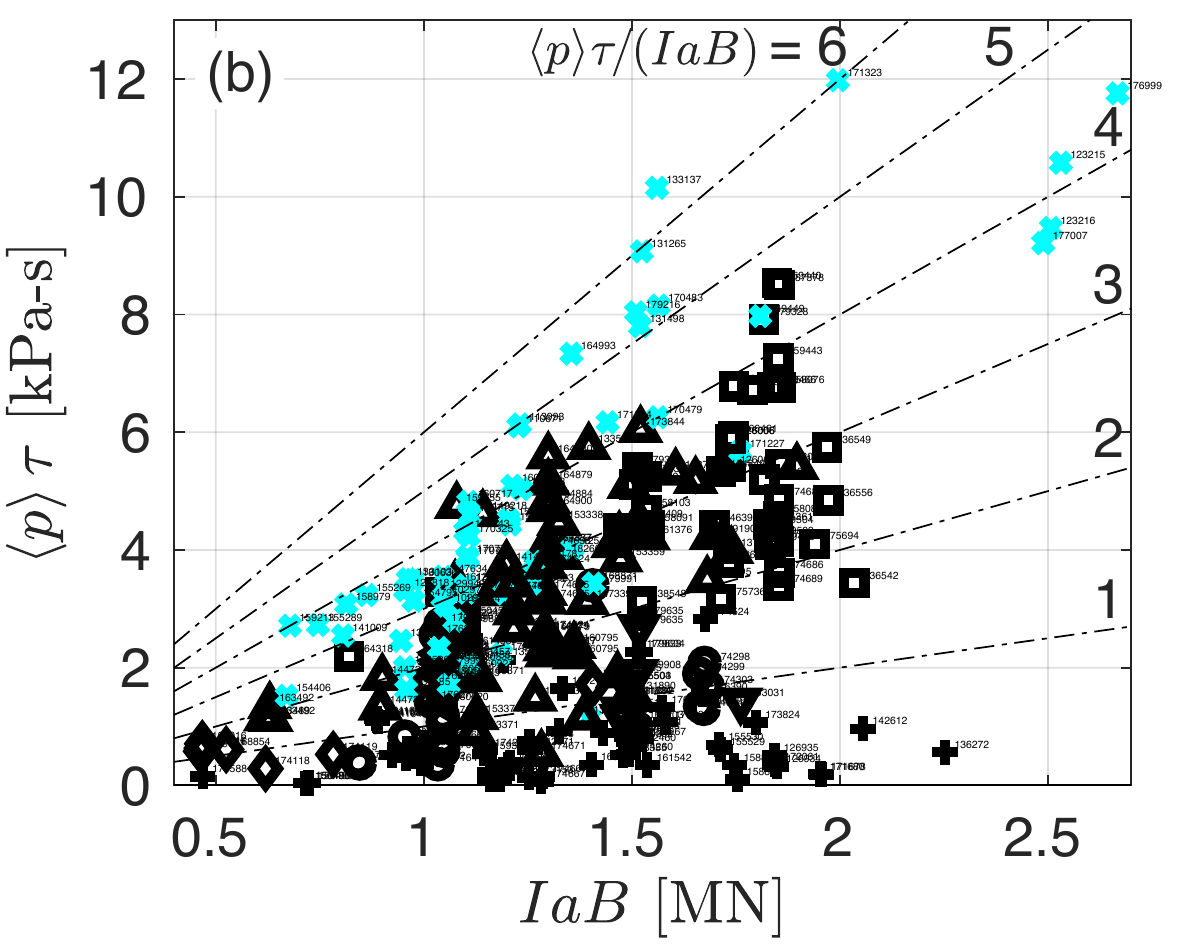}
\end{subfigure}
\begin{subfigure}{0.325\textwidth}
\centering
\includegraphics[width=1\textwidth]{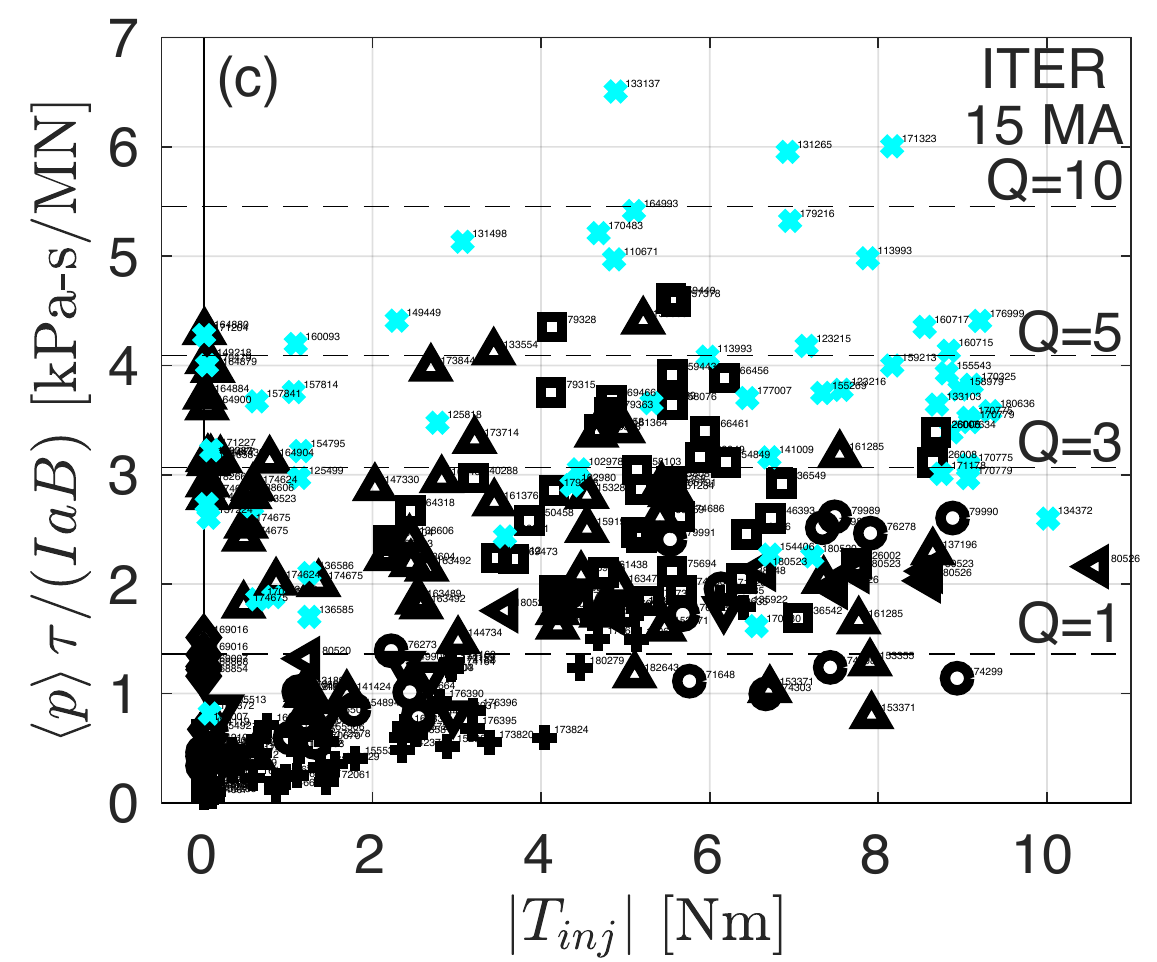}
\end{subfigure}
\end{subfigure}
\vspace{-5 pt}
\caption{Comparison of (a) total pressure \pres{} and confinement time (\taue{}), (b) triple product (\lawson{}) and \IaB{}, (c) normalized triple product (\snyder{}) and torque (\Tinj{}) for ELMing and no-ELM plasmas.}
\label{fig:ELM-abs}
\end{figure*}

% ELMing absolute performance
Improved absolute performance is also found in ELMing plasmas, as shown in Fig. \ref{fig:ELM-abs} (to be compared with no-ELM Fig. \ref{fig:abs}). ELMing plasmas reach higher \taue{}, higher \pres{}, and higher \lawson{}. The highest ELMing \lawson{} is found for advanced inductive and super-H plasmas, which as shown in Fig. \ref{fig:ELM-abs}(b) operate at high \IaB{}. Improved ELMing performance to no-ELM performance is found at all \IaB{}. A such, normalized triple product (\snyder{}) as shown in Fig. \ref{fig:ELM-abs}(c) is robustly higher by about a third, and meets ITER 15 MA Q=10 levels. Plotting \snyder{} against \Tinj{} reveals a performance degradation in ELMing plasmas more severe than no-ELM plasmas [compare to Fig. \ref{fig:abs}(c)]. The performance at \Tinj{}=0 quantified by \snyder{} is actually the same with and without ELMs, though \qnf{} is quite different.

\begin{figure*}[!h]
\begin{subfigure}{1\textwidth}
\begin{subfigure}{0.325\textwidth}
\centering
\includegraphics[width=1\textwidth]{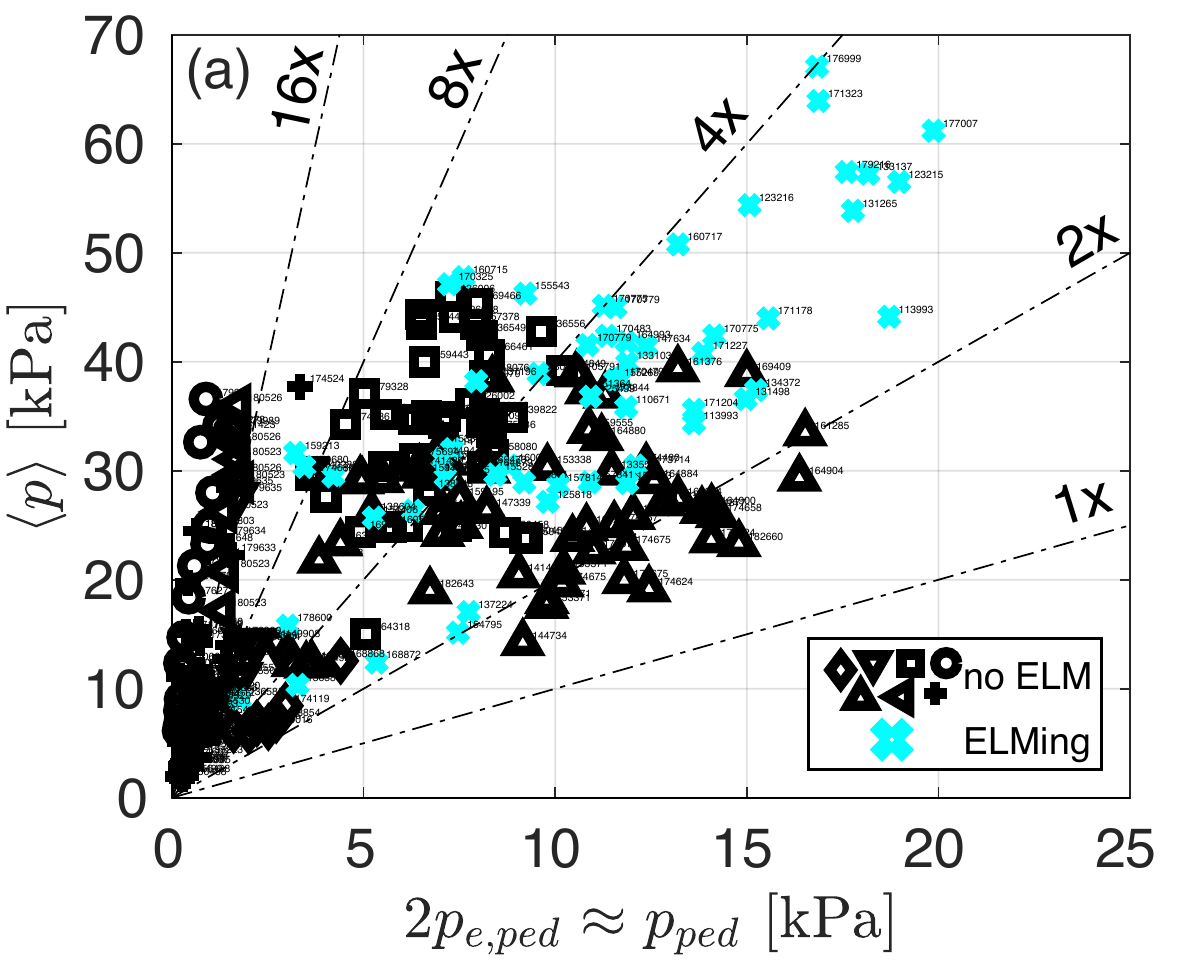}
\end{subfigure}
\begin{subfigure}{0.325\textwidth}
\centering
\includegraphics[width=1\textwidth]{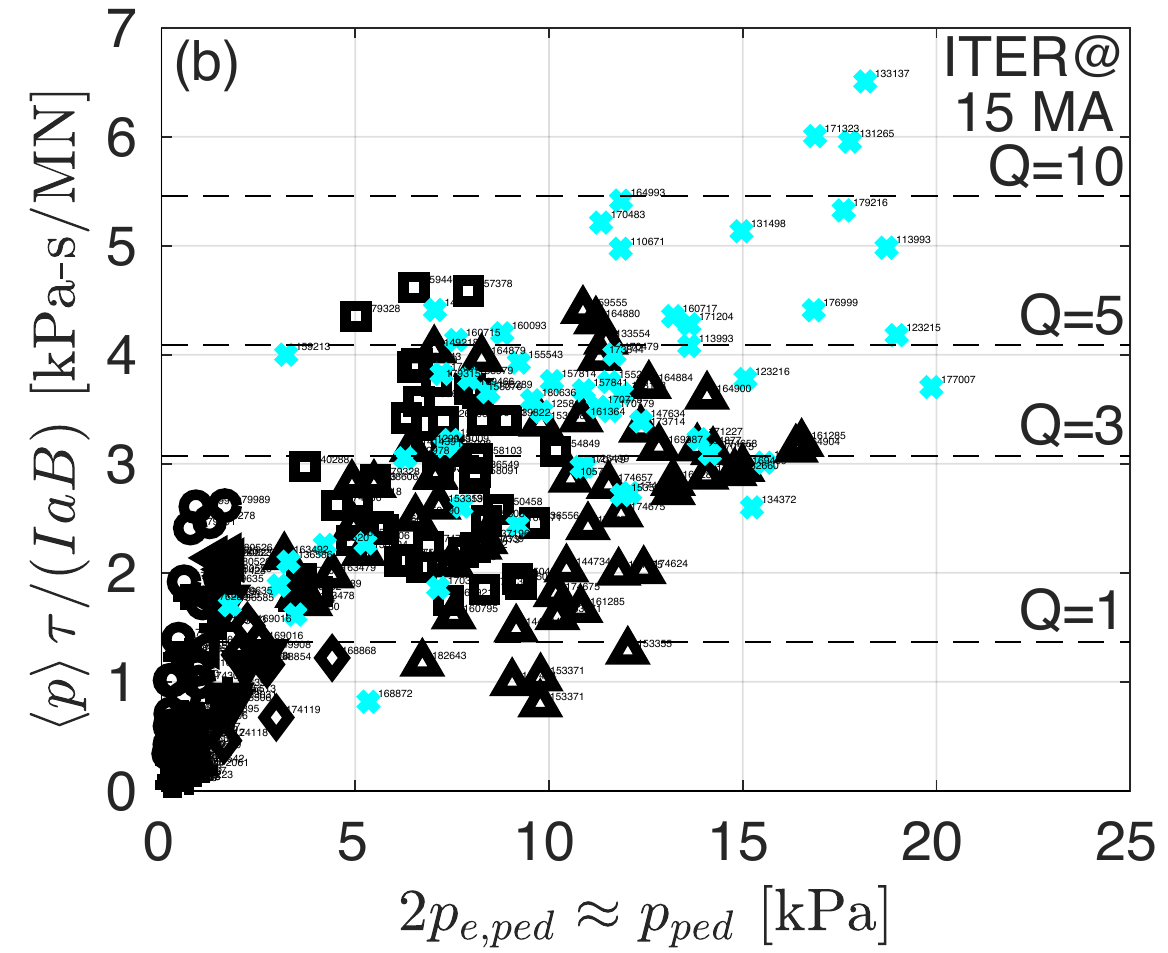}
\end{subfigure}
\begin{subfigure}{0.325\textwidth}
\centering
\includegraphics[width=1\textwidth]{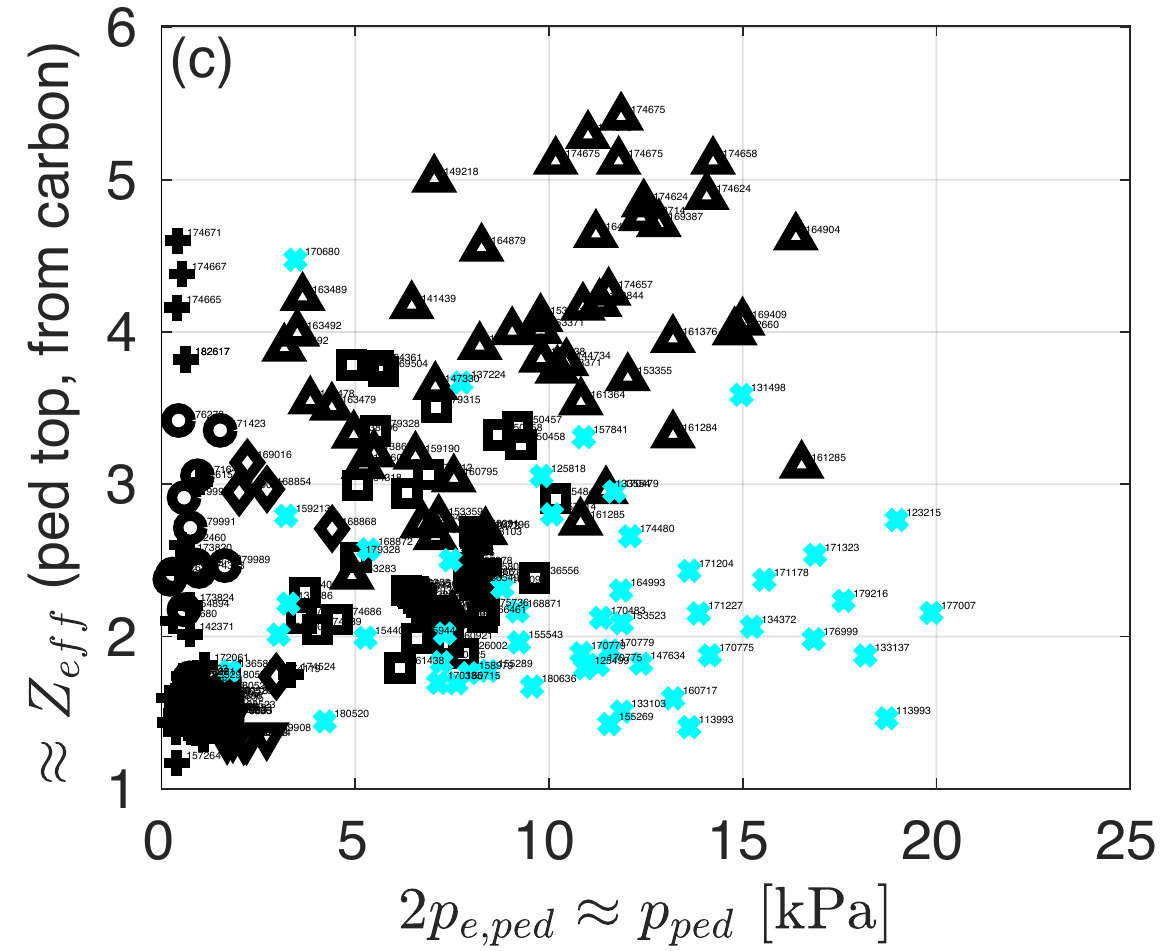}
\end{subfigure}
\end{subfigure}
\vspace{-10 pt}
\caption{Comparison of (a) pedestal (\pped{}) contribution to \pres{}, (b) \pped{} dependence on \snyder{}, and (c) impurity content \Zeff{} correlation with \pped{} for ELMing and no-ELM plasmas.}
\label{fig:ELM-ped}
\end{figure*}

% ELMing pedestal
The pedestal pressure (\pped{}) contribution to ELMing plasma performance is summarized in Fig. \ref{fig:ELM-ped} (to be compared with no-ELM Fig. \ref{fig:coreped}). The pressure fraction from the pedestal [Fig. \ref{fig:ELM-ped}(a)] is on the high end as compared to no-ELM plasmas, with \ppedratio{} between 2 and 4, and is in a narrower range than no-ELM plasmas. About 50\% higher \pped{} can be achieved in stationary ELMing conditions as compared to no-ELM conditions, and a good correlation with \pped{} and \snyder{} is found for ELMing data as shown in Fig. \ref{fig:ELM-ped}(b). Unlike no-ELM plasmas [Fig. \ref{fig:coreped}(c)], no positive \pped{} correlation with \Zeff{} is found for ELMing plasmas, shown in Fig. \ref{fig:ELM-ped}. Instead, the highest \pped{} ($>$15 kPa) ELMing plasmas are almost all found below \Zeff{}$<$3. This is consistent with the hypothesis that the ELM plays an important role in exhausting impurities.

\begin{figure*}
\begin{subfigure}{1\textwidth}
\begin{subfigure}{0.325\textwidth}
\centering
\includegraphics[width=1\textwidth]{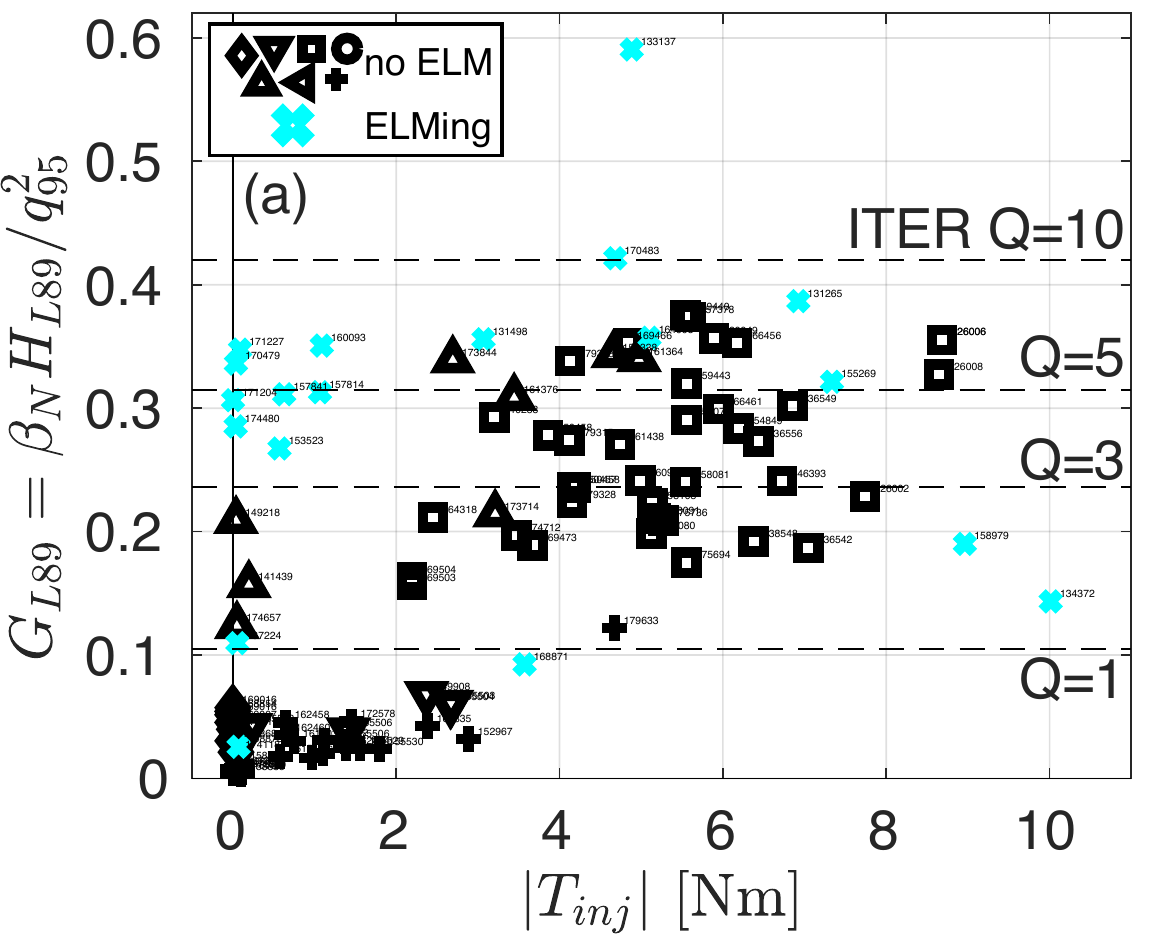}
\end{subfigure}
\begin{subfigure}{0.325\textwidth}
\centering
\includegraphics[width=1\textwidth]{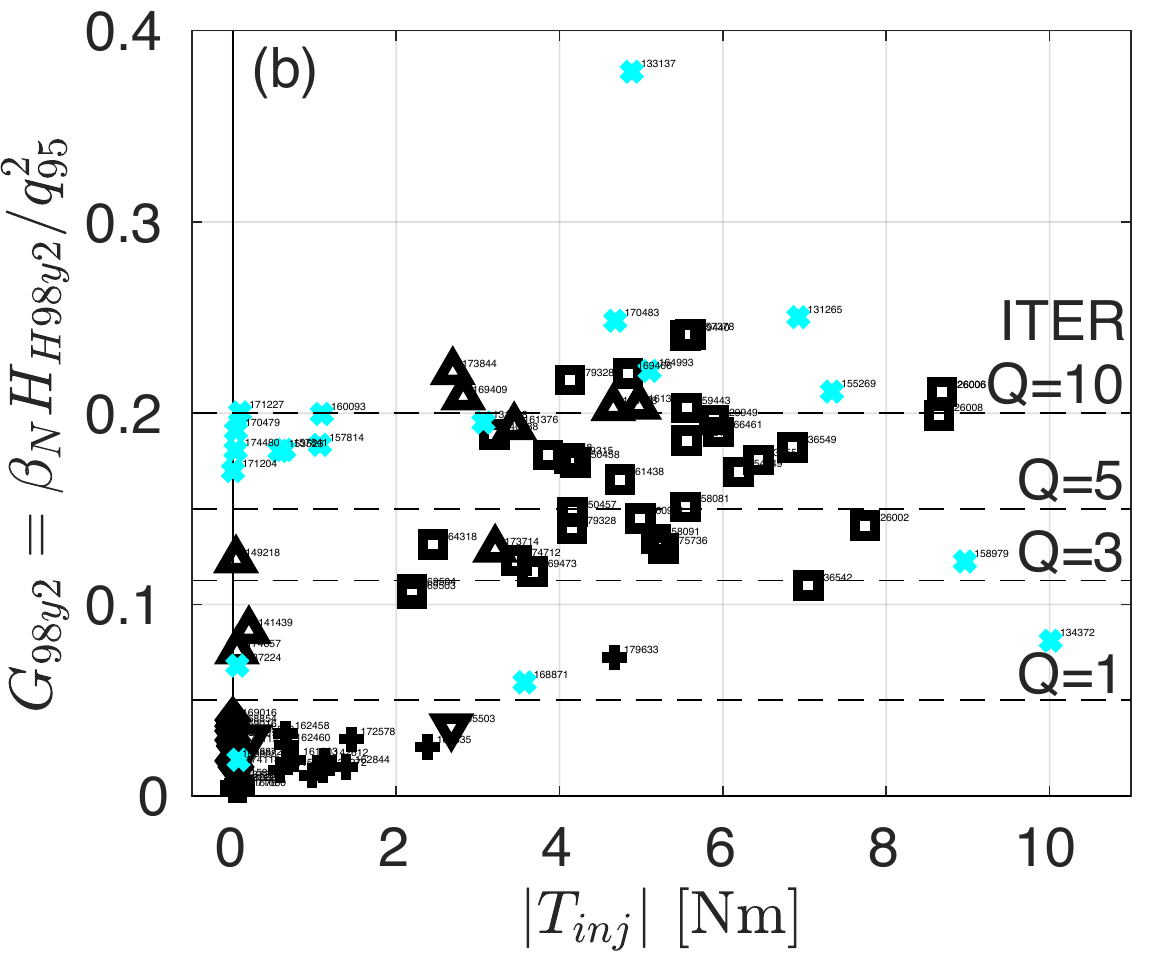}
\end{subfigure}
\begin{subfigure}{0.325\textwidth}
\centering
\includegraphics[width=1\textwidth]{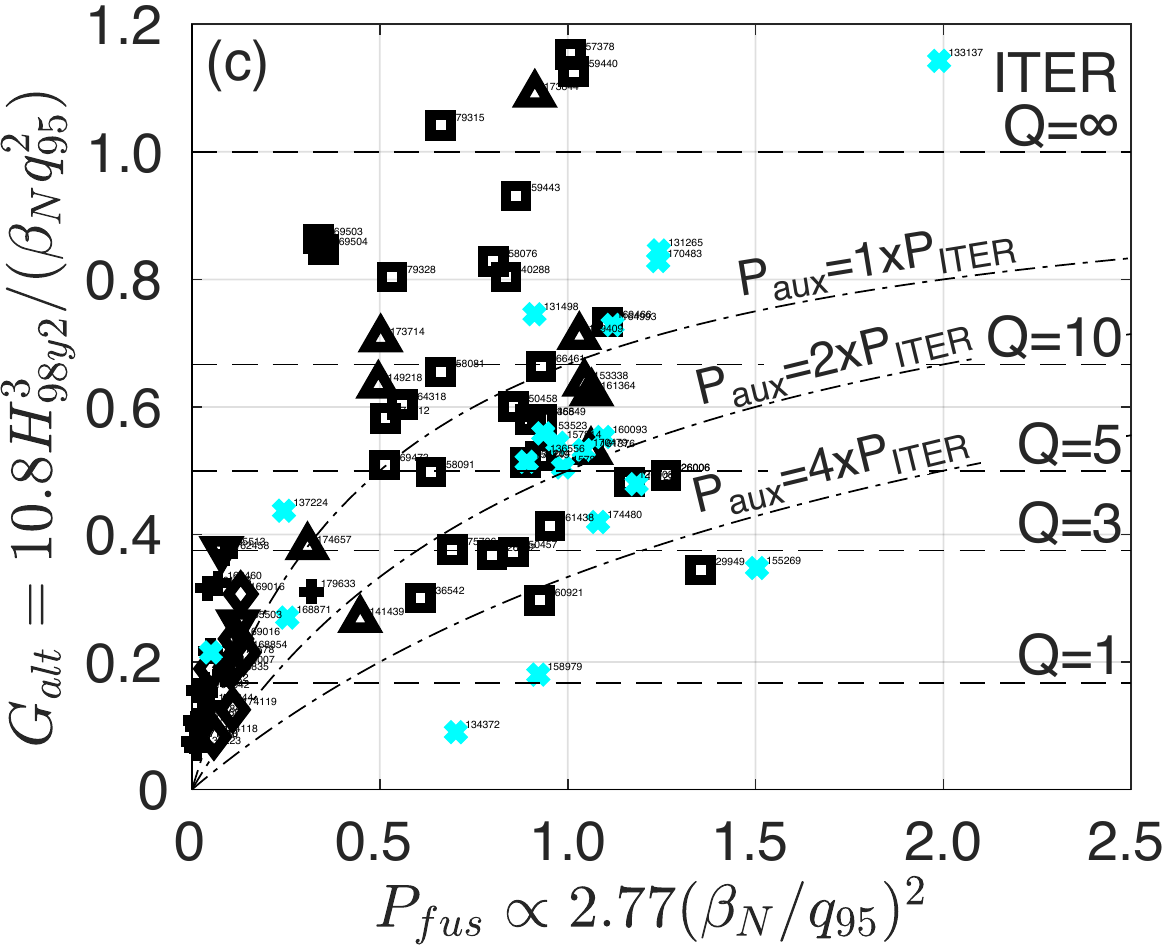}
\end{subfigure}
\end{subfigure}
\vspace{-5 pt}
\caption{Comparison of ELMing and no-ELM plasmas in terms of figures of merit for fusion performance isolating discharges with ITER-like plasma shape. (a) Gain factor based on L-mode scaling (\GH{}) and (b) L-mode scaling (\GH{}) plotted against torque \Tinj{}, and (c) alternate gain factor (\Galt{}) against scaling of fusion power.}
\label{fig:ELM-ITER}
\end{figure*}

% ELMing ITER
The performance of ELMing plasmas within ITER shape constraints is shown in Fig. \ref{fig:ELM-ITER} (to be compared with no-ELM Fig. \ref{fig:ITER}). Performance measured by \GL{} [Fig. \ref{fig:ELM-ITER}(a)] and \GH{} [Fig. \ref{fig:ELM-ITER}(b)] is found to decrease with \Tinj{}, as was found with \snyder{} in Fig. \ref{fig:ELM-abs}(c). Interestingly, at low/zero \Tinj{}, the ITER Q=10 criterial is met according to \GH{} but not \GL{}, which falls just short. Exceptionally high \GH{}/\GL{} discharges are found at high \Tinj{}, corresponding to the advanced inductive scenario \cite{Doyle2010}. Re-visiting the analysis of Ref. \cite{Peeters2007} for ELMing discharges [Fig. \ref{fig:ELM-ITER}(c)] reveals many datapoints that extrapolate to Q=10 or greater within ITER's \Paux{} constraints. However, the high \GL{}/\GH{} datapoints at \Tinj{} $<$2 Nm do not meet the ITER \Paux{} criteria.

% ---- ---- ---- ---- -------- ----  ---- ---- ---- ---- ---- ---- ---- ---- ----
% ---- ---- ---- ---- -------- ----    INTEGRATION   ---- ---- ---- ---- ----
% ---- ---- ---- ---- -------- ----  ---- ---- ---- ---- ---- ---- ---- ---- ----

\section{No-ELM Plasma Integration with Electron Heating and Dissipative Divertors}
\label{sec:integ}

Viable tokamak scenarios without ELMs must also be integrated with two essential features of burning plasmas: auxiliary electron heating and a dissipative divertor state. The former models the heating from fusion-produced alpha particles, while the latter is mandatory to manage the heat flux to the material boundary. While sophisticated modeling (of heat flux channels and atomic dissipation) is necessary to define the operating point of a regime against burning plasma requirements, a few simple and robust metrics can be considered. For electron heating, the ratio of electron cyclotron heating (ECH) power delivered by high power microwaves to the total power (\echratio{}), and for divertor dissipation the fraction of radiated power to total power (\frad{}) and the ratio of the density at the separatrix to the Greenwald density (\nsepoverng{}). The motivation for these metrics and how no-ELM plasmas perform against them will now be described in turn.

\begin{figure}
\begin{subfigure}{0.325\textwidth}
\begin{subfigure}{1\textwidth}
\centering
\includegraphics[width=1\textwidth]{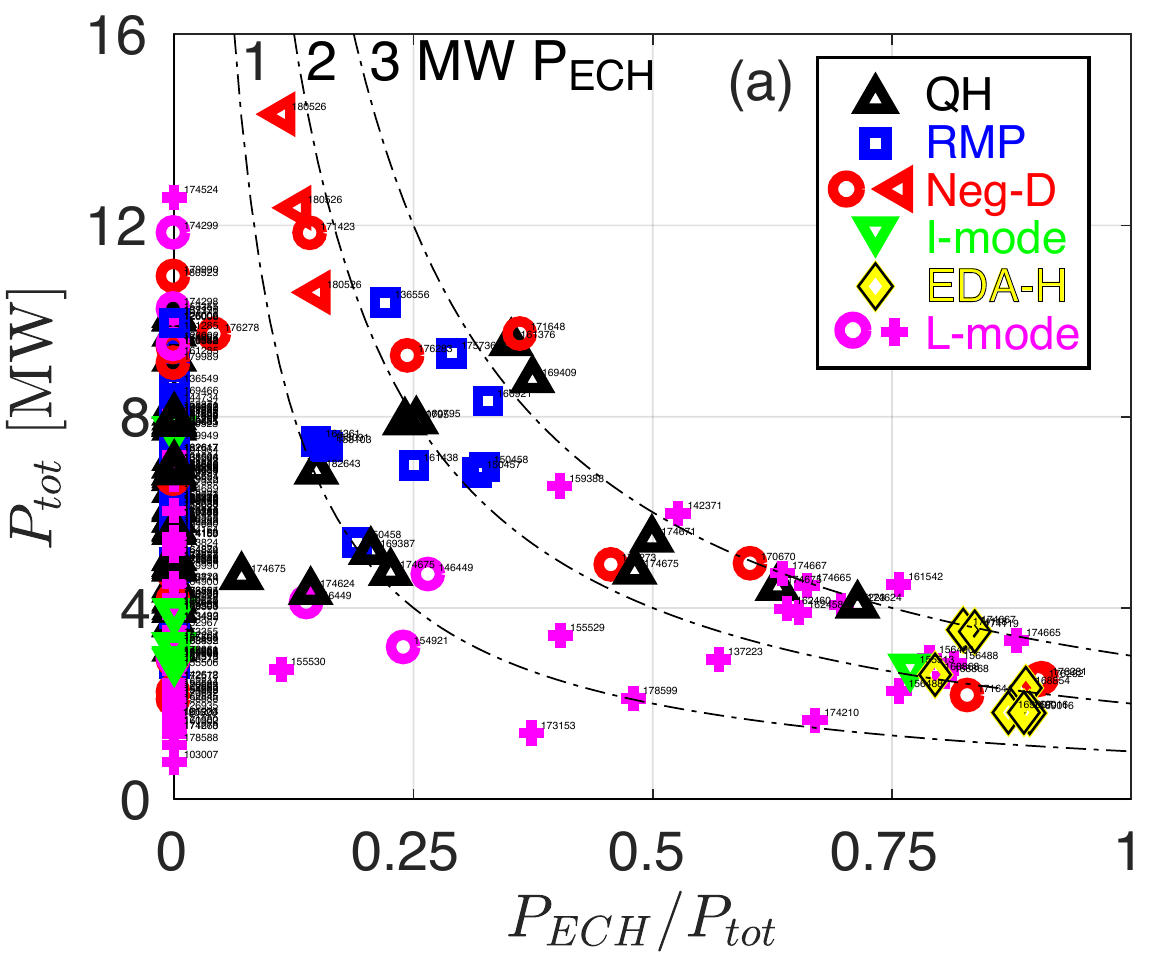}
\end{subfigure}
\begin{subfigure}{1\textwidth}
\centering
\includegraphics[width=1\textwidth]{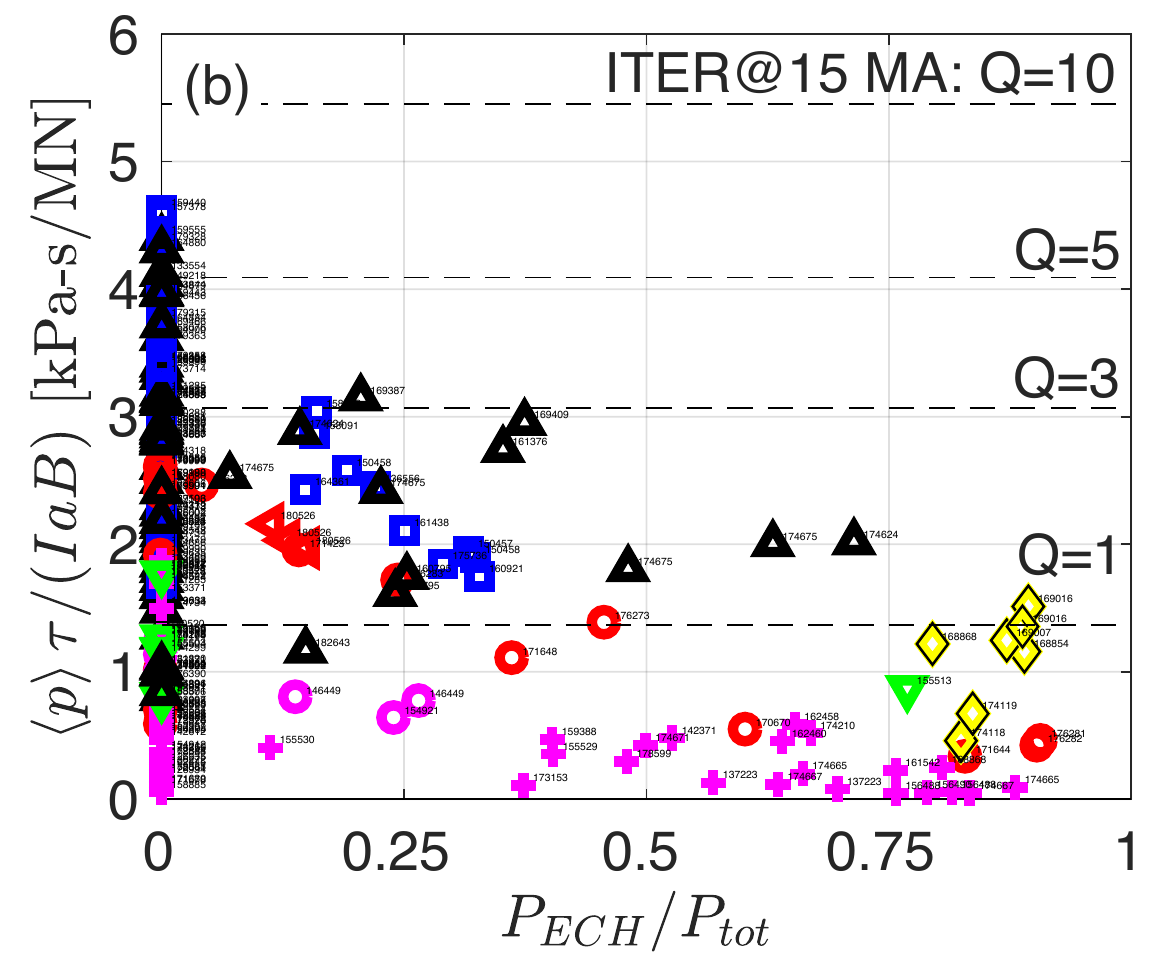}
\end{subfigure}
\end{subfigure}
\vspace{-5 pt}
\caption{Integration with electron heating: (a) operational space in terms of \echratio{} and \Ptot{}, with dashed lines indicating constant \Pech{}. (b) Plasma performance (\snyder{}) is found to fall with increasing \echratio{}.}
\label{fig:ECH}
\end{figure}

% ---- ECH ----
\subsection{Boundaries and Performance towards Dominant Electron Heating}
\vspace{-10 pt}
% ECH intro
Burning plasmas self-heat through the collisional slowing down of MeV alpha particles, and at this high energy the dominant heat deposition is to the electrons, modifying the plasma turbulent transport properties \cite{Petty1999,Prater2004,Kaye2013,Howard2016,Angioni2017,Grierson2018}. DIII-D has long been equipped with $\approx$ 3 MW of ECH power (\Pech{}) to mimic this heating channel, though this is a small fraction of the available NBI power ($\approx$ 16 MW). NBI heats ions preferentially and depending on the NBI configuration can also impart significant \Tinj{}, rotation, and ExB shear, as described in Sec. \ref{sec:corr}. The mix of ion and electron heating required to mimic a burning plasma depends on the detailed kinetic profiles expected in the plasma core.

% Operation
 Plotting \Ptot{} against \echratio{} in Fig. \ref{fig:ECH}(a) makes clear the dominant operational consideration is a reduced access to high \Ptot{} at high \echratio{} on DIII-D, entirely due to limited available \Pech{}. \rmp{} plasmas are observed across the full range of \Pech{}, though interestingly only at fairly high \Ptot{} and low \echratio{}. This operating range is found to be limited by the return of ELMs at low rotation and pressure, though this threshold has not yet been explored in detail. \QH{} plasmas are also observed across the full range of \Pech{}, and they clearly access higher \echratio{} (lower \Ptot{}) \cite{Ernst2018IAEA}. For the \negd{}, \Imode{}, \QH{}, and \EDA{} regimes no operational limit in \echratio{} is found if \Lmode{} is avoided. Indeed, the \EDA{} in DIII-D is thus far {\it{only}} identified at high \echratio{}.

% ECH performance
Plasma performance quantified by \snyder{} is plotted against \echratio{} in Fig. \ref{fig:ECH}(b). The loss of \Ptot{} access is clearly manifest in a reduced \snyder{}. The performance degradation for \rmp{} plasmas is particularly severe, supporting the particularly important role of rotation shown in Fig. \ref{fig:altcorr}(b). Unlike past \QH{} results at high \Tinj{} \cite{Ernst2016}, recent experiments in low \Tinj{} \QH{} plasmas (specifically the wide-pedestal variant at high \qnf{}) have found a performance improvement with \echratio{} at fixed \Ptot{} \cite{Ernst2018IAEA}. This indicates the higher performance \echratio{}=0 \QH{} plasmas are likely benefiting from other factors such as higher \Ptot{} or lower \qnf{}. Interestingly, against this metric the two regimes found to have the highest performance are \QH{} and \EDA{}, which as shown in Fig. \ref{fig:coreped}(a) also have the highest fractional \ppedratio{}. As such, a relatively strong pedestal is correlated with an improved \snyder{} at high \echratio{} and low \Ptot{}. The observed performance at high \Ptot{} and high \echratio{} is a highly relevant and compelling question \cite{Ernst2020APS}, awaiting increased \Pech{} to be available in DIII-D and has motivated similar research programs elsewhere \cite{Stober2020}.

\begin{figure*}
\begin{subfigure}{1\textwidth}
\begin{subfigure}{0.325\textwidth}
\centering
\includegraphics[width=1\textwidth]{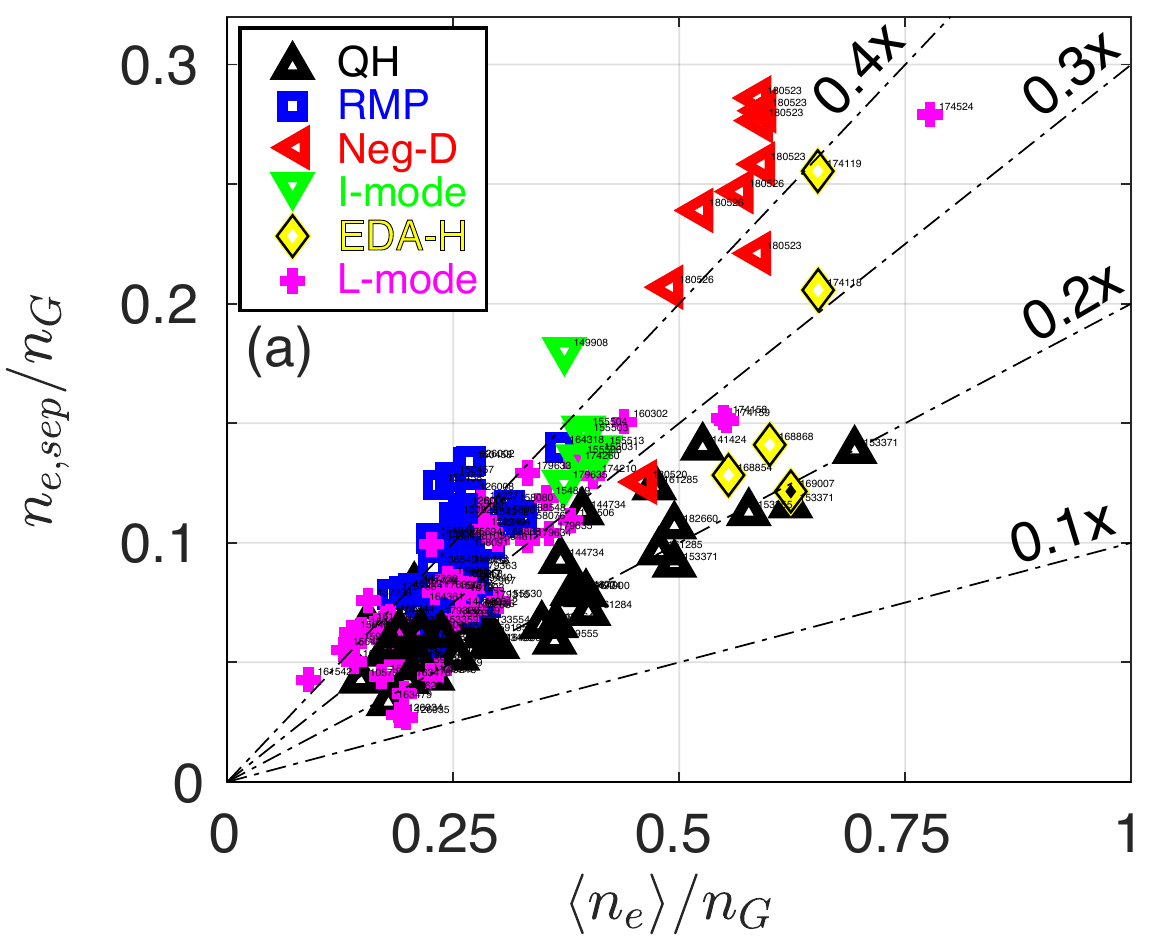}
\end{subfigure}
\begin{subfigure}{0.325\textwidth}
\centering
\includegraphics[width=1\textwidth]{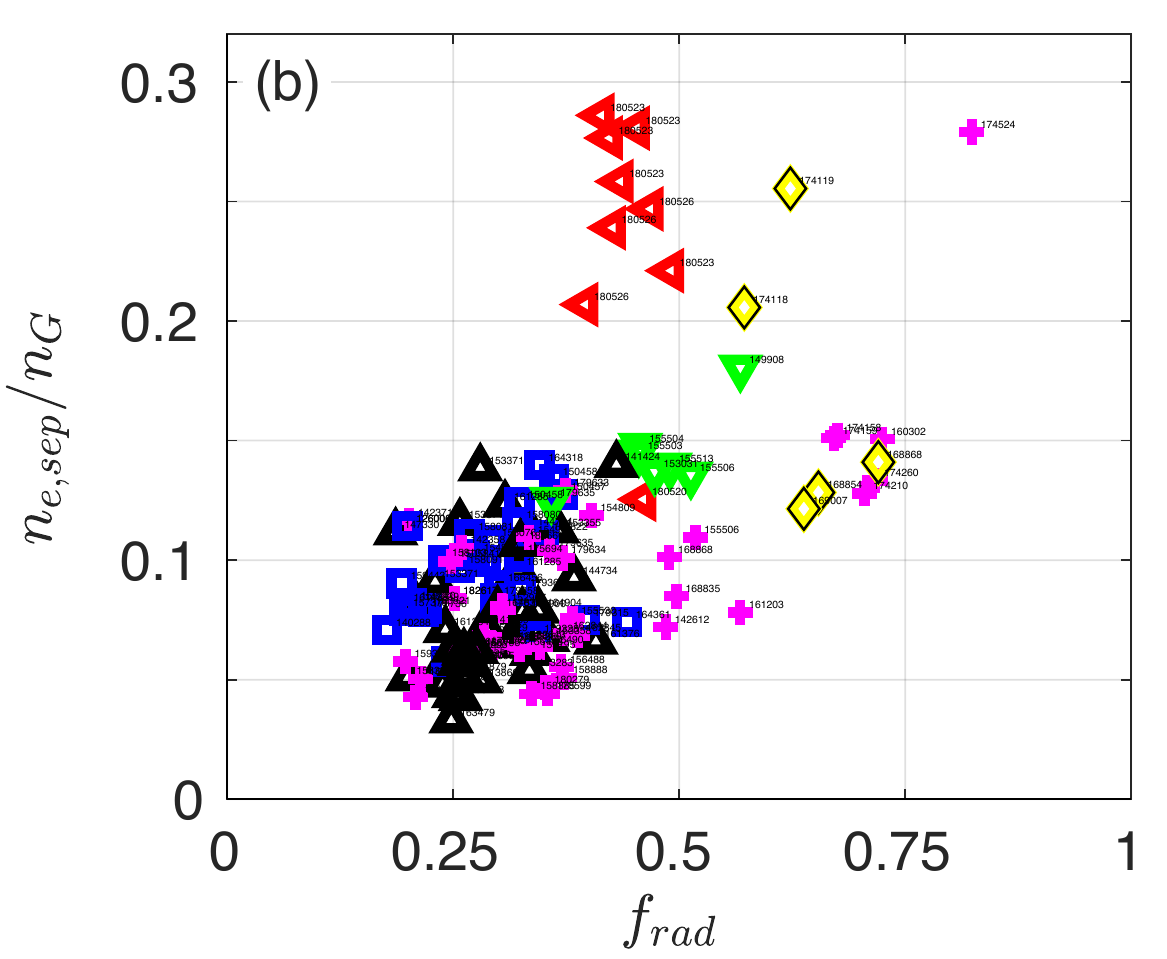}
\end{subfigure}
\hspace{1 in}
%\vspace{-5 pt}
\begin{subfigure}{0.325\textwidth}
\centering
\includegraphics[width=1\textwidth]{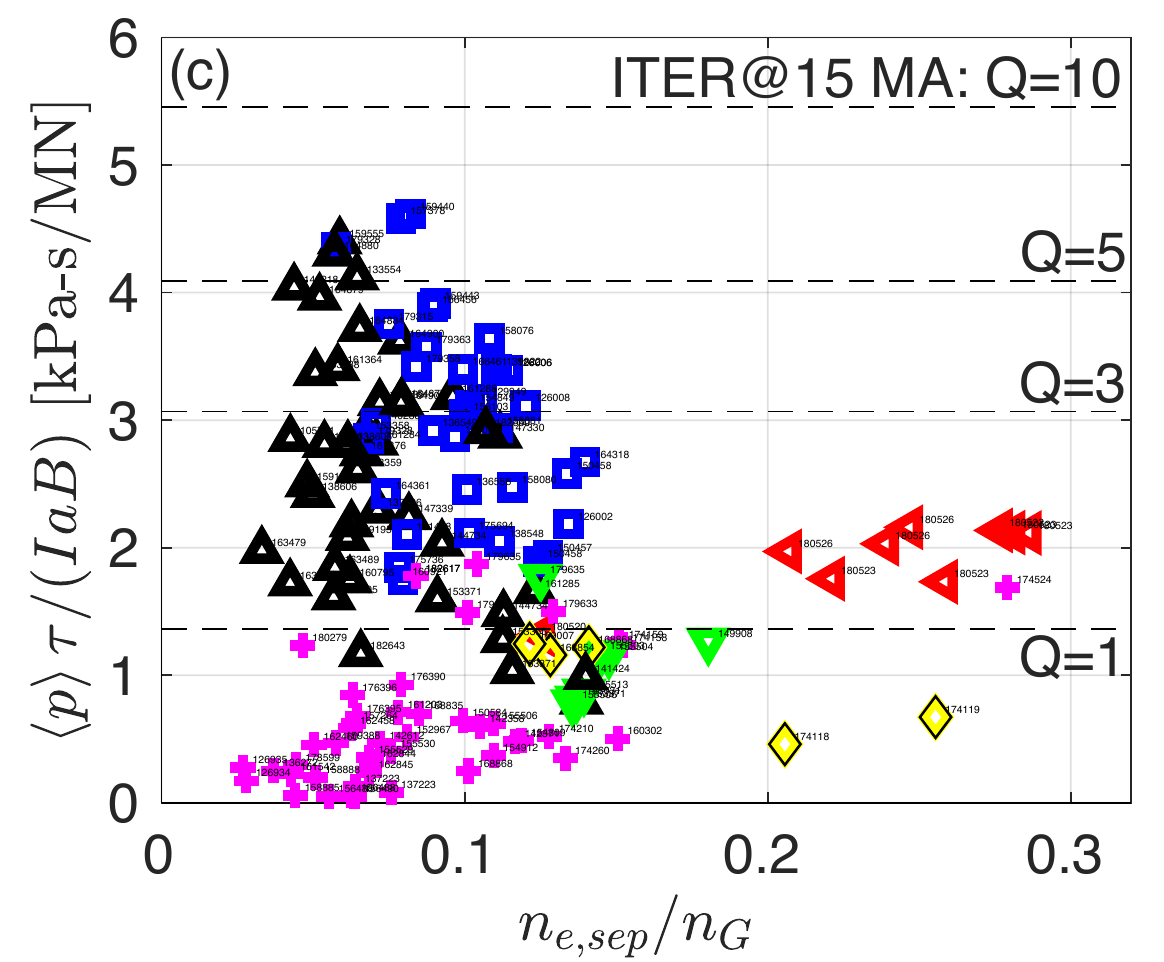}
\end{subfigure}
\begin{subfigure}{0.325\textwidth}
\centering
\includegraphics[width=1\textwidth]{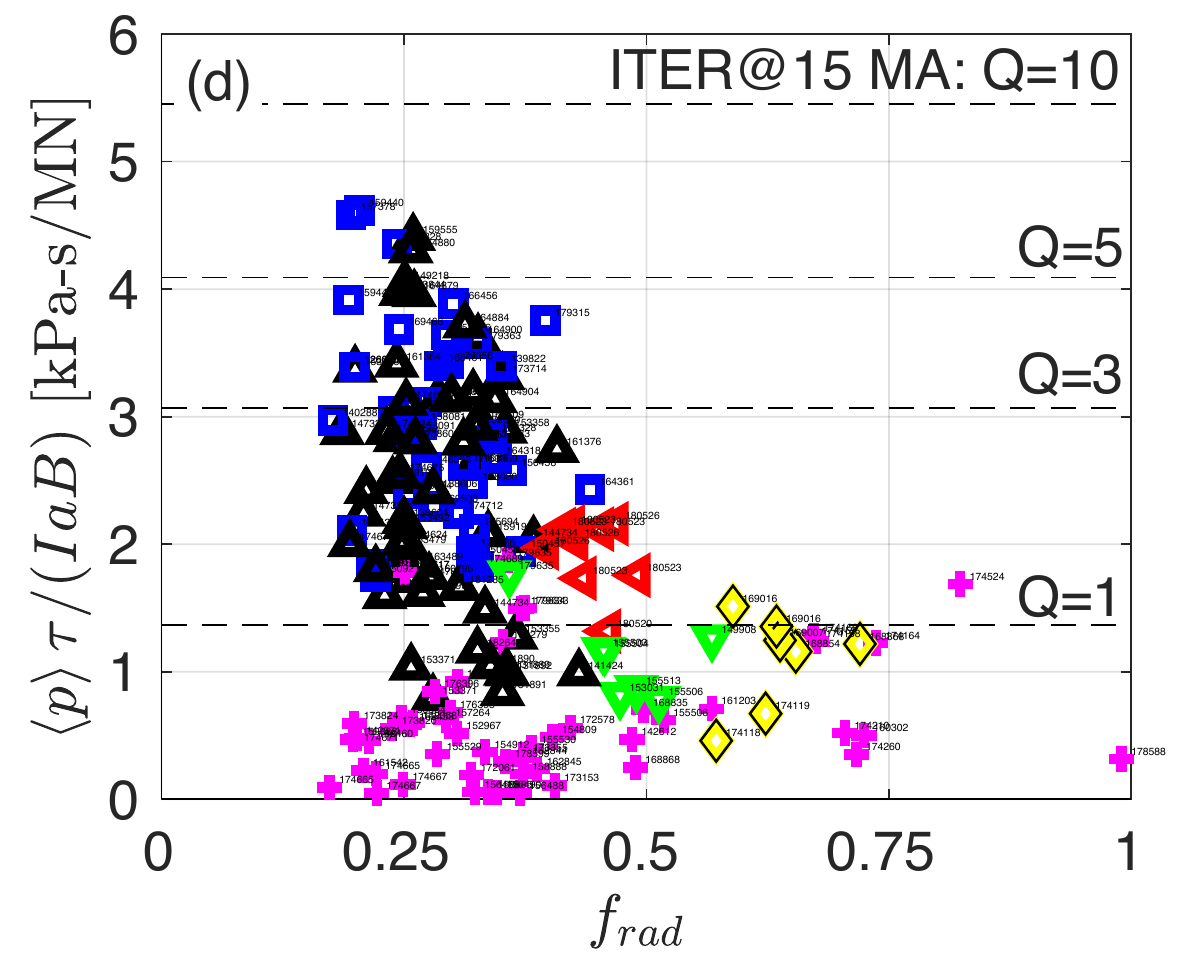}
\end{subfigure}
\end{subfigure}
\vspace{-5 pt}
\caption{Integration with dissipative divertor: (a) comparison of line-averaged \noverng{} to separatrix density \nsepoverng{}, (b) operational space in terms of \frad{} and \nsepoverng{}. Plasma Performance measured by \snyder{} against (c) \nsepoverng{} and (d) \frad{}, indicating decreasing performance with increasing divertor compatibility.}
\label{fig:div}
\end{figure*}

\subsection{Boundaries and Performance towards Dissipative Divertor Conditions}

% Divertor Intro
The final topic discussed is integration with dissipative divertor conditions. This is a uniquely challenging direction for no-ELM regimes at least in present tokamaks, owing to the inability to simultaneously access low collisionality (favorable to access the current-limited pedestal physics of interest) together with high density (favorable to access highly dissipative divertor regimes).  Results will be presented here with additional interpretation discussed in Sec. \ref{sec:disc}.

% nesep
Access and performance towards divertor integration is presented in Fig. \ref{fig:div}, with limiter plasmas (circle symbols) now excluded. \nsepoverng{} is now used as a density metric instead of \noverng{}\cite{Goldston2017}. This is because the interface of the divertor solution to the plasma occurs at the separatrix (not at the core or pedestal top), though as with \Zeff{} it is a more challenging measurement with increased experimental uncertainty. Here \nsep{} is extracted from automated hyperbolic tangent fits to the electron pedestal profiles. Predictions for the required \nsepoverng{} in a reactor depend sensitively on the divertor geometry and impurity seeding strategy, but in the case of ITER recent modeling \nsepoverng{} values in the range of 0.4-0.7 are used \cite{Kukushkin2013}, which is off the chart in Fig. \ref{fig:div}. Considering \nsepoverng{} in Fig. \ref{fig:div}(a) presents an important distinction between no-ELM regimes. While as shown in Fig. \ref{fig:basics}(d) several regimes access high \noverng{} \cite{Garofalo2015a}, \QH{} plasmas uniquely follow a trajectory of low \nsep{}/\navg{} (consistent with its low-recycling wall operational preference), challenging prospects for divertor integration in this regime \cite{Baylor2005}. \rmp{} plasmas operate at higher \nsep{}/\navg{}, but still limit to a similarly low \nsepoverng{} owing to the low pedestal-top density (\nped{}) limit \cite{Evans2006,Suttrop2018,PazSoldan2019a}. Only the \negd{} and \EDA{} regimes access relatively high \nsepoverng{} in DIII-D, indicating a clear advantage in terms of dissipative divertor integration. Considering observed plasma performance at high \nsepoverng{} in Fig. \ref{fig:div}(c) the \negd{} regime is superior owing to its unique tolerance of both high density and high power.

% f_rad
A second dissipative divertor metric is the radiation fraction (\frad{}=\Prad{}/\Ptot{}, where \Prad{} is the total core and divertor radiated power), as high \frad{} implies less conducted power for the divertor first-wall to handle \cite{Reinke2017}. Again, extrapolation of this parameter to a reactor requires specifying the divertor geometry and \nsepoverng{}, but ITER is expected to require \frad{} $\approx$ 0.8 with about 0.3 inside the separatrix \cite{Shimada2009}. As shown in Fig. \ref{fig:div}(b), \Imode{}, \EDA{}, \Lmode{}, and \negd{} plasmas reach 0.5 \frad{} or greater, while \rmp{} and \QH{} plasmas have not yet reached this level with stationarity at the level defined in Sec. \ref{sec:criteria} (higher has been achieved transiently). Considering observed plasma performance at high \frad{} ($>$0.5) shown in Fig. \ref{fig:div}(d), there is no clear separation observed between the \EDA{}, \Imode{}, and \Lmode{} plasmas, and indeed performance is overall severely degraded.

% Future
Unlike other axes, it should be noted high \frad{} is relatively underexplored in DIII-D no-ELM studies. There have been some dedicated attempts in \rmp{} \cite{Petrie2011,Orlov2018IAEA} with mid-Z radiators such as Ne and Ar and initial attempts in \QH{} plasmas. The collisionless \QH{} and \rmp{} regimes are challenged on these axes by a return of the ELM found at high density and collisionality, which as described may be qualitatively different when extrapolating to a reactor. However, high \frad{} at relatively low \nsepoverng{} may be accessible by using efficient high-Z radiators (Xe, Kr, etc), motivating an interesting future experimental direction. \EDA{} plasmas, as reported in other devices\cite{Hughes2011}, appear to be well-suited to divertor integration, reaching both high \nsepoverng{} and high \frad{}, and indeed \EDA{} is challenged instead by pedestal compatibility towards low \nustar{} operation \cite{Oyama2006}. \Imode{} appears comparably attractive in Fig. \ref{fig:div}(b), though recent work casts doubt on its ability to tolerate the impurity seeding needed to increase \frad{} \cite{Reinke2019}.  Advancing \negd{} plasmas towards higher \frad{} and \nsepoverng{} stands out as a fertile ground for future study, with little experimental data yet also no known mechanism to prohibit robust divertor integration with a relatively high performance core, as will be further discussed.

%(though still below recently proposed stability limits \cite{Eich2018}.
% discusses ITER

% ---- ---- ---- ---- -------- ----  ---- ---- ---- ---- ---- ---- ---- ---- ----
% ---- ---- ---- ---- -------- ----  Discussion    ---- ---- ---- ---- ---- ----
% ---- ---- ---- ---- -------- ----  ---- ---- ---- ---- ---- ---- ---- ---- ----

\section{Discussion and Conclusions}
\label{sec:disc}

\renewcommand{\arraystretch}{1.2}
\begin{table}
\begin{tabular}{|r|c|c|c|c|c|l|}
\hline
                            & \EDA{}             & \Imode{}                               & \negd{}                                &  \QH{} & \rmp{}             & Figure                                 \\                             \hline
High \Ptot{} [MW]       & \cellcolor{red}4    & \cellcolor{yellow}7                       & \cellcolor{green}15                    & \cellcolor{green}10                        & \cellcolor{green}11 & \ref{fig:basics}(d)   \\                             \hline
Low \PoverPLH{}    & \cellcolor{green}$<$1  & \cellcolor{green}1                      & \cellcolor{green}$<$1                      & \cellcolor{green}1 & \cellcolor{yellow}2 & \ref{fig:basics}(d)   \\                            \hline
Low \Tinj{} [Nm]       & \cellcolor{green}0  & \cellcolor{green}0                      & \cellcolor{green}0                      & \cellcolor{green}0                          & \cellcolor{red}2    & \ref{fig:basics}(c,f) \\                            \hline
Low \qnf{}         & \cellcolor{red}4.5    & \cellcolor{yellow}3.5 & \cellcolor{green}$<$3 & \cellcolor{green}3                          & \cellcolor{green}3  & \ref{fig:basics}(a,e) \\                            \hline
Low \nustar{}      & \cellcolor{red}3    & \cellcolor{yellow}1.5 &   \cellcolor{red}2                                               & \cellcolor{green}$<$.1                          & \cellcolor{green}$<$.1  & \ref{fig:basics}(e) \\                            \hline
%High \noverng{} & \cellcolor{green}  & \cellcolor{yellow}                     & \cellcolor{green}                      & \cellcolor{green}                            & \cellcolor{red}    & \ref{fig:basics}(c)    \\                            \hline
High \betan{}      & \cellcolor{red}1.4    & \cellcolor{yellow}1.6 & \cellcolor{green}3                      & \cellcolor{green}2.7                          & \cellcolor{green}2.5  & \ref{fig:Hfactors}(a) \\                            \hline
Low \Zeff{}        & \cellcolor{green}1.5 & \cellcolor{green}1.3                      & \cellcolor{green}1.5                   & \cellcolor{red}$>$2                            & \cellcolor{yellow}1.8 & \ref{fig:altcorr}(c) \\                            \hline
High \echratio{}   & \cellcolor{green}.9  & \cellcolor{green}.8                      & \cellcolor{green}.9                      & \cellcolor{green}.7                          & \cellcolor{yellow}.4 & \ref{fig:ECH}(a)  \\     \hline
High \nsepoverng{} & \cellcolor{yellow}.25  & \cellcolor{yellow}.2                     & \cellcolor{yellow}.29                      & \cellcolor{red}.15                            & \cellcolor{red}.15    & \ref{fig:div}(a,b) \\                            \hline
High \frad{}       & \cellcolor{green}.7  & \cellcolor{yellow}.6                     & \cellcolor{yellow}.5                     & \cellcolor{red}.4                            & \cellcolor{red}.4    & \ref{fig:div}(b)  \\ \hline
\hline
High \snyder{}       & \cellcolor{red}$<$1.5  & \cellcolor{red}$<$1.8                     & \cellcolor{yellow}$<$2.6                     & \cellcolor{green}$<$4.4                            & \cellcolor{green}$<$4.6    & \ref{fig:abs}(c)  \\ \hline
\end{tabular}
\caption{Operational boundaries and peak performance for no-ELM regimes as observed in DIII-D, with numeric limits included alongside figures from which these values are extracted. Colors qualitatively indicate presently demonstrated access: red for no access, yellow for marginal access, and green for robust access. Access is not necessarily achieved simultaneously.}
\label{tab:access}
\end{table}

% Intro / summary
This work has endeavored to document the observed operational boundaries as well as the plasma performance within those boundaries for stationary plasmas without ELMs in DIII-D. These operational boundaries and their compatibility with various regimes are summarized in Tab. \ref{tab:access}. Almost all criteria listed in Tab. \ref{tab:access} must be simultaneously achieved in an integrated burning plasma scenario, with few exceptions: perhaps \qnf{} need not be so low, perhaps \betan{} or \PoverPLH{} need not be so high, perhaps \nustar{} need not be low in a \negd{} scenario, etc. However, it can be clearly seen that no regime is yet able to demonstrate comprehensive reactor compatibility. While this table does not detail the observed performance, it is generally found that where access is readily achieved (green color in Tab. \ref{tab:access}), relatively high performance (as compared to other no-ELM regimes) is also generally found. It also bears repeating that the identified correlations of performance with carbon content and rotation (Fig. \ref{fig:altcorr}) challenge extrapolability of all regimes and motivate further targeted studies. Also worth noting is that assessment of non-inductive fraction needed for steady-state tokamak operation is not in the scope of this work, largely because extracting the non-inductive fraction is considerably more challenging than the metrics already included. With this as pre-amble, a few salient further issues are now discussed.

% Density / nustar (boring)
\paragraph*{Pedestal-Divertor Integration:} Challenging reactor-relevance of some no-ELM regimes is the well-known inaccessibility (in present tokamaks) of a high-performance (low \nustar{}) pedestal simultaneous with a highly dissipative (high \nsep{} \& \frad{}) divertor state. Indeed, a key divergence is found between high-performance pedestal regimes (\QH{}, \rmp{}) vs. regimes that favor dissipative divertors (\EDA{}). Connecting these two extremes requires no-ELM regime study in high \pped{} conditions, which in principle enables low \nustar{} ($\propto$ \nped{}$^3$/\pped{}$^2$) at high \nped{}. The connection of \nped{} to \nsep{} is indirect, and sensitive to the no-ELM regime [as evidenced by Fig. \ref{fig:div}(a)] as well as the opacity of the pedestal to neutral fueling\cite{Mordijck2020}. High \pped{} is predicted (based on peeling-ballooning stability calculations \cite{Snyder2011}) to be accessible via large size, high-field, and/or strong shaping. However, the interaction of these engineering parameters with the specifics of each no-ELM regime access criteria may complicate this picture, as seen for example with \Bt{} thresholds for \Imode{} plasmas\cite{Hubbard2017} or shape thresholds for \rmp{} plasmas\cite{Shafer2020IAEA}. As such, pursuing no-ELM regime integration studies at maximal \pped{} is highlighted as an important pedestal physics research direction that motivates upgrades to existing facilities \cite{Buttery2019}. This question also forms a key scientific motivation of the ITER project, whose primary goal will be integrating a high-performance burning plasma core (high \lawson{}, low \nustar{}) with a dissipative divertor solution (high \nsep{} \& \frad{}).

% Power vs LH and Near-threshold Regimes
\paragraph*{Extrapolation of Power and the Role of \Bt{}:} A second challenge lies with uncertainty in how to extrapolate \Pnet{} after a pedestal is established. \PLH{}-like normalizations ($\propto$\navg{}$S$\Bt{}, where $S$ is the plasma surface area) well-predicts \Hmode{} access, qualitatively consistent with the need for a critical ExB velocity to trigger the shear-suppression of edge turbulence \cite{Ryter2014,Cavedon2020}. However, it is not clear whether this scaling is appropriate after the pedestal is formed, as opposed to alternate scalings such as \Pnet{}/$S$ (heat flux), or \Pnet{}/\navg{}$S$ (heat flux per particle). This question directly relates to the impact of \Bt{} in the extrapolation of no-ELM regime solutions, and more broadly in how to predict the pedestal transport without ELMs. For example, it is well documented that an upper limit in \Pnet{} exists where the \EDA{}, \Imode{}, and \Lmode{} regimes transition to ELMing \Hmode{} \cite{Hughes2011,Hubbard2017,Martin2008}. This is troubling for \EDA{} and \Imode{} since regime access is meant to increase performance and increase the fusion power contribution to \Pnet{}, thus risking a dynamically unstable situation. Furthermore, much of the \EDA{}, \Imode{}, and \Lmode{} regimes findings of this study can be linked to the intolerance to \Pnet{} of these regimes in DIII-D, where \PLH{} is only about 2-3 MW. In contrast, the tolerance of \QH{}, \rmp{}, and \negd{} plasmas to \Pnet{} significantly opens their operating domain in many directions. If \PoverPLH{} is the relevant power normalization, then the high \Pnet{} regimes are clearly overpowered (\PoverPLH{}$\gg$1) in DIII-D. However, if the relevant normalization is \Pnet{}/$S$, the high \Pnet{} regimes are in-line with expectations in future tokamaks. The key difference between these extremes is thus whether \Bt{} enters in the normalization of power. These questions motivate dedicated study of no-ELM regimes over as wide a range of \Bt{} as possible, to decouple the no-ELM regime access from \PLH{} as much as possible, though naturally indirect effects (such as changes in \nustar{}) will complicate these studies. Once again, ITER will provide an excellent platform to resolve these issues as it will operate at high \Pnet{}/$S$ yet low \PoverPLH{}. Similarly, the mid-size (low $S$) yet very high-\Bt{} SPARC tokamak\cite{Creely2020} will provide a yet clearer separation of these competing normalizations, with DIII-D like high \Pnet{}/$S$ yet ITER-like low \PoverPLH{}.

% Negative triangularity discussion
\paragraph*{Negative Triangularity:} The above challenges are not resolvable with presently operating tokamaks. Interestingly, the \negd{} regime side-steps the aforementioned integration issues and in principle can achieve fully and unambiguously integrated no-ELM scenarios in mid-scale tokamaks, though likely at the expense of lower absolute performance. The low \nustar{} (high \pped{}) challenge is simply abandoned, with the core contribution to \pres{} instead relied on to recover high performance. Questions about how to normalize power and the importance of \PLH{} are similarly side-stepped, because there is no desire for \Hmode{}. The only `access' question is whether \Hmode{} will be robustly avoided, and to what maximum \PoverPLH{}. Emerging work indicates the \Hmode{} suppression is robust in \negd{} due to MHD ballooning edge stability considerations \cite{Saarelma2020}, supported by the observation of \PoverPLH{}$>$6 shown in Fig. \ref{fig:basics}(d). The \negd{} regime thus relieves the physics risk from integration issues but increases the burden on the core confinement to achieve sufficiently high plasma performance despite low \pped{}. This work shows that considering normalized performance (Fig. \ref{fig:Hfactors}), \negd{} plasmas are already comparable to the other no-ELM regimes. Absolute performance (Fig. \ref{fig:abs}) is considerably lower, but points to raising \elong{} and \Ip{} to access high \IaB{}, benefitting from the tolerance to \Ptot{} expected. Operation at high \In{} and high \Ptot{} is also a compelling direction to further raise normalized performance without ELMs. This work provides ample basis to quantify any future no-ELM scenario achievements in performance and integration, \negd{} or otherwise.
\vspace{-10 pt}
\section*{Acknowledgments}

This work summarizes and reviews progress over nearly two decades of DIII-D experimental work. As such, the central acknowledgement is to the many talented individuals comprising the DIII-D team who worked diligently to create the plasmas here documented. To the degree possible, the constituent studies and authors have been recognized via citation. Useful targeted discussions relating to the scope of this study were provided by: P. Snyder, D. Ernst, M. Knolker, A. Garofalo, B. Grierson, A. Jarvinen, F. Laggner. Constitutent databases incorporated into this study were provided by: T. Evans, K. Burrell, T. Osborne, M. Austin. Further discharges of interest were highlighted by: X. Chen, T. Abrams, D. Ernst, J. Hughes, A. Marinoni. D. Ernst improved the impurity concentration analysis in wide-pedestal QH mode discharges.

{\small{
This material is based upon work supported by the U.S. Department of Energy, Office of Science, Office of Fusion Energy Sciences, using the DIII-D National Fusion Facility, a DOE Office of Science user facility, under Award(s) DE-FC02-04ER54698. Disclaimer: This report was prepared as an account of work sponsored by an agency of the United States Government. Neither the United States Government nor any agency thereof, nor any of their employees, makes any warranty, express or implied, or assumes any legal liability or responsibility for the accuracy, completeness, or usefulness of any information, apparatus, product, or process disclosed, or represents that its use would not infringe privately owned rights. Reference herein to any specific commercial product, process, or service by trade name, trademark, manufacturer, or otherwise does not necessarily constitute or imply its endorsement, recommendation, or favoring by the United States Government or any agency thereof. The views and opinions of authors expressed herein do not necessarily state or reflect those of the United States Government or any agency thereof.}}

% ---- ---- ---- ---- -------- ----  ---- ---- ---- ---- ---- ---- ---- ---- ---- ----
% ---- ---- ---- ---- -------- ----  Metrics Appendix   ---- ---- ---- ---- ---- ----
% ---- ---- ---- ---- -------- ----  ---- ---- ---- ---- ---- ---- ---- ---- ---- ----

\appendix
\section{Comparison of Performance Metrics}
\label{sec:metrics}

\begin{figure*}
\begin{subfigure}{1\textwidth}
\begin{subfigure}{0.325\textwidth}
\centering
\includegraphics[width=1\textwidth]{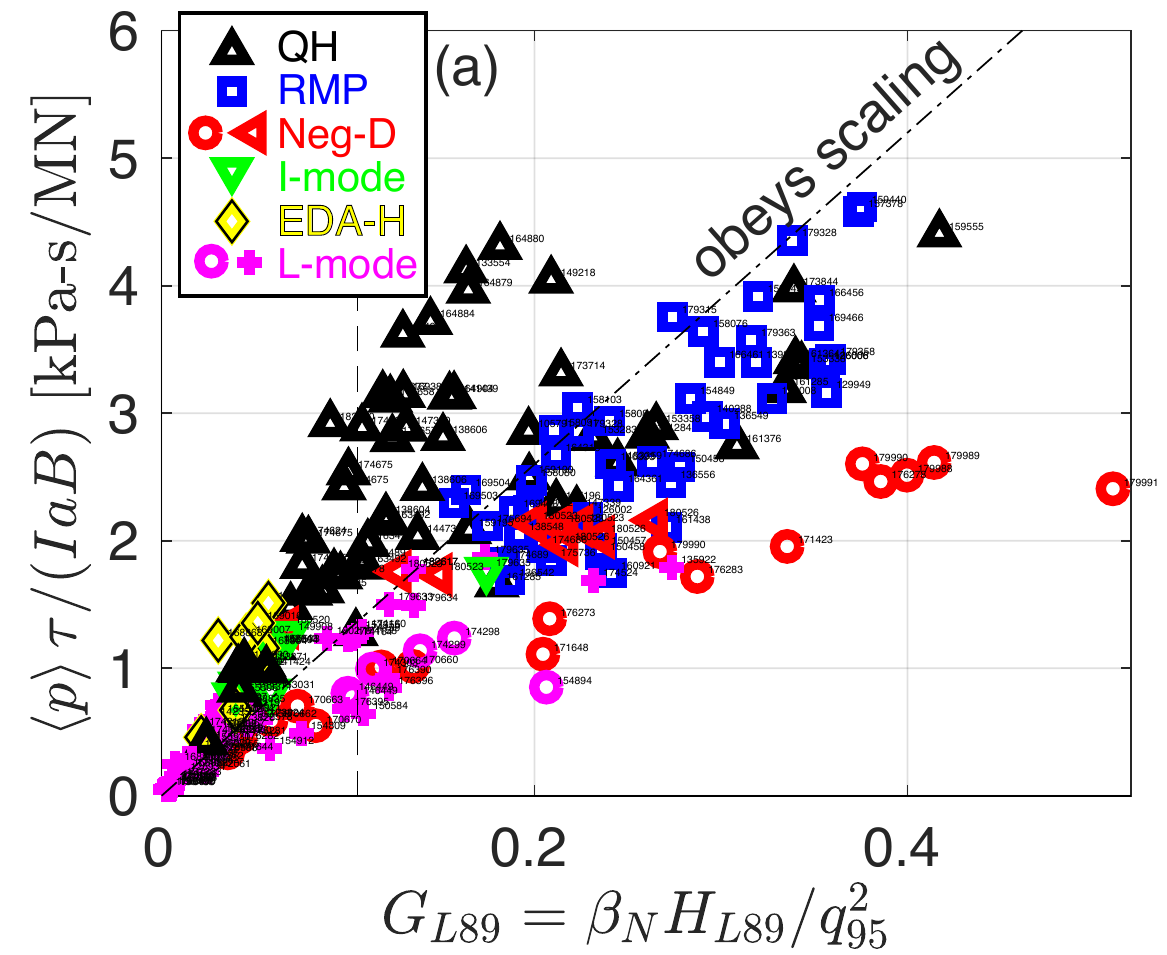}
\end{subfigure}
\begin{subfigure}{0.325\textwidth}
\centering
\includegraphics[width=1\textwidth]{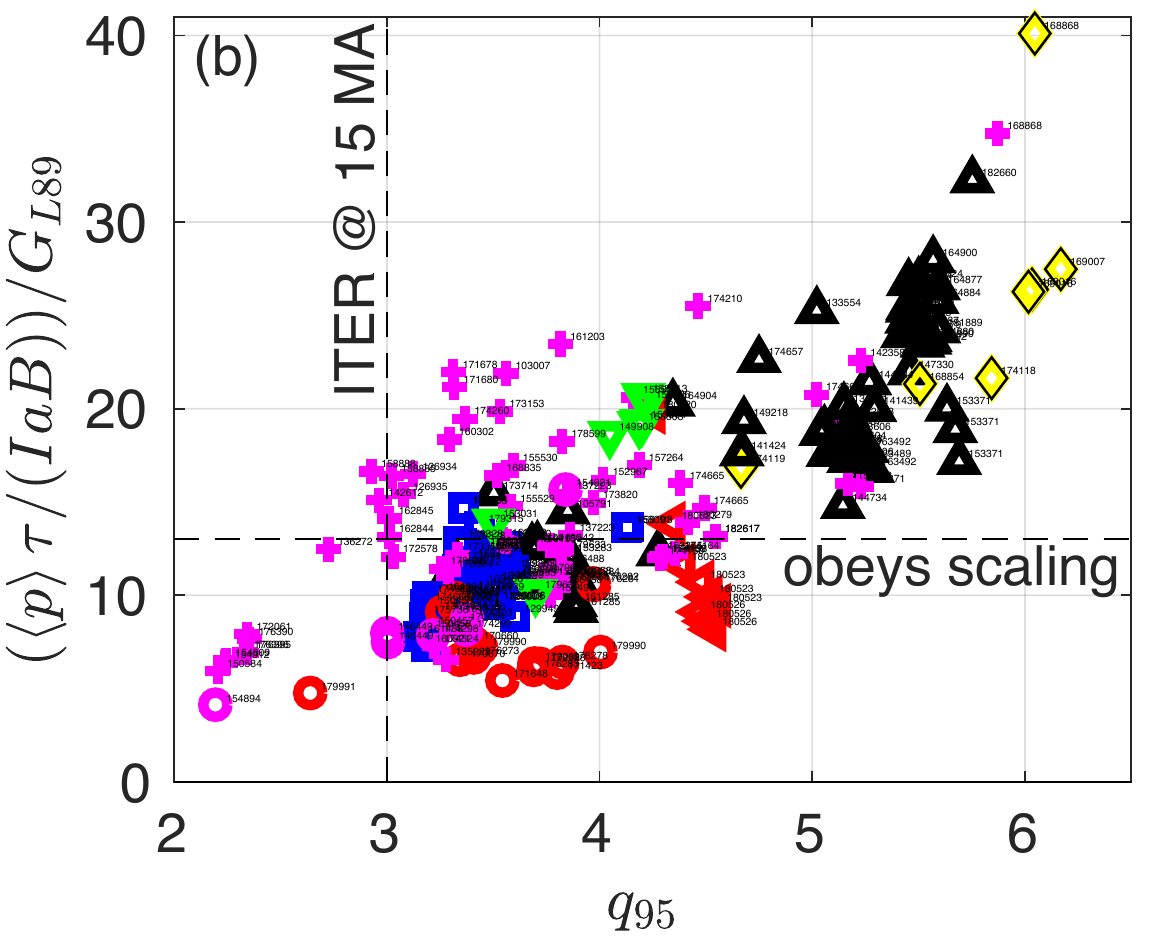}
\end{subfigure}
\begin{subfigure}{0.325\textwidth}
\centering
\includegraphics[width=1\textwidth]{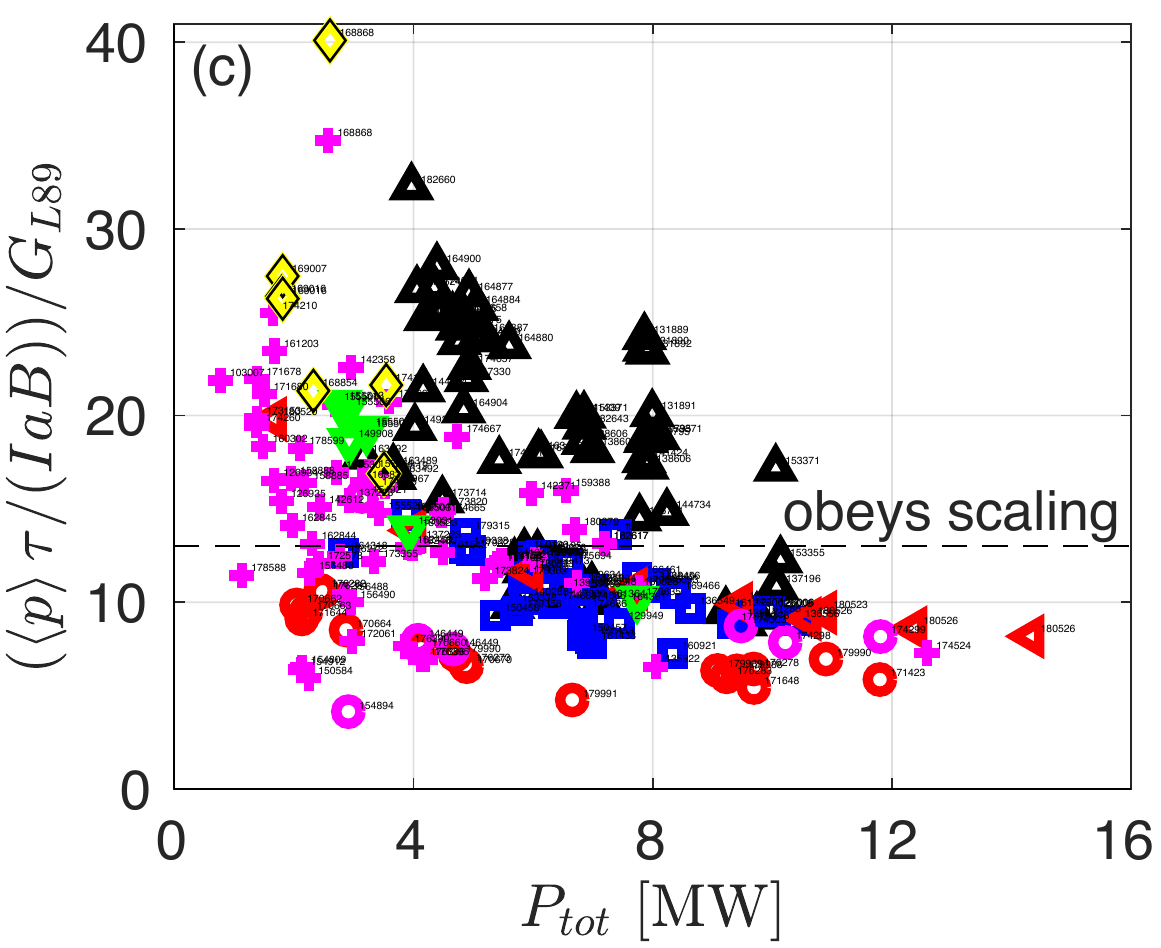}
\end{subfigure}
\end{subfigure}
\vspace{-5 pt}
\caption{Comparison of fusion performance metrics, comparing (a) \HL{} scaling law (\GL{}) against normalized triple product (\snyder{}. The ratio of these metrics is seen to correlate with (b) the safety factor (\qnf{}) and (c) the total injected power (\Ptot{}).}
\label{fig:metrics}
\end{figure*}

A short discussion of the observed differences between plasma performance metrics discussed in this work is appended, focusing on the difference between fusion performance metrics based on the \HL{} scaling law (\GL{}) and the normalized triple product (\snyder{}). Note a similar study of \GH{} and \GL{} (not shown) found no meaningful distinction between these two performance metrics. In contrast, imporatnt distinctions between \GL{} and \snyder{} are pictured in Fig. \ref{fig:metrics}. Comparing \GL{} directly against \snyder{} in Fig. \ref{fig:metrics}(a), each regime population is found to be largely co-linear, indicating both metrics are capturing relative trends well. However, some no-ELM regime distinction is found. \negd{} plasmas perform very well in \GL{} as compared to \snyder{}, but this is merely representative of the penality in \qnf{} to be paid for \negd{} plasmas, which artificially inflates \GL{} as it is proportional to $q_{95}^{-2}$, thus further motivating the use of \snyder{} throughout this work. Conversely, strongly shaped \QH{} plasmas are penalized in \GL{} as compared to \snyder{}. The ratio of \snyder{} to \GL{} is found to be correlated with \qnf{}, shown in Fig. \ref{fig:metrics}(b), which shows the positive effect of strong shape as well as provides evidence that low \qnf{} is over-rewarded in the \GL{} (and \GH{}) scaling laws. A second correlation with \Ptot{} is highlighted, which indicates a gradually decreasing ratio of \snyder{} to \GL{} with increasing \Ptot{}, suggesting that high power is over-rewarded in \GL{}. These findings highlight the value of \snyder{}, which does not impose a scaling law to interpret the plasma performance but rather lets the triple product  (\lawson{}) speak for itself.

\section*{References}
% RUN WITH APS STYLE TO GENERATE FILE
\bibliographystyle{iopart-num}

% APS:
%\bibliography{library}

\providecommand{\newblock}{}

% ---- IOP CHANGE
%\section*{References}

% COPY HERE

\end{document}